\documentclass[aps,amsmath,amssymb,pra,twocolumn,footinbib,superscriptaddress,floatfix,10pt]{revtex4-2}

\usepackage{amsthm}

\usepackage{soul}

\usepackage{xcolor}
\usepackage{color}
\usepackage{graphicx}% Include figure files
\usepackage{dcolumn}% Align table columns on decimal point
\usepackage{amssymb}     
\usepackage{bm}% bold math
\usepackage{footmisc}
\usepackage{mathtools}
\usepackage{framed}
\usepackage{mathrsfs}
\usepackage{lipsum}
\usepackage{makecell}
\usepackage{physics}
\usepackage{braket}
\usepackage{diagbox}
\usepackage{hyperref}
\usepackage{multirow} 
\definecolor{darkblue}{rgb}{0,0,0.5}
\hypersetup{
	colorlinks=true,
	linkcolor=black,
	filecolor=blue,
	citecolor=darkblue,  
	urlcolor=black,
}

\newcommand{\N}{\mathbb{N}}
\newcommand{\Z}{\mathbb{Z}}
\newcommand{\R}{\mathbb{R}}
\newcommand{\C}{\mathbb{C}}

\newcommand{\calH}{{\cal H}}

\def\d{{\rm d}}

\def\>{\rangle}
\def\<{\langle}

\newcommand{\bs}[1]{\boldsymbol{#1}}
\newcommand{\map}[1]{\mathcal{#1}}
\usepackage{chngcntr}   

\newcommand{\beginsupplement}{
	\setcounter{section}{0}  \renewcommand{\thesection}{Supplemental\ Note\ \Roman{section}}
	\renewcommand{\appendixname}{} 
	\counterwithout{equation}{section} 
	\setcounter{equation}{0}   \renewcommand{\theequation}{S\arabic{equation}}
	\counterwithout{theorem}{section}
	\setcounter{theorem}{0}
	\renewcommand{\thetheorem}{S\arabic{theorem}}
}

\newtheorem{theorem}{Theorem}
\renewcommand{\thetheorem}{\arabic{theorem}}

\newtheorem{conjecture}[theorem]{Conjecture}
\newtheorem{corollary}[theorem]{Corollary}

\newtheorem{definition}[theorem]{Definition}

\newtheorem{lemma}[theorem]{Lemma}

\newtheorem{proposition}[theorem]{Proposition}
\newtheorem{remark}[theorem]{Remark}

\newcommand*{\XZ}[1]{{\color{blue} [XZ: #1]}}

\newcommand{\QZ}[1]{{{\textcolor{red}{#1}}}}

\begin{document}

	\preprint{APS/123-QED}
	
	%Paradigms, resource theory, and operational advantages of genuine non-Gaussian entangled states
	\title{Complexity of quantum tomography from genuine non-Gaussian entanglement
		%: quantum correlations beyond Hong--Ou--Mandel and boson sampling
	}
	
	%Complexity from quantum correlations beyond Hong-Ou-Mandel
	%Complexity from genuine non-Gaussian entanglement: quantum correlations beyond Hong-Ou-Mandel
	
	\author{Xiaobin Zhao}
	\email{xzhao721@usc.edu}
	\affiliation{Ming Hsieh Department of Electrical and Computer Engineering,
		University of Southern California, Los Angeles, CA 90089, USA}

	\author{Pengcheng Liao}
	\affiliation{Ming Hsieh Department of Electrical and Computer Engineering,
		University of Southern California, Los Angeles, CA 90089, USA}

	\author{Francesco Anna Mele} 
	\affiliation{Scuola Normale Superiore, Piazza dei Cavalieri, 7, Pisa, 56126, Italy}
	
	\author{Ulysse Chabaud}
	\affiliation{DIENS, \'{E}cole Normale Sup\'{e}rieure, PSL University, CNRS, INRIA, 45 rue d’Ulm, Paris, 75005, France}
	
	\author{Quntao Zhuang}
	\email{qzhuang@usc.edu}
	\affiliation{Ming Hsieh Department of Electrical and Computer Engineering,
		University of Southern California, Los Angeles, CA 90089, USA}
	\affiliation{Department of Physics and Astronomy, University of Southern California, Los Angeles, CA 90089, USA}
	
	\nopagebreak
	
	\begin{abstract}

		Quantum state tomography, a fundamental tool for quantum physics, usually requires a number of state copies that scale exponentially with the system size, owing to the intricate quantum correlations between subsystems. We show that, in bosonic systems, the nature of correlations indeed fully determines this scaling.  Motivated by the Hong–Ou–Mandel effect and Boson-sampling, we define Gaussian-entanglable (GE) states, produced by generalized interference between separable bosonic modes.  GE states greatly extend the Gaussian family, encompassing arbitrary separable states, multi-mode Gottesman–Kitaev–Preskill codes, entangled cat states, and Boson-sampling outputs---resources for error correction and quantum advantage.  Nonetheless, we prove that an $m$-mode pure GE state is learnable with only \textbf{poly}$(m)$ copies, by providing an explicit protocol involving only heterodyne detection and classical post-processing. For states outside GE, we introduce an operational monotone---the minimum number of ancillary modes required to render them GE---and prove that it exactly captures the exponential overhead in tomography.  As a by-product, we show that deterministic generation of NOON states with $N\ge3$ photons by two-mode interference is impossible.

	\end{abstract}
	
	\maketitle

	\section{Introduction}
	
	%{\noindent\bf \Large Introduction}\\
	Quantum state tomography~\cite{smithey1993measurement,lvovsky2009continuous,anshu2024survey} aims to provide a classical description of quantum states from experimental data. As a fundamental task in physics and quantum information processing, it not only facilitates the tests of quantum physics, but also enables quantum device certification and benchmarking~\cite{eisert2020quantum}. However, quantum state tomography is a challenging task in general---each measurement requires a new copy of the state as measurement collapses the original quantum state. To learn an approximate description of a quantum system with $n$ qubits, the required number of samples is exponential in $n$~\cite{o2016efficient,kueng2017low}. In the case of bosonic systems, the infinite-dimensional Hilbert space cast additional challenge to the problem of tomography~\cite{mele2024learning}.

	\begin{figure*}[t]
		\centering
		\includegraphics[width=0.69\linewidth,trim=0 0 0 0,angle=0,clip]{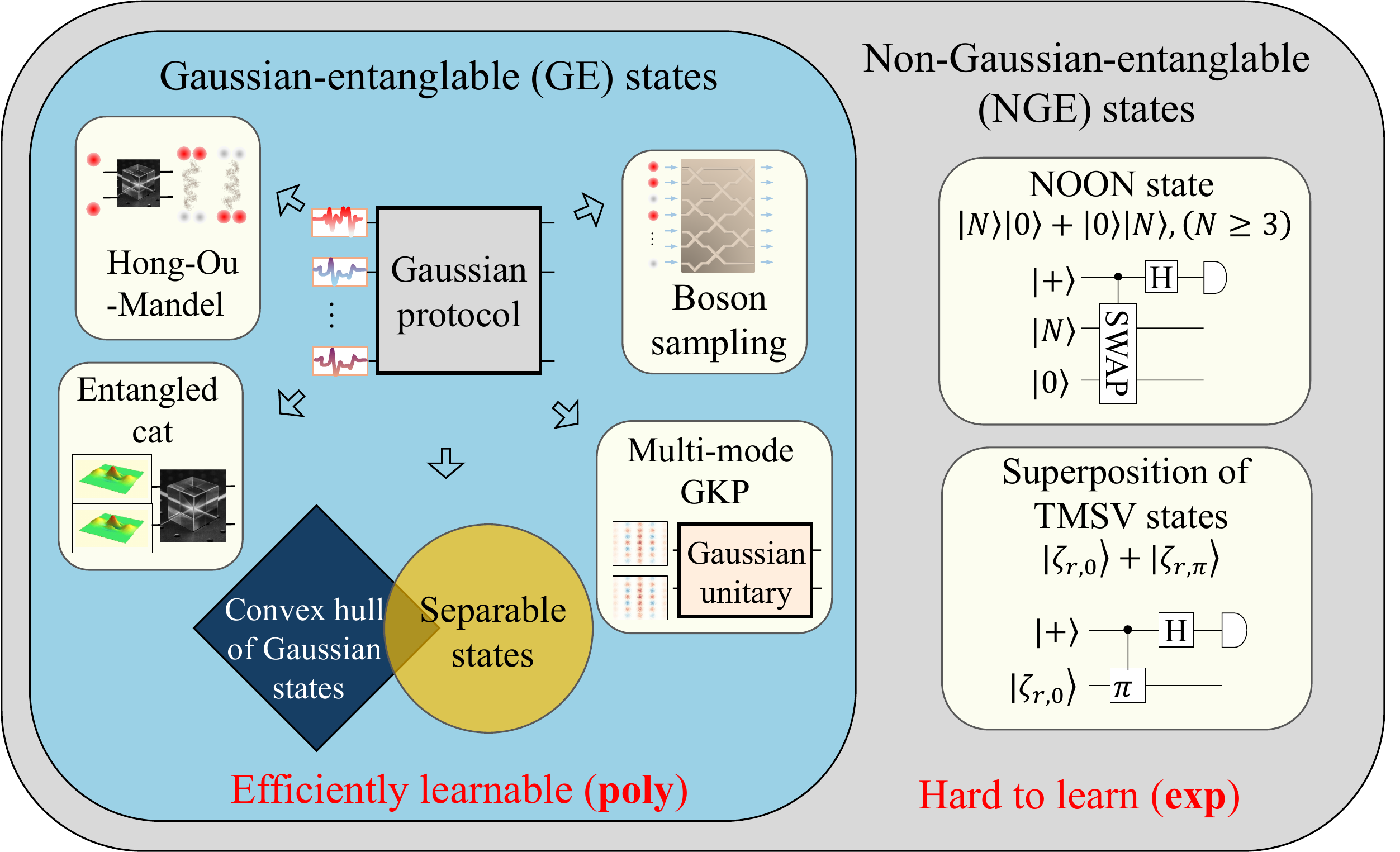}
		\caption{\textbf{Quantum correlations and the complexity of tomography.} Gaussian-entanglable (GE) state are states generated by performing Gaussian protocols on (possibly non-Gaussian) separable input states. The class of GE states includes the output state of the Hong--Ou--Mandel experiment, i.e.~two photons input to a beam-splitter where photon bunching effect emerges, and the output states of the Boson sampling protocol, where single photons together with vacuum states are input into a multi-mode linear interferometer. Other examples include entangled cat states, obtained by applying a beam-splitter to two identical cat states, and multi-mode GKP states, generated by performing Gaussian unitaries on single-mode GKP states. In addition, the class of GE states encompasses both the convex hull of Gaussian states and the entire set of separable states. Here we prove that the NOON state $|N\>|0\>+|0\>|N\>$ with $N\ge 3$, superposition of TMSV states $|\zeta_{r,0}\>+|\zeta_{r,\pi}\>$, and the two-mode arithmetic progression state (see Eq.~(\ref{NGE_state:example1})) do not belong to the set of GE states. These states are conventionally generated by applying controlled gates followed by post-selection. 
			At last, we show that the sample complexity in learning pure GE states scales as $\sim\textbf{poly}(m)$ {\color{black} regarding number of modes $m$;} %regarding system partition number $n$; 
			while NGE state generally requires an exponential overhead $\sim \textbf{exp}(m)$. 
			\label{fig:concept}}
	\end{figure*}

	%Photons play an important role in revealing fundamental physics. In Young's 1801 double slit experiment, the concept of wave-particle duality is demonstrated, via interfering beams of photons.  
	
	The challenge in quantum state tomography arises from the complex nature of quantum correlations between subsystems: highly entangled states involving superpositions of a large number of states are generally hard to learn, while separable states with classical correlations are naturally easier to learn. At the same time, one may identify subsets of non-separable quantum states that are easier to learn. For instance, Refs.~\cite{mele2024learning,fanizza2025efficienthamiltonianstructuretrace,bittel2025optimalestimatestracedistance} developed efficient learning protocols for Gaussian bosonic quantum states~\cite{weedbrook2012gaussian,serafini2017quantum}---a family of experimentally friendly states---and its generalization of $t$-doped Gaussian states. Thus, the interplay between properties of quantum correlation and learning sample complexity in bosonic quantum systems is unclear. In this work, we approach the problem by considering how quantum correlations are generated from separable states.
	
	We begin by considering the example of the Hong--Ou--Mandel (HOM) effect~\cite{HOM_paper}, where two incoming identical and separable single photons interfere on a beam-splitter to deterministically generate an entangled output, manifesting a bunching effect between indistinguishable photons. The output state $\sim \ket{20}+\ket{02}$ is the $N=2$ case of the family of NOON states $\sim\ket{N0}+\ket{0N}$, which are maximally entangled $N$-photon states useful to enhance quantum metrology~\cite{boto2000quantum,mitchell2004super}. Similar to HOM, in Boson sampling experiments~\cite{aaronson2011computational}, multiple modes of photons interfere on a network of beam-splitters creating highly multipartite entangled states, whose measurement statistics are hard to simulate classically, creating an opportunity to demonstrate quantum advantage~\cite{zhong2020quantum,madsen2022quantum}. 
	
	Inspired by the HOM effect~\cite{HOM_paper} and Boson sampling experiments~\cite{aaronson2011computational}, we %formulate
	{\color{black} define }Gaussian-entanglable (GE) states {\color{black} as states} produced by generalized interference described by Gaussian processes~\cite{weedbrook2012gaussian} between separable bosonic modes (see Fig.~\ref{fig:concept}). Despite the fact that GE states %enabling 
	{\color{black} enable }quantum computational advantage, we prove that an $m$-mode pure GE state is learnable with only \textbf{poly}$(m)$ samples, by providing an explicit protocol involving only heterodyne detection and classical post-processing. For non-GE states, we derive an operation-driven monotone---the number of additional modes necessary to reduce them to GE states. Moreover, we prove that this monotone exactly captures the exponential overhead in sample complexity of learning the states. 
	Besides their clear implications for the sample complexity of learning, our results also settle a long-standing open problem, namely that NOON states with $N\ge3$ photons cannot be generated deterministically by interfering an arbitrary product state, as they are not in the GE class. {\color{black} Earlier studies addressed only restricted cases involving number state inputs in linear optical settings~\cite{vanmeter2007general,parellada2023no,parellada2024lie}}.

	\ 
	
	\section{Results}
	
	%{\noindent\bf \Large Results}\\
	To characterize quantum correlations beyond HOM and Boson sampling in bosonic systems, we introduce non-Gaussian-entanglable (NGE) states: these are the quantum states which cannot be generated via a generalized interference experiment (see Fig.~\ref{fig:concept}), where separable quantum states are input to Gaussian protocols~\cite{weedbrook2012gaussian,serafini2017quantum}. Conversely, quantum states generated via generalized inteference experiments are Gaussian-entanglable (GE) states.
	GE states are operationally meaningful, especially considering the recent progress of multi-mode quantum state engineering~\cite{diringer2024,brady2024advances}. By definition, GE states include all Gaussian states and their convex mixtures~\cite{marian2013relative,genoni2008quantifying,genoni2010quantifying,zhuang2018resource,takagi2018convex,albarelli2018resource}. 
	In addition, they include states that are neither separable nor convex mixtures of Gaussian states, such as the output states of Boson sampling experiments~\cite{aaronson2011computational}, which involve identical photons scattering through a linear optical interferometer and whose measurement statistics is computationally hard to classically simulate. Moreover, the recently proposed multi-mode Gottesman--Kitaev--Preskill (GKP) states, with superior error correction performance over single-mode versions~\cite{royer2022encoding,wu2023optimal,conrad2022gottesman,brady2024safeguarding}, are GE---they can be engineered via applying Gaussian unitaries on single-mode GKP states~\cite{conrad2022gottesman,brady2024safeguarding}, reducing the challenge of multi-mode state engineering to preparing single-mode GKP states. On the contrary, NGE states will generally require a multi-mode non-Gaussian operation---or equivalently realized by Gaussian operations assisted with a sufficient amount of non-Gaussian ancilla~\cite{gottesman2001encoding}, creating a large overhead in quantum information processing. 
	
	%Therefore, The characterization of GE and NGE states is important for understanding multi-mode continuous-variable quantum systems.

	%\section{Review of previous approaches}

	%, which include all Gaussian protocols---a probabilistic mixture of Gaussian unitaries and Gaussian channels that are joint between the subsystems
	
	To sharpen the distinction between GE and NGE states, we develop a convex resource theory~\cite{chitambar2019} of genuine non-Gaussian entanglement. With GE states as the free states, the free operations are Gaussian protocols. Besides proving that NOON states with $N\ge 3$ are NGE, we provide other examples of NGE states, such as superposition of two-mode squeezed vacuum states.
	%Generated by applying Gaussian protocols on general separable states, the free resource of GE states is convex and includes all separable (potentially non-Gaussian) states as well as all Gaussian (potentially entangled) states (see Fig.~\ref{fig:schematic}c), and therefore represent a wide class of states beyond what is considered in previous resource theory of non-Gaussianity~\cite{zhuang2018resource}. 
	%As concrete examples, we show that the NOON state \cite{boto2000quantum} with a photon number above two, as well as the superposition of two two-mode squeezed vacuum states (TMSV), are resource states.  As a by-product, this formally proves that NOON state beyond $N=2$ cannot be generated by a beam-splitter on product states, making the $N=2$ case of the Hong–Ou–Mandel effect a special case. 
	Furthermore, we introduce and analyze two monotones to quantify genuine non-Gaussian entanglement, one based on entanglement entropy (the \textit{NG entanglement entropy}) and the other one based on the minimal number of ancillary modes necessary to convert an NGE state into a GE state (the \textit{GE cost}). We show that the sample complexity of tomography for pure GE states grows polynomially with the number of modes, while NGE states generally require an exponential overhead, quantified by their GE cost. This implies that output states in Boson sampling experiments are easy to learn, despite their measurement statistics being hard to sample from with a classical computer~\cite{aaronson2011computational}. 
	We also provide numerical verification of the tomography process. {\color{black} The main results of this work are summarized in Table \ref{tab:results}.}
	
	%\QZ{**small table in overview for main results; while states can be included in Fig 1 c}

	\begin{table}[t]
		\centering
		\color{black}
		\setlength{\tabcolsep}{4pt} 
		\renewcommand{\arraystretch}{1.6} 
		\begin{tabular}{|>{\raggedright}p{2.2cm}|>{\raggedright}p{2.7cm}|>{\centering\arraybackslash}p{2.5cm}|}
			\hline 
			\makecell{\\[-2.8em]Classification of \\ quantum states\\} & 
			\multicolumn{2}{c|}{\makecell{\\[-0.5em] Main results\\ }} 
			\\[-1.5em]
			\hline
			\hline 
			\multirow{5}{*}{\ \ \,\, GE states} 
			& \centering  Decomposition of pure GE states (Theorem \ref{lem:pureGEstate}) 
			& \multirow{6}{*}{ \makecell{Sample-efficient \\ algorithm for
					\\ learning \\ GE states\\ 
					(Theorem~\ref{theo:sample complexity of GE state})\\[2mm]\\  
			}} \\
			\cline{2-2}
			& \centering  Bounding the volume \\of GE state space \\(Corollary  \ref{theo:computational complexity}) & \\
			\cline{1-3}
			\multirow{6}{*}{\makecell{\\[-0.5em]
					\ \,\, NGE states}} 
			& \centering  NOON states are NGE (Supplemental note \ref{app:pure nge state}) 
			& \multirow{6}{*}{\makecell{     Exponential \\
					overhead \\in learning \\NGE states \\(Theorem~\ref{lem:connection GE cost and learning} and\\ Proposition~\ref{lem:sample complexity upper bound})}}\\
			\cline{2-2}
			&  \makecell[c]{  \\[-0.5em]  Establishment of \\resource\\monotones\\(Eqs.~(\ref{NG entropy}), (\ref{GE cost 5.1}))\\[0.5em]  }
			& \\
			\hline
		\end{tabular}
		\caption{\textbf{Main results of this work.}} 
		\label{tab:results}
	\end{table}

	\section{Continuous-variable systems and state learning}
	
	%\bigskip{\noindent{\bf Continuous-variable systems and state learning}}\\
	%\noindent
	Continuous-variable quantum systems model continuous degrees of freedom, such as optical fields~\cite{walls2007quantum}, mechanical motions of ions~\cite{fluhmann2019encoding}, and microwave cavities~\cite{blais2020quantum}. Owing to their information-rich infinite-dimensional Hilbert space, these systems are crucial for quantum communication~\cite{giovannetti2014ultimate}, quantum sensing~\cite{lawrie2019quantum,ganapathy2023broadband}, quantum error correction~\cite{gottesman2001encoding,brady2024advances} and quantum computing~\cite{larsen2021fault} (see Supplementary Note \ref{app:general introduction to CV systems}).
	
	While general states of continuous-variable quantum systems are unbounded in energy, we are interested in practically relevant states with energy constraints. In particular, we consider $m$-mode quantum states with finite energy moment constraint, $[{\rm Tr}{\hat{E}_m^k\hat{\rho}}]^{1/k}\le m E$, where the energy operator $\hat{E}_m=\frac{1}{4}\sum_{\ell=1}^m (\hat{q}_\ell^2+\hat{p}_\ell^2)$ is a quadratic function of the position and momentum quadratures $\hat{q}_\ell$ and $\hat{p}_\ell$ in natural units $\hbar=2$. We denote the set of $m$-mode pure states satisfying this $k$-th energy moment constraint as $\calH_{m,E}^k$.
	
	%Motivate volume here, 
	Learning a state within %an error 
	{\color{black} a trace distance error} of $\epsilon$ is equivalent to pinpointing it in state space within an $\epsilon$-ball {\color{black} with respect to the trace distance metric}. The number of such $\epsilon$-balls---the volume---of the state space that the unknown state belongs to can manifest the challenge of state learning. It has been shown in Ref.~\cite{mele2024learning} that, for pure states subject to $k$-th moment constraint, the volume of the state space grows double-exponentially with the system size, scaling as $\textbf{exp}\!\left[\left(\frac{E}{12\epsilon^{2/k}}\right)^m\right]$. 
	
	%it has been shown that the general moment-constrained state space has a large $\textbf{ exp}\left( \mathcal O\left[\left(\frac{E}{12\epsilon^{2/k}}\right)^m\right]\right)$ volume, leading to exponential sample complexity in general. 
	
	\section{Gaussian state and protocols}
	
	%\bigskip{\noindent{\bf Gaussian state and protocols}
		%\\
		%\noindent
		Among bosonic states and operations, Gaussian states (e.g., coherent states and squeezed states), and operations (e.g., beam-splitters, phase rotations and squeezing) are not only analytically tractable but also experimentally friendly~\cite{weedbrook2012gaussian,serafini2017quantum}. 
		
		%Here $D(\bs \xi ):=\exp(i\bs r^T \Omega \bs \xi)$ denotes the displacement operator, ${\bs r} =[\cdots,a_j+a^\dag_j, i(a^\dag_j-a_j),\cdots]^{\rm T}$ is the vector of quadrature operators with $a_j$ and $a_j^\dag$ being the annihilation and creation operator of the $j$-th mode, respectively, $\Omega=\left(\begin{matrix}0&1\\-1&0\end{matrix}\right)\otimes I$ is the symplectic form. 

		Gaussian states are specific types of quantum states with their Wigner characteristic function in a Gaussian form~\cite{holevo1975some,holevo2011probabilistic}; therefore, a Gaussian state $\hat{\rho}$ is uniquely specified by its mean 
		$\bs \xi:=\Tr[\hat{\bs r} \hat\rho ]$ and covariance matrix $
		V:=\frac 1 2 \Tr [\{(\hat{\bs r}-\bs \xi),(\hat{\bs r}-\bs \xi)^{\rm T}\}\hat \rho]$~\cite{Weedbrook_2012,serafini2017quantum}, where $\hat{\bs r} =[\hat q_1, \hat p_1,\cdots,\hat q_m,\hat p_m]^{\rm T}$ is the vector of quadrature operators. For instance, the %maximally entangled two-mode state---
		two-mode squeezed vacuum states (TMSV), defined as $\ket{\zeta_{r,\phi}}_{AB}=(\cosh r)^{-1}\sum_{k=0}^\infty e^{ik\phi}(\tanh r)^k \ket{k}_A\ket{k}_B$, have zero-mean and are uniquely determined by a covariance matrix that depends on the squeezing strength $r$ and phase angle $\phi$. Here $\ket{k}$ is the Fock state with photon number $k$. Gaussian quantum channels are completely positive trace-preserving maps that transform all Gaussian states into Gaussian states (see Supplementary Note \ref{app:property of Gaussian unitary}). In the following text, we shall use the notation $\mathbb G(\map H)$ and $\mathbb {GU}(\map H)$ to represent the set of all Gaussian quantum channels and unitaries on the Hilbert space $\map H$.  %An arbitrary $m$-mode Gaussian unitary, as indicated by Lemma \ref{lem:Bloch Messiah decomposition}, will be expressed as the product of two operations: $U^g=U_{\bs \alpha}  U_S $, where $U_{\bs \alpha}=\bigotimes_{j=1}^{m} D(\alpha_j)$ represents a multi-mode displacement operation with displacements $\bs \alpha=(\alpha_1,\cdots,\alpha_m)^T$, $U_S$
		%=U_{O_1}\left(\bigotimes_{j=1}^{m+m'} S(z_j) \right)U_{O_2}$ 
		%is a multi-mode Gaussian unitary corresponding to symplectic transform $S$ that does not change the displacement. We will use $U_O$ to represent a multi-port beam-splitter operation associated with symplectic orthogonal transform $O$ in phase space. 
		More generally, a Gaussian protocol~\cite{takagi2018convex} includes the following set of quantum operations: (1) Gaussian unitaries on the system or part of the system, (2) partial trace, (3) composition with ancillary vacuum states, (4) probabilistically applying the above operations. {\color{black} Note that we exclude Gaussian measurements followed by the conditional operations (1)–(4) from the set of Gaussian protocols we consider, because these are more challenging from an experimental standpoint and the resulting feed‑forward scheme might not reduce to a convex mixture of Gaussian channels.}
		%\label{def_Gaussian_protocol}
		%\end{definition}
		%\XZ{Here, $\map R_k$ forms up a GE cost vector $\map R=(\map R_1,\cdots,\map R_n)$ (that will be shown in Definition \ref{defi:GE cost}), which is related to the sample complexity of learning the corresponding set of states (to be shown in Theorem \ref{lem:connection GE cost and learning}).    }

		%Note that we excluded homodyne detection mentioned in  Ref.~\cite{takagi2018convex} as we focus on the deterministic production of quantum states, specifically looking at Gaussian protocols with identical input and output mode numbers. %Extension towards conditional state production can be considered in future works. On this ground, we can define bipartite GE states as follows: 

		Gaussian protocols (1)–(4) alone, without non-Gaussian ancilla, are limited in power---they are not universal for various tasks~\cite{eisert2002distilling,giedke2002characterization,fiuravsek2002gaussian,niset2009no,banaszek1998nonlocality,banaszek1999testing,lloyd1999quantum,bartlett2002universal}. As it turns out, even when allowing for non-Gaussian input states, the generalized Gaussian interference process in Fig.~\ref{fig:concept} can only generate a limited class of states, as we define below.
		
		%Therefore, characterizing quantum states and operations beyond the Gaussian class is important for understanding the resource needed for these tasks. Towards this end, various types of resource theories of non-Gaussianity have been developed~\cite{marian2013relative,genoni2008quantifying,genoni2010quantifying,zhuang2018resource,takagi2018convex,albarelli2018resource,chabaud2020stellar,chabaud2021}. 

		\iffalse 
		\subsection{Entanglement measure}
		
		Given a bi-partite state $\rho_{AB}$, the entanglement
		entropy is defined as the von Neumann entropy associated with this reduced state $\rho_A=\Tr_B [\rho_{AB}]$ \cite{bennett1996concentrating}, i.e., $\map E(\rho_{AB})= -\Tr \rho_A \log_2 \rho_A $.
		\fi

		%{\color{red}$\emph{\textbf{SEP}}(\map H_1\otimes \map H_2)$ denotes the set of separable states between systems 1 and 2
			
			\section{Gaussian-entanglable state}
			
			%\bigskip
			%{\noindent{\bf Gaussian-entanglable state}}
			%\\
			%\noindent
			Consider a bi-partition of the Hilbert space $\map H=\map H_A\otimes \map H_B$. A quantum state $\hat \rho_{AB}$ is (bi-partite) GE if it can be obtained by probabilistically applying a Gaussian protocol $\mathfrak G_x$ on an ensemble of (possibly non-Gaussian) separable state $\hat \sigma_{AB,y}\in \mathbb{{SEP}}(\map H_{A}\otimes \map H_{B})$ as $\hat \rho_{AB}= \sum_{x',y} p_{x',y}' \,\mathfrak G_{x'}\left(\hat \sigma_{AB,y}\right)$ with $p_{x',y}'$ being a joint probability distribution.
			Equivalently, a GE state can be expressed as: 
			\begin{align}\label{eq1}
				\hat \rho_{AB}=\sum_{x,y} p_{x,y} \,\map G_{x} \left(\hat \sigma_{AB,y}\right),
			\end{align}
			where $p_{x,y}$ refers to a joint probability distribution and $\map G_x\in \mathbb G(\map H_A\otimes \map H_B)$ denotes a Gaussian channel.%, $\mathbb{SEP}(\map H_A\otimes \map H_B)$ denotes the set of all separable states shared by systems  $A$ and $B$.
			
			%\XZ{We should highlight that, even with such a general operation, we still identify a pure NGE state. This might be the meaning of having Eq.~(1). In this sense, we don't need to go through all properties of all mixed states for Eq.~(1). }

			%\QZ{Allowing ancilla+Gaussian protocol is too general, because non-Gaussian gate can be implemented via Gaussian operations+non-Gaussian ancilla, if we allow too many nonGaussian ancilla, we can do universal gates. }\XZ{Ok, I agree. Besides, a restriction to state extension/purification and $m\to2$ Gaussian operations might make the story too complicated. }\QZ{yes, also Gaussian protocol allows teleportation, so one can locally have the targeted state and teleport the ancilla through.}

			%Note that given a GE state, the decomposition in Eq.~(\ref{eq1}) is not unique. 
			In practice, the GE state in Eq.~(\ref{eq1}) can be obtained by sampling from the joint probability distribution $p_{x,y}$, preparing the separable input state $\hat \sigma_{AB,y}$, and implementing the channel $\map G_x$. When $\map G_x$ is unitary, it corresponds to randomly applying Gaussian unitaries on separable states.
			%The sampling requires classical coupling with a set of classical  registers $\{|y'\>\<y'|\}$ in the form $\overline \sigma_{y'}=\sum_{y'} \sigma_{y'}\otimes |y'\>\<y'|$. Then, by applying a measure-and-operate process, one can realize a state in the form of Eq.~(\ref{eq1}). 
			By definition, the set of GE states is convex---a mixture of GE states is still GE. 
			%For instance, the above definition of GE state also includes the state $\rho_{AB}=\sum_{xy} p_{x,y} U^g_x \sigma_{AB,y} U^{g\dagger}_x$, obtained probablistically applying random Gaussian unitaries $U_x^g$.

			%\noindent Note that the state in Eq.~(\ref{eq6}) has a block-diagonal covariance matrix with non-degenerate symplectic eigenvalues. But it is a GE state.  
			
			As sketched in Fig.~\ref{fig:concept}, GE states can be non-Gaussian, and even not a convex mixture of Gaussian states. For instance, all entangled states generated by beam-splitter interference of non-Gaussian states such as number states are GE, e.g., the state $(\ket{0}_A\ket{2}_B+\ket{2}_A\ket{0}_B)/\sqrt{2}$ from the HOM effect~\cite{HOM_paper}. Similarly, the entangled cat state~\cite{diringer2024} $\propto \ket{\underline{\alpha}}_A\ket{\underline{\alpha}}_B+\ket{\underline{-\alpha}}_A\ket{\underline{-\alpha}}_B$ is GE, as it can be generated using a balanced beam-splitter on a product of cat state and vacuum, $\propto
			%(\ket{\underline{\sqrt{2}\alpha}}+\ket{\underline{-\sqrt{2}\alpha}})_A
			|{\rm cat}\>_{\sqrt 2\,\alpha,A}\otimes \ket{0}_B$. Here $\ket{\underline{\alpha}}$ is a coherent state with amplitude $\alpha$~\footnote{We have introduced the underline to distinguish between coherent states and number states, omitted in the special case of the vacuum state $\ket{0}$.}, and $|{\rm cat}\>_\alpha\propto |\underline{\alpha}\>+|\underline{-\alpha}\>$ denotes a cat state. Moreover, one can obtain a state $\propto |{\rm cat}\>_{\sqrt 2\,\alpha,A}|0\>_B+|0\>_A|{\rm cat}\>_{\sqrt 2\,\alpha,B}$ via interference between two identical cat states $|{\rm cat}\>_{\alpha}$.  Other examples include GE states generated by the multi-mode GKP encoding process: $|\psi\>_{m\to m+m'}=\hat U^g(|\psi\>_m\otimes |\text{GKP}\,\>^{\otimes m'})$,
			where $\hat U^g$ is a multi-mode Gaussian unitary, $|\psi\>_m$ is an arbitrary $m$-mode state, $|\text{GKP}\>$ is the canonical GKP state. Any multi-mode GKP state is also GE~\cite{royer2022encoding,wu2023optimal,conrad2022gottesman,brady2024safeguarding}. %{\color{blue} shall we replace by TABLE 1? }

			As we show, one can obtain any pure GE states from Gaussian unitary on a product state, as shown below. 
			\begin{theorem}[Decomposition of pure GE states]\label{lem:pureGEstate}
				Any pure GE state can be expressed in the following form: 
				\begin{align}\label{eq3}
					|\psi\>_{AB}= \hat U^{ g} |\phi\>_A|\phi'\>_B,
				\end{align}
				where $\hat U^{g}$ is a Gaussian unitary, $|\phi\>$ and $|\phi'\>$ are local states of systems $A$ and $B$. 
				
				\iffalse 
				Denote the covariance matrix of $|\psi\>_{AB}$ as $V_{AB}$. For any symplectic matrix $S$ such that $SV_{AB}S^T=\Lambda$ is the Williamson diagonal form, the corresponding Gaussian unitary $\hat U_S$ is equivalent to $\hat U^{ g}$ in Eq.~\eqref{eq3} up to a  beam-splitter $\hat U_O$ for an orthogonal matrix $O$ and a displacement operation $\hat D$ \cite{Weedbrook_2012}, i.e., $\hat U^g=\hat U_S \hat U_O \hat D$. If the symplectic eigenvalues of $V_{AB}$ are non-degenerate, we have 
				%\begin{align}\label{eq5}
				$\hat U^g = \hat U_S (\hat U^g_A\otimes \hat U^g_B)\hat D$,
				%\end{align} 
				where $\hat U^g_A\otimes \hat U^g_B$ %are local passive Gaussian unitaries. 
				{\color{black}is a tensor product of phase shifters. }
				\fi 
			\end{theorem}

			% \begin{proof}

				% Given Eq.~(\ref{eq1}), the GE state is defined as:
				% \begin{align}
					% \sigma_{AB}:= \sum_x \sum_j p_x  q_j U_x^{\rm g }\left(\rho_A\otimes \rho_B\right) U_x^{{\rm g}\dag }.
					% \end{align}
				% where $\{p_j\}$ and $\{q_j\}$ are probability distributions. The GE state is pure if the relation $\Tr[\sigma_{AB}^2]=1$ is fulfilled. In this case, the GE state must be a rank-one matrix: 
				% \begin{align}\label{eqa2a}
					% \sigma_{AB}&= U^{\rm g} \left(|\psi\>\<\psi|_A\otimes |\phi\>\<\phi|_B\right)U^{{\rm g}\dag}. 
					% \end{align}
				
				% Since $U^g$ and additional local Gaussian unitary symplectic diagonalize $V_{AB}$ from definition, then from Lemma~\ref{lemma_nmode_Williamson}, if the symplectic eigenvalues are different, $U^g$ (plus the local Gaussian unitaries) and $U_S$ are the same up to local phase rotations. Then we have proven the Lemma.

				% \end{proof}
			
			% For pure state $\sigma_{AB}$ produced from Eq.~\eqref{eq1}, we have
			% \begin{align}
				% 0=S(\sigma_{AB})\ge  \sum_{x,y} P_{x,y} S(\Phi_x(\rho_{AB,y})),
				% \end{align}  
			% and therefore $S(\Phi_x(\rho_{AB,y}))=0$ for all $x$ and $y$. Moreover, equality of concavity is only achieved when all states are equal $\Phi_x(\rho_{AB,y})=\ketbra{\phi}_{AB}$ is independent of $x$ or $y$. Then we have for pure GE state, $\ketbra{\phi}_{AB}=\Phi(\rho_{AB})$ can be produced by a single Gaussian protocol on a single state.
			
			% \QZ{can one prove that the state can also be generated from Gaussian unitary on separable state?}

			% Rely on the lemma in Ref.~\cite{mari2014quantum}, the only state that remains pure under pure loss is coherent state. 

			The proof, as elaborated in Supplementary Note~\ref{app:proof of the pure state theorem}, is based on the concavity of entropy, Gaussian unitary decompositions, and more importantly the fact that the only states that remain pure under multi-mode pure loss are coherent states, which generalizes a single-mode result that is crucial for solving the minimum entropy output conjecture and the Gaussian state majorization conjecture~\cite{mari2014quantum,de2018gaussian} (see Lemma \ref{lem:13_main} in Methods). 
			%\QZ{move lemma to method}

			%On this account, the proof of Eq.~(\ref{eq3}) can be briefly summarized as follows: Based on the concavity and non-negativity of quantum entropy, an arbitrary pure GE can be written in the form $|\Phi\>\<\Phi|_{AB}=\map G(|\psi\>\<\psi|_A\otimes |\phi\>\<\phi|_B)$ where $\map G$ is a Gaussian channel. Given the premise of pure output and the intrinsic properties of Gaussian unitary (shown in Supplementary Note \ref{app:property of Gaussian unitary}), the Gaussian unitary purification of $\map G$ must be a beam-splitter operation followed by local Gaussian unitaries on the $AB$ mode and the ancilla. Then, from Lemma \ref{lem:13_main}, either the condition for the initial state to be a coherent state or the requirement for the beam-splitter to act trivially as a product on AB and ancilla must be fulfilled. In both cases, the GE state can be written as a Gaussian unitary on pure product state. The proof of Eq.~(\ref{eq5}) can be quickly proved based on certain properties of Gaussian unitary (see Supplementary Note \ref{app:property of Gaussian unitary}). 

			GE states possess several distinctive properties (see details in Supplementary Note 
			\ref{app:general properties of GE states}). {\color{black}For a pure Gaussian state with a non‑degenerate symplectic spectrum, the condition that its covariance matrix is block‑diagonal (linear uncorrelated) is equivalent to the state being a product state (independent). This equivalence fails when the symplectic eigenvalues are degenerate.}
			%In particular, linear uncorrelatedness ({\color{blue}block-diagonal covariance matrix}) and independence ({\color{blue}product state}) are equivalent for pure GE states with non-degenerate symplectic eigenvalues, although this equivalence does not hold in the degenerate case.
			In addition, Gaussian protocols map GE states to GE states. At last, one can extend the definition of bi-partite GE states in Eq.~(\ref{eq1}) and Theorem \ref{lem:pureGEstate} to a multipartite $m$-mode case. 
			
			{\color{black}
				Furthermore, one can denote the covariance matrix of $|\psi\>_{AB}$ as $V_{AB}$. For any symplectic matrix $S$ such that $SV_{AB}S^T=\Lambda$ is the Williamson diagonal form, the corresponding Gaussian unitary $\hat U_S$ is equivalent to $\hat U^{ g}$ in Eq.~\eqref{eq3} up to a  beam-splitter $\hat U_O$ for an orthogonal matrix $O$ and a displacement operation $\hat D$ \cite{Weedbrook_2012}, i.e., $\hat U^g=\hat D\,\hat U_S \hat U_O $. If the symplectic eigenvalues of $V_{AB}$ are non-degenerate, we have $\hat U^g = \hat D\,\hat U_S (\hat U^g_A\otimes \hat U^g_B)$, where $\hat U^g_A\otimes \hat U^g_B$ is a tensor product of phase shifters. Hence, by first measuring the displacement vector and covariance matrix, then applying the partial inverse $\hat U_S^\dagger\hat D^\dagger$ of the Gaussian unitary, any GE state can be transformed into a passive-separable state. The next section outlines a tomography protocol for GE states that exploits this counter‑rotation step. }

			\section{Efficient learning of pure GE states}
			
			%\bigskip{\noindent{\bf Efficient learning of pure GE states}}\\
			As anticipated, pure GE states have reduced sample complexity for quantum state tomography, since one only needs to estimate the corresponding Gaussian unitary and local states by Theorem \ref{lem:pureGEstate}. In particular, we obtain the following result: 
			\begin{theorem}[Sample complexity of pure GE states] 
				\label{theo:sample complexity of GE state}
				Consider the set of $m$-mode pure states $\{|\psi\>\}$ obtained by applying a Gaussian unitary to a tensor product of single-mode states $\{\otimes_{j=1}^m|\psi_j\>\}$, where the states satisfy energy moment constraint, %$\psi_j\in \calH_{1,E}^2,\forall j$ and 
				$|\psi\>\in \calH_{m,E}^2$. 
				% Meanwhile, the state satisfy the following moment constraint: 
				% \begin{align}
					% \begin{cases}
						% \sqrt{\Tr\left[\left(\sum_{j=1}^m a^\dag_j a_j+\frac m 2 I \right)^2|\psi\>\<\psi|\right]}&\le m E_{\rm II}\\
						% \sqrt{\Tr\left[\left( a^\dag_j a_j+\frac{1}{2}\right)^2|\psi_j\>\<\psi_j|\right]}&\le E_{\rm II},\forall j. 
						% \end{cases}.
					% \end{align}
				Then, the tomography of such states requires $M$ copies such that:
				\begin{align}
					&\Theta\!\left[\frac{1-\delta}{\log E}\frac{mE}{\epsilon}\right]\le M\le \mathcal O \!\left[\textbf{\em poly}\!\left(m,E,\frac{1}{\epsilon},
					\log\frac{1}{\delta}\right)\right],
					\label{M_bounds_GE}%\\
					%&\mathcal O\left[\textbf{poly}\left(m^5, E_{\rm II}^2,\left(\frac 1 {\epsilon_{\rm ps}}\right)^6,\log \frac 1 {\delta_{\rm ps}} \right)\right]
				\end{align}
				where $\delta$ denotes the failure probability and $\epsilon$ refers to the trace distance between the input and the reconstructed state. 
			\end{theorem}
			The upper bound in Theorem \ref{theo:sample complexity of GE state} is demonstrated in Supplemental Note \ref{app:proof of theorem 2 for pure GE states}, while the lower bound is determined by the sample complexity of product states with moment constraints \cite{mele2024learning}. In particular, the tomography of pure GE states is sample-efficient.

\begin{figure}[t]
				\centering
				\includegraphics[width=0.94\linewidth,trim=11 80 0 0,angle=0,clip]{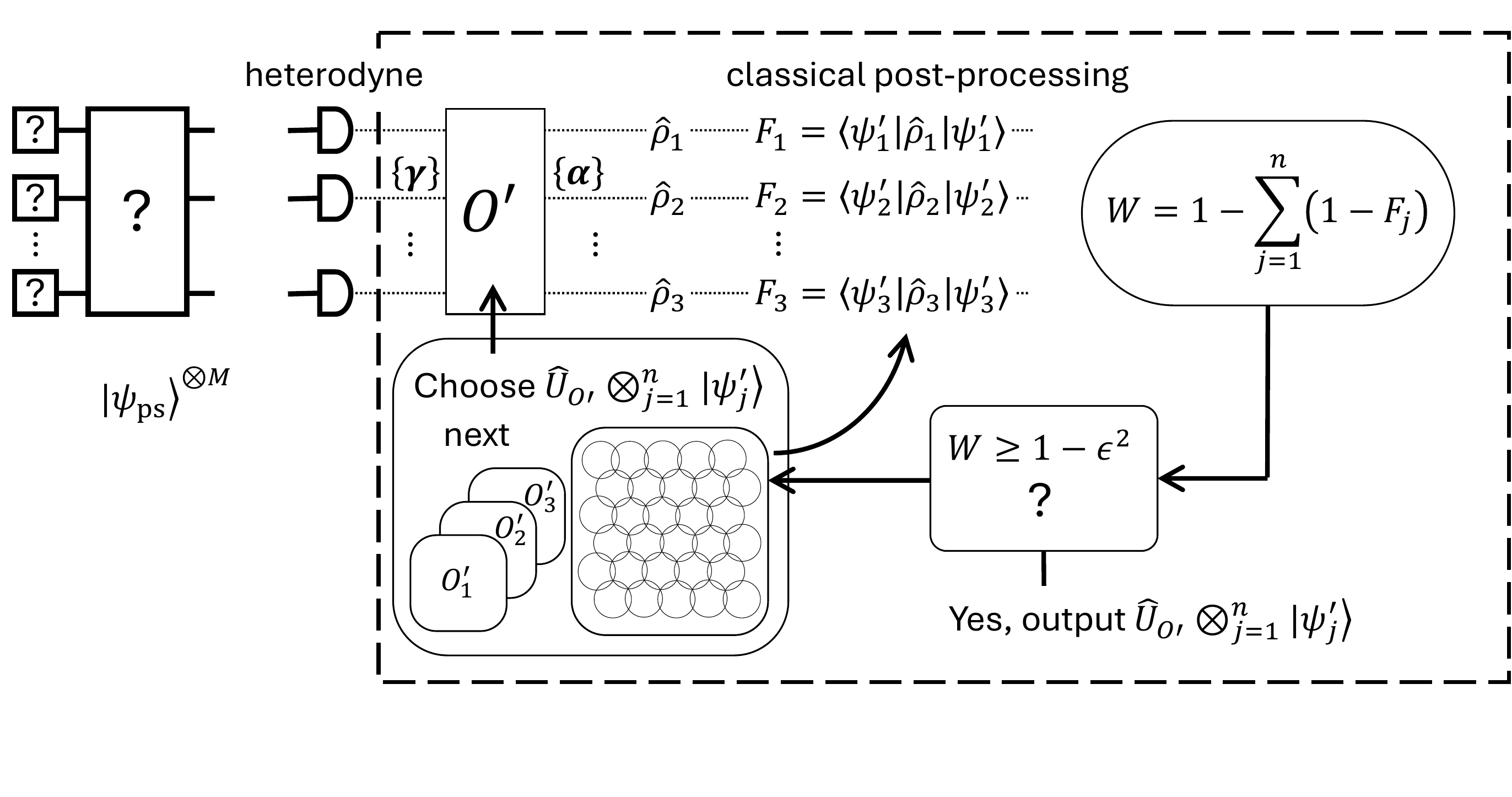}
				\caption{\textbf{Sample-efficient protocol for learning Boson sampling output states. }{\color{black} It follows a non‑adaptive procedure that uses only local measurement data: (i) Perform heterodyne measurement on all $M$ state copies to collect phase-space samples $\{\gamma_j\}$. (ii) Iterate over an $\epsilon-$net of passive-separable states $|\psi_{\rm ps}\>=\hat U_O \otimes_{j=1}^n |\psi_j\>$, reconstruct the corresponding local states $\{\hat \rho_j\}$ after every passive operation $U_{O'}^\dag$, and evaluate a local fidelity $F_j=\<\psi_j'|\hat\rho_j|\psi_j'\>$ for each possible local state $|\psi_j'\>$.   (iii) The first candidate with $W:=1-\sum_{j=1}^n(1-F_j)\ge 1-\epsilon^2$ is accepted as the reconstructed state; otherwise the protocol aborts. }
					\label{fig:scheme_learning}}
			\end{figure}

			As a concrete example, consider the output state of an $m$-mode Boson sampling process, known as passive-separable states $\{|\psi_{\rm ps}\>\}$ \cite{chabaud2023resources}, generated by applying a passive Gaussian unitary $\hat U_O$ to a product state $\{\otimes_{j=1}^n|\psi_j\>\}$, where $|\psi_j\>$ is a single-mode state (for multi-mode subsystem we refer to Supplemental Note \ref{app:tomography algorithm}). We present a tomography algorithm that takes as input $M$ copies of the unknown state $|\psi_{\rm ps}\>=\hat U_O\otimes_{j=1}^n|\psi_j\>$, output the classical description of a Boson sampling output state $|\widetilde{\psi_{\rm ps}}\>$ with smaller than $\epsilon^2$ fidelity error with the input state $|\psi_{\rm ps}\>$ (leading to a trace distance error smaller than $\epsilon$ by the Fuchs--van de Graaf inequality \cite{nielsen2002quantum}) and a failure probability lower than $\delta$. As shown in Fig.~\ref{fig:scheme_learning}, without relying on any adaptive strategy, the algorithm first performs heterodyne measurement on all $M$ copies of the quantum states and then performs classical post-processing to obtain an estimate of the state. 
			
			To describe the post-processing algorithm, we denote the heterodyne measurement result $q_k^{(\ell)}+ip_k^{(\ell)}$ for the $k$-th mode in the $\ell$-th copy. We collect all results as vectors
			$\bs \gamma^{(\ell)}=(q_1^{(\ell)},p_1^{(\ell)},\cdots,q_m^{(\ell)},p_m^{(\ell)})^T, 1\le \ell \le M$.
			Given the measurement results $\bs \gamma^{(\ell)}$, the algorithm searches over all possible passive operations $\hat U_{O'}$ and goes over all possible local states $|\psi_j'\>$. For each choice of $\hat U_{O'}$ and $\{|\psi_j'\>\}$, we perform classical processing as detailed hereafter to obtain a fidelity lower bound estimate $W$. The exhaustive search ends when the fidelity lower bound estimate satisfies $W\ge 1-\epsilon^2$, outputting the description of the corresponding estimate $|\widetilde{\psi_{\rm ps}}\>=\hat U_{O'}\otimes_{j=1}^m |\psi_j'\>$. Otherwise a failure is declared.
			
			As sketched in Fig.~\ref{fig:scheme_learning}, to obtain the fidelity lower bound estimate, we first calculate $\bs \alpha^{(\ell)}=O' \bs \gamma^{(\ell)}$ and reconstruct all local states $\{\widehat \rho_j\}_{j=1}^m$, whose matrix element in the number basis is
			\begin{align}
				\hat\rho_{j,kk'}=%& \frac{1}{M}\sum_{\ell=1}^M\int \frac{\d^{2} \bs u}{\pi} e^{  \frac {\|\bs u\|_2^2}8-\frac{i\bs u^T \Omega \left[\bs \alpha^{(\ell)}_{2j},\bs \alpha^{(\ell)}_{2j+1}\right]^T} 2 }  \chi_{kk'}(\bs u),
				& \frac{1}{M}\sum_{\ell=1}^M\int_{\mathbb C}\frac{\d^{2} \bs u}{\pi} e^{  \frac {\|\bs u\|_2^2}8+\frac{i \left[\bs \alpha^{(\ell)}_{2j},\bs \alpha^{(\ell)}_{2j+1}\right]\Omega \bs u} 2 }  \chi_{kk'}(\bs u),
			\end{align} 
			where $\Omega$ denotes the symplectic form (see Methods) and $\chi_{kk'}(\bs u)$ refers to the characteristic function of the operator $|k\>\<k'|$. The above reconstruction is a standard heterodyne tomography~\cite{becker2024classical}, but other reconstruction algorithms can also be adopted \cite{MAURODARIANO2003205,chabaud2020building}. Finally, following \cite{chabaud2021efficient}, we can compute the fidelity witness estimate 
			\begin{align}
				W&=1-\sum_{j=1}^n (1-F_j),
			\end{align}
			where each $F_j= \<\psi_j'|\widehat \rho_j|\psi_j'\>$ is the local state fidelity for the current estimate.
			
			A challenge in the classical processing is that it needs to go over all possible guesses of $\hat U_{O'}$ and local states, which are parameterized by continuous parameters with infinite possibilities. The way out of this dilemma is that we only require a finite error in the state learning: the finite-error relaxation allows one to establish a discrete set of $\epsilon$-balls of the possible $\hat U_{O'}$ and local states. Essentially, learning the parameters up to a discrete set of values suffices for meeting the $\epsilon$ error requirement. Accordingly, for each of the $\epsilon$-balls, we take the mean values as the choice and perform a fidelity test, which also has a failure probability of not satisfying the needed error tolerance. The failure probability then builds up with the number of $\epsilon$-balls---generally a huge number. On the other hand, the failure probability decreases exponentially with the number of samples $M$; therefore, the required number of samples is logarithmic of the number of $\epsilon$-balls. As a result, the key part of the proof of Theorem~\ref{theo:sample complexity of GE state} is to upper bound the number of $\epsilon$-balls (the volume) required for Boson sampling states, as we analyze in the Methods section in detail.
			
			With the sample-efficient learning of passive-separable states introduced, the generalization to arbitrary GE states can be obtained by applying a Gaussian unitary to reduce a GE state to a passive-separable state. To do so, we first estimate the mean and covariance matrix of the input state to estimate a symplectic matrix $S$ which diagonalizes the covariance matrix. Based on $S$, we apply a Gaussian unitary to reduce the input GE state to a zero-mean passive-separable state, which then can be sample-efficiently learned via the algorithm sketched in Fig.~\ref{fig:scheme_learning}. The details can be found in Supplemental Note \ref{app:learning GE states}.

		\section{Non-Gaussian-entanglable states}
		
		%\bigskip{\noindent{\bf Non-Gaussian-entanglable states}}\\
		%\noindent
		While we have shown that GE states are efficiently learnable, we know that general quantum states are hard to learn~\cite{mele2024learning}. This immediately indicates the existence of states that do not belong to the GE class defined in Eq.~\eqref{eq1}, which we refer to as non-Gaussian-entanglable (NGE) states. By definition, these states feature genuine non-Gaussian entanglement, which cannot be produced by Gaussian protocols. In practice, we show that NGE states can be produced by a multi-mode non-Gaussian unitary, or by a local non-Gaussian unitary acting on a non-trivial GE state. Furthermore, entanglement-breaking channels~\cite{horodecki2003entanglement}, followed by Gaussian protocols, can destroy genuine non-Gaussian entanglement. 
		%(ii) {\color{red} The Gaussianification process that produces Gaussian states with the same second moments.} 
		Moreover, %a connection exists between Gaussian unitary operations and the set of NGE states:
		%\QZ{words}
		%\begin{proposition}[Gaussian unitary on NGE state] \label{lem:Gaussian unitary preserves NGE state}
		it can be quickly verified that a Gaussian unitary maps NGE states to NGE states (see Supplementary Note \ref{app:onservation law of GE and NGE states}). 
		%\end{proposition}
		%\noindent The proof can be found 
		
		From Theorem~\ref{lem:pureGEstate}, we have an efficient protocol to identify pure NGE states (see Methods). 
		%the following corollary for  identification of NGE states.
		%\QZ{words, maybe method}
		%\begin{corollary}[Verification of pure NGE state]\label{cor:nge}
		%For a bipartite pure state $\ket{\psi}_{AB}$. Denote the symplectic matrix that diagonalizes the covariance matrix as $S$. 
		%(1) If the symplectic eigenvalues are non-degenerate, and $U^g_S\ket{\psi}_{AB}$ is not a product state, then $\ket{\psi}_{AB}$ is NGE. 
		%(2) If there is no beam-splitter network $U^g_O$ with orthogonal $O$ such that $U^g_O U^{g }_{S^{-1}}\ket{\psi}_{AB}$ is a product state, then $\ket{\psi}_{AB}$ is NGE.
		%\end{corollary}
		As a result, we provide a few examples of NGE states (see Fig.~\ref{fig:concept}), whose membership proofs can be found in Supplemental Note~\ref{app:pure nge state}. Firstly, we prove that the NOON states $\ket{\rm NOON}\propto \ket{N}_A\ket{0}_B+\ket{0}_A\ket{N}_B$ are NGE for $N>2$. This means that, aside from the trivial $N=1$ case, the $N=2$ case from the HOM effect is the only nontrivial instance of a beam-splitter generating NOON state---those with $N>2$ cannot be generated deterministically by any Gaussian protocol acting on two-mode separable states, including linear optical operations. Note that, for this specific problem, previous approaches have only established such a no-go result for number state inputs in a linear optical setting~\cite{vanmeter2007general,parellada2023no,parellada2024lie}.

		In addition, we show that the superposition of a TMSV state and a $\pi$-phase rotated TMSV state,
		$
		\ket{\Psi_{\rm sTMSV}}\propto \ket{\zeta_{r,0}}+\ket{\zeta_{r,\pi}}
		$
		with $r>0$, is also NGE. Such states are along the line of recent single-mode experiments generating non-classical superposition of Gaussian states in quantum systems with universal control~\cite{saner2024generating}.
		
		%\QZ{**whether mention ref\cite{saner2024generating} superposition of non-classical states or not}

		%\ref{app:degenerate_NGE}.

		% \XZ{Maybe $\ket{\alpha}\ket{\alpha^*}+\ket{-\alpha^*}\ket{\alpha}$ or equivalently $\ket{\alpha_1}\ket{\alpha_2}+\ket{-\alpha_2}\ket{-\alpha_1}$ is NGE given beam-splitter can not transform $\alpha$ into $\alpha^*$. TMSV case is a Gaussian superposition of that. }

		% \XZ{It is true. On the other hand, for the degenerate case, beam-splitter can generate correlation even if it acts on product state with a diagonal covariance matrix.   }
		
		% \begin{proposition}
			% Consider a GE state generated by a single Gaussian channel with a  transformation on covariance matrix $V_A\oplus V_B\to X(V_A\oplus V_B)X^T +Y$ where $X=\left(\begin{matrix}
				% X_{11}&X_{01}\\
				% X_{10}&X_{11}
				% \end{matrix}\right)$ and $Y=Y^T:=\left(\begin{matrix}
				% Y_{11}&Y_{01}\\
				% Y_{01}^T&Y_{11}
				% \end{matrix}\right)$ are real matrices satisfying the complete positivity condition $Y+i\Omega-iX\Omega X^T\ge 0$. Linear-uncorrelatedness implies:
			% \begin{align}
				% X_{00}V_A X_{10}^T+X_{01}V_B X_{11}^T+Y_{01}=0
				% \end{align}
			% \end{proposition}

		\section{Resource measures of genuine non-Gaussian entanglement}
		
		%\bigskip{\noindent{\bf Resource measures of {\color{black} genuine non-Gaussian entanglement}
				%NGE}}
		%\\
		To quantify genuine non-Gaussian entanglement, we further employ the quantum resource theory framework~\cite{chitambar2019}, which has been successful in quantifying quantum entanglement, quantum coherence and quantum non-Gaussianity. Here, we model all GE states as free states. To be general and also to provide insight into multi-mode state learning, we consider the case of multi-partite GE states, with $n$ partitions $\bm A=\{A_1,\cdots,A_n\}$, and define the Hilbert space $\map H_{\bm A}$ accordingly. On this account, Gaussian protocols are free operations as they map free states to free states (see Supplementary Note \ref{app:onservation law of GE and NGE states}). However, due to the multipartite nature of entanglement, the general quantification of genuine non-Gaussian entanglement faces similar challenges to the case of multipartite entanglement~\cite{walter2016multipartite}. 

		\begin{figure}[t]
			\centering
			\includegraphics[width=0.9\linewidth,trim=1 1 1 1,angle=0,clip]{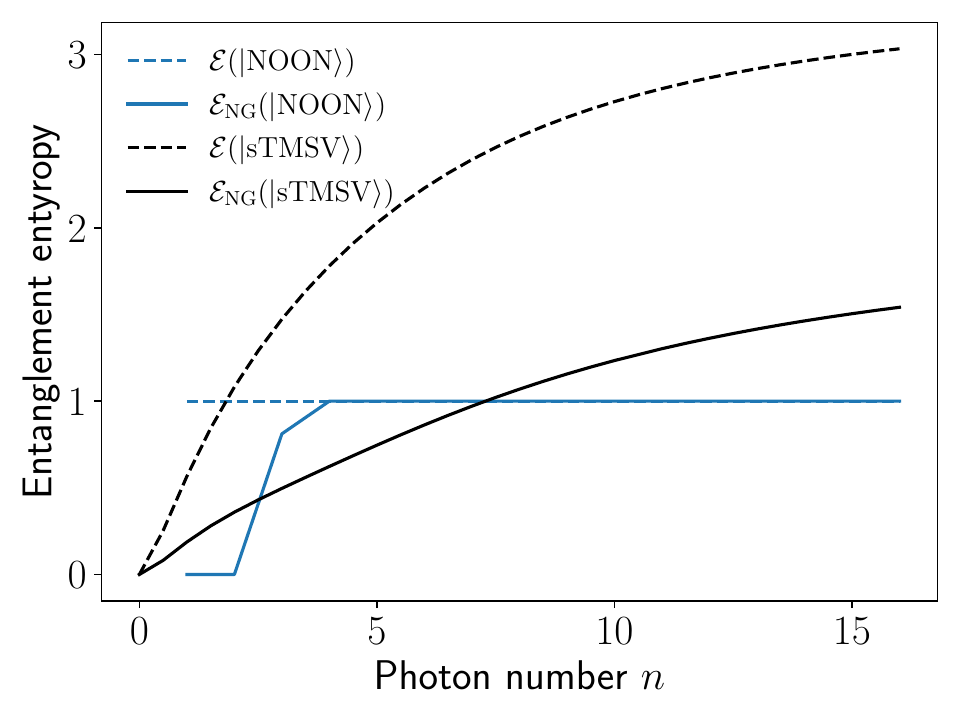}
			\caption{\textbf{Comparison between NG entropy and entanglement entropy.} The $x$-axis denotes the photon number, dashed lines indicate the entanglement entropy of the NOON state $|\text{NOON}\>$ and a superposition of two TMSV states $|\text{sTMSV}\>$, while the solid lines reflect their NG entropy. 
				\label{fig:NG entropy}
			}
		\end{figure}
		
		%Intuitively, one can specify the NG entanglement entropy for pure states 
		
		{\color{black}To tackle this issue, we define the non-Gaussian entanglement entropy, abbreviated simply as the NG entropy
			(see Supplementary Note \ref{app:resource theory of NGE}). } Given an arbitrary $n$-partite pure state $|\psi\>_{\bm A} \in\map H_{\bm A}$, denote $\hat \rho_{\bm A}(\hat U^g)=\hat U^g|\psi\>\<\psi|_{\bs A}\hat  U^{g\dag}$ and its reduced state as $\hat \rho_{A_j}(\hat U^g)=\Tr_{k\ne j}[\hat \rho_{\bm A}(\hat U^g)]$; the NG entropy is defined as:
		\begin{align}\label{NG entropy}
			& \map E_{\rm NG}(|\psi\>\<\psi|_{\bs A})\coloneq\min_{\hat U^g\in \mathbb{GU}\left( \map H_{A}\right)} \frac{1}{n}\sum_{j=1}^n  H \left(\hat \rho_{A_j}(\hat U^g)\right), 
		\end{align}
		with $ H(\hat \rho)=-\Tr[\hat \rho \log_2 \hat \rho]$ being the von Neumann entropy  of the state $\hat \rho$. Here we minimize over all Gaussian unitaries. Given an arbitrary pure GE state $|\psi\>$, the resource measure $\map E_{\rm NG}$ satisfies the following criteria: (1) $\mathcal{E}_{\rm NG}(|\psi\>\<\psi|)=0$ if and only if $\psi$ is GE; (2) $\mathcal{E}_{\rm NG}(|\psi\>\<\psi|)$ is invariant under Gaussian unitaries. %(3) $\mathcal{E}_{\rm NG}(|\psi\>\<\psi|)$ is nonzero if the state $|\psi\>$ is NGE. 
		A numerical evaluation of the NG entropy for pure NGE states is shown in Fig.~\ref{fig:NG entropy}. %\QZ{figure to be updated}
		% which relies on counter-rotation and the concept of entanglement entropy \cite{bennett1996concentrating}. Here, we introduce the following Computationally-simple NG entropy:
		% \noindent\emph{Definition 4. Given an arbitrary  $n$-partite pure state $|\psi\>_{A_1,\cdots A_n} \in\otimes_{j=1}^n \map H_{A_j}$, the computationally-simple NG entropy (CS-NG entropy) is defined as follows: 
			% \begin{align}
				% &\overline{\map E}_{\rm NG}(|\psi\>\<\psi|)\nonumber \\=&\min_{\substack{U^g\in \mathbb{GU}_{|\psi\>}\left(\otimes_{j=1}^n \map H_{A_j}\right)} }\sum_{j=1}^n  H \left(\Tr_{k\neq j}\left[U^g|\psi\>\<\psi| U^{g\dag} \right]\right)
				% \end{align}
			% with $\mathbb{GU}_{|\psi\>}\left(\otimes_{j=1}^n \map H_{A_j}\right)$ being the set of Gaussian unitaries that transforms $|\psi\>$ into a state with Williamson form of covariance matrix.  
			% }
		%In a similar way, but with more operational meaning,  %\XZ{We could give an numerical example. }
		For operational purposes, we may also define a measure as the maximum overlap between a given pure state and an arbitrary pure GE state (see Methods). %This measure can be used to distinguish GE and NGE states: the $\map P_{\rm NG}$ of any pure GE state is zero, while it has a nonzero value is the state is NGE. 
		This measure distinctly separates GE and NGE states: it is zero for every pure GE state and non‑zero for any NGE state, and invariant under Gaussian unitaries. 
		%\QZ{PNG can be max overlap of psi to any pure GE state, move the defitinos to supp}

\begin{figure}[t]
			\centering
			\includegraphics[width=\linewidth]{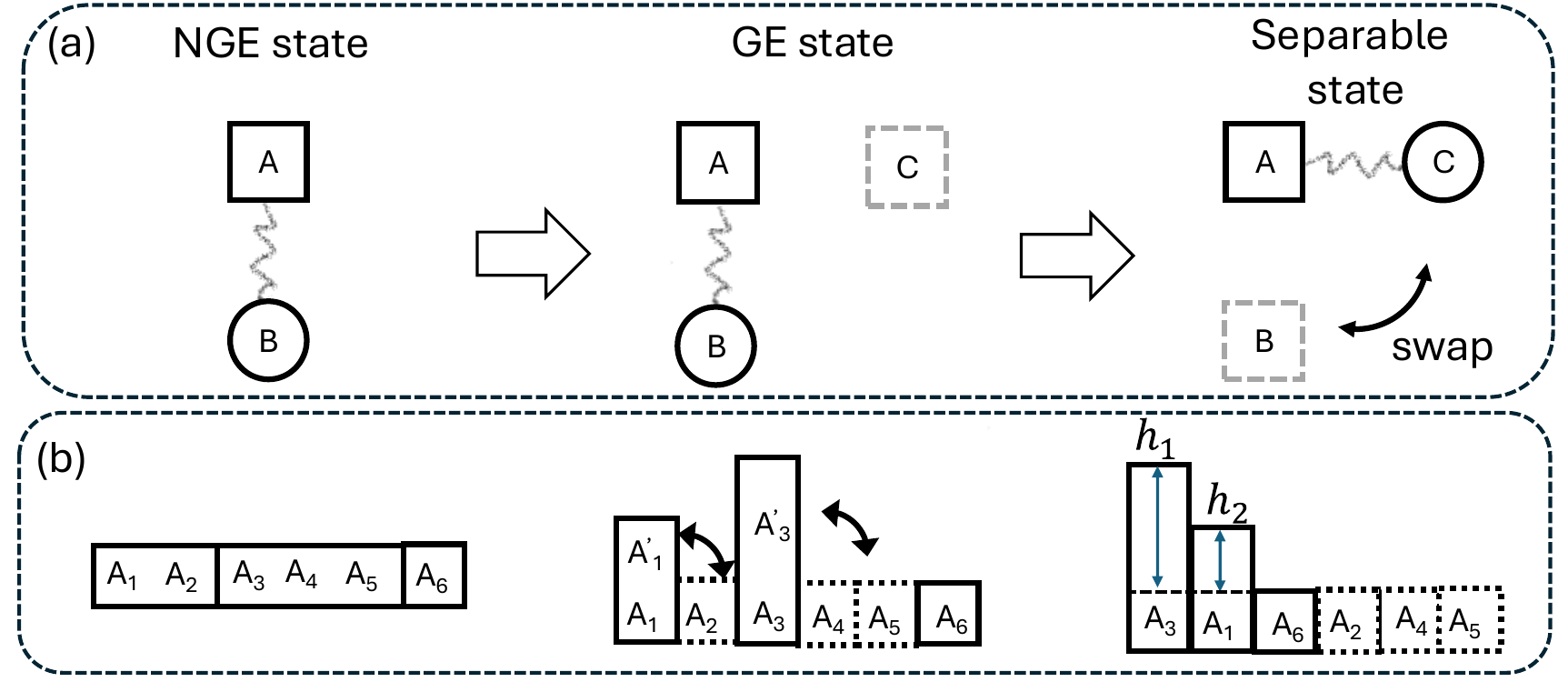}
			\caption{ (a) A bi-partite NGE state $\hat \rho_{AB}$ can be extended to a tri-partite GE state $\hat \rho_{AB}\otimes \hat \sigma_C$. 
				%Through a fully transmissive beam-splitter swapping system B and C, one can achieve a separable state with system partition $AC-B$. 
				(b) Young diagram representation of a multipartite case of six modes. The states of subsystems $A_1A_2$ and $A_3A_4A_5$ are both NGE. By introducing the ancilla $A_1^\prime$ and $A_3^\prime$, the state is GE between $A_1A_1^\prime$, $A_2$, $A_3A_3^\prime$, $A_4$, $A_5$ and $A_6$. In addition, we can permute local systems to have a sorted column length and $\map R(\psi)=(h_1,h_2,0,0,0,0)$. %\XZ{There is a PRL \cite{ren2021metrological}}
				\label{fig:Young diagram}
			}
		\end{figure}

		%\subsubsection{Special case: pure states}
		%We can find upper bound easily by specifying a particular Gaussian unitary\begin{align}\mathcal{E}_{NG}(\psi_{AB})\le E(U_{S^\star}\psi_{AB} U_{S^\star}^\dagger),\end{align}where $S^\star$ symplectic diagonalize the covariance matrix of $\psi_{AB}$.
		%For pure GE state with unique symplectic eigenvalues, the upper bound equals zero and is therefore tight. 
		%For pure GE state with degeneracy, we can optimize over block diagonal beam-splitters to obtain the best upper bound, and equals zero therefore tight.
		% \subsection{Generating power}
		% \XZ{We should consider non-Gaussian compression in Ref.~\cite{mele2024learning}}
		% \begin{lemma}
			% An arbitrary bipartite NGE state can be generated by a tripartite 1-separable state. 
			% \end{lemma}
		% \begin{proof}
			% Consider a separable state regarding the $AB$-mode, where $B$-mode is entangled with an ancilla:  $\rho_{ABC}:=\rho_A\otimes \sigma_{AB}$. One can implement a Gaussian unitary to swap the $A$-mode with the $BC$-modes. Then, by discarding the $C$-mode, one can obtain an arbitrary bipartite state $\sigma_{AB}$.  
			% \end{proof}
		As no single `golden unit' of multiparite entanglement exists due to the complex structure of multi-partite entanglement, we further introduce a vector resource measure to characterize multipartite genuine non-Gaussian entanglement. This resource measure is based on the number of ancillary modes necessary to make an NGE state GE: the GE cost defined below. To get a clearer picture, let us begin with a simple bi-partite system of two modes $A$ and $B$. As shown in Fig.~\ref{fig:Young diagram}a, any NGE state $\hat \rho_{AB}$ can extended into a tri-partite state $\hat \rho_{AB}\otimes \hat \sigma_C$ by introducing an ancilla $C$. Meanwhile, this state is naturally a GE state between $AC$ and $B$ because it can be generated from  a separable state $\hat \rho_{AC}\otimes \hat \sigma_B$ between $AC$ and $B$ via a fully transmissive beam-splitter, i.e., a Gaussian unitary operation swapping systems $B$ and $C$. %any NGE state $\rho_{AB}$ can be prepared if any of the modes has access to a local ancilla $C$. Consider the extended bipartite system $AC-B$, now GE state defined in Eq.~\eqref{eq1} between $AC-B$ allows one to start with a separable state $\sigma_{AC-B}=\rho_{AC}\otimes \sigma_B$. Then, one can apply a fully transmissive beam-splitter to swap system $C$ and $B$ to obtain the state $\rho_{AB}\otimes \sigma_C$. Finally, a partial trace gives the state $\rho_{AB}$. 
		Likewise, for $m$ single-mode subsystems, %any NGE state can be generated if $m-1$ ancilla modes are provided to the first system. 
		{\color{black} any NGE state can be produced by introducing an $(m-1)-$mode non-Gaussian ancillary state entangled with the first mode.  }

		%\QZ{shall we present the simple single-mode case definition while put a general case in Supplementary Note of main text? needs a figure}

		For simplicity of analysis, let us consider an $n$-mode state with $n$-partition, while the more general case can be found in Supplementary Note~\ref{supp:GE_cost}. The GE states are those generated by applying $n$-mode Gaussian protocols on a $n$-mode fully separable state. Then, the corresponding GE cost is given as follows:

		\begin{definition}[GE cost vector and GE cost function]\label{defi:GE cost}
			Given an arbitrary $n$-mode density matrix $\rho$ over $\map H_{\bs A}$, a {\color{black} GE cost vector}
			%genuine NGE of formation %can be
			is defined as a list of ancillary-mode numbers in a decreasing order:  
			\begin{align} 
				\label{GE cost 5.1}
				\map R(\rho)=\{|A_1'|,\cdots,|A_n'|\},\ \ (|A_1'|\ge |A_2'|\ge \cdots \ge |A_n'|), 
			\end{align}
			%for each subsystem needed %to generate the state via Gaussian protocols---
			of an extension $\rho_{\rm ext}$ satisfying $\rho:=\Tr_{A_1',\cdots,A_n'}[\rho_{\rm ext}]$, such that $\rho_{\rm ext}$ is a GE state that can be obtained by applying a Gaussian protocol on a separable state between $A_1A_1'$, $A_2A_2'$,... and $A_nA_n'$. We will denote the $k$-th entry of $\map R$ as $\map R_k$ and the number of nonzero entries of $\map R$ as $\ell$. 
			{\color{black}
				The GE cost function is then defined as:
				\begin{align}
					\map {R}_{f}(\rho)&=\min_{\rho_{\rm ext}} \map R_1 
				\end{align}
				where the minimization is taken over all possible GE extensions $\rho_{\rm ext}$ of the original state $\rho$. 
			}
			% \begin{align}\label{GE cost 5.1}
				% \map R(\Psi) &=  \left(|A_1'|,|A_2'|,\cdots, |A_n'|\right)^T\\[0.5em]
				% \emph{s.t.\ \ }\nonumber &\begin{cases}
					% \widetilde\Psi\in \mathbb{GE}\left(\bigotimes_{j=1}^n \map H_{A_j}\otimes \map H_{A_j'};\right.\\
					% \left.\,\ \ \left\{A_1A_1',A_2A_2',\cdots,A_nA_n'\right\}\right)\\[0.5em]
					% \Psi=\Tr_{A_j' } \left[\,\widetilde\Psi\,\right]
					% \end{cases}.
				% \end{align}
			%Further, the GE cost defined in Eq.~(\ref{GE cost 5.1}) can be represented in a decreasing order: $\map R^{\downarrow}(\psi)=(h_1,\cdots,h_n)^T\text{\em such\ that } h_1\ge h_2\ge \cdots,\ge h_n$. %In this case, one can define the minimum of sorted GE cost as: \begin{align}\underline{\map R^{\downarrow}}(\Psi)&=\min_{h_n}\cdots \min_{h_2}\min_{h_1}\map R^{\downarrow}(\Psi).\end{align}
		\end{definition}
		
		%\QZ{define GE cost vector and GE cost function. cost function is non-unique, but non-increasing under Gaussian protocol.}
		
		Fig.~\ref{fig:Young diagram}b shows an example of six modes, where ancilla modes $A_1^\prime$ and $A_3^\prime$ are introduced to extend the state to a GE state. The definition of a GE cost vector is closely related to the concept of Young diagram, as it is connected to irreducible representations of the permutation group.
		Note that although we have introduced the extension procedure with swap operations, general Gaussian protocols can be applied.
		
		Note that the GE cost vector is not unique, but the GE cost function is. {\color{black} Moreover, the GE cost function $\map R_f$ (as well as the one-norm of the GE cost vector $\|\map R\|_1:=\min_{\rho_{\rm ext}}\sum_{j=1}^\ell\map R_j$) are non-increasing under Gaussian protocols. In addition, the GE cost function $\map R_f$ can also be bounded as $\|\map R\|_1/\ell\le \map R_f\le \|\map R\|_1$, %For instance, the resourcefulness given by $\map R^{\downarrow}_1:=(5,2,1,1,0,\cdots,0)^T$ is neither larger or smaller than $\map R^{\downarrow}_2:=(3,3,2,1,0,\cdots,0)^T$ even if the relation $\|\map R^{\downarrow}_1\|_1=\|\map R^{\downarrow}_2\|_1=9$ is satisfied. 
			%In addition, the GE cost of a state is not unique--a single state can admit several valid values under the definition. Nevertheless, 
			In the next section, we will show that the GE cost function provides useful prior information for learning, as it guides how resources must be allocated in a tomography protocol.}

		\section{Learning pure states}
		
\begin{figure}[t]
			\centering
			\includegraphics[width=1\linewidth,trim=5 5 5 5,angle=0,clip]{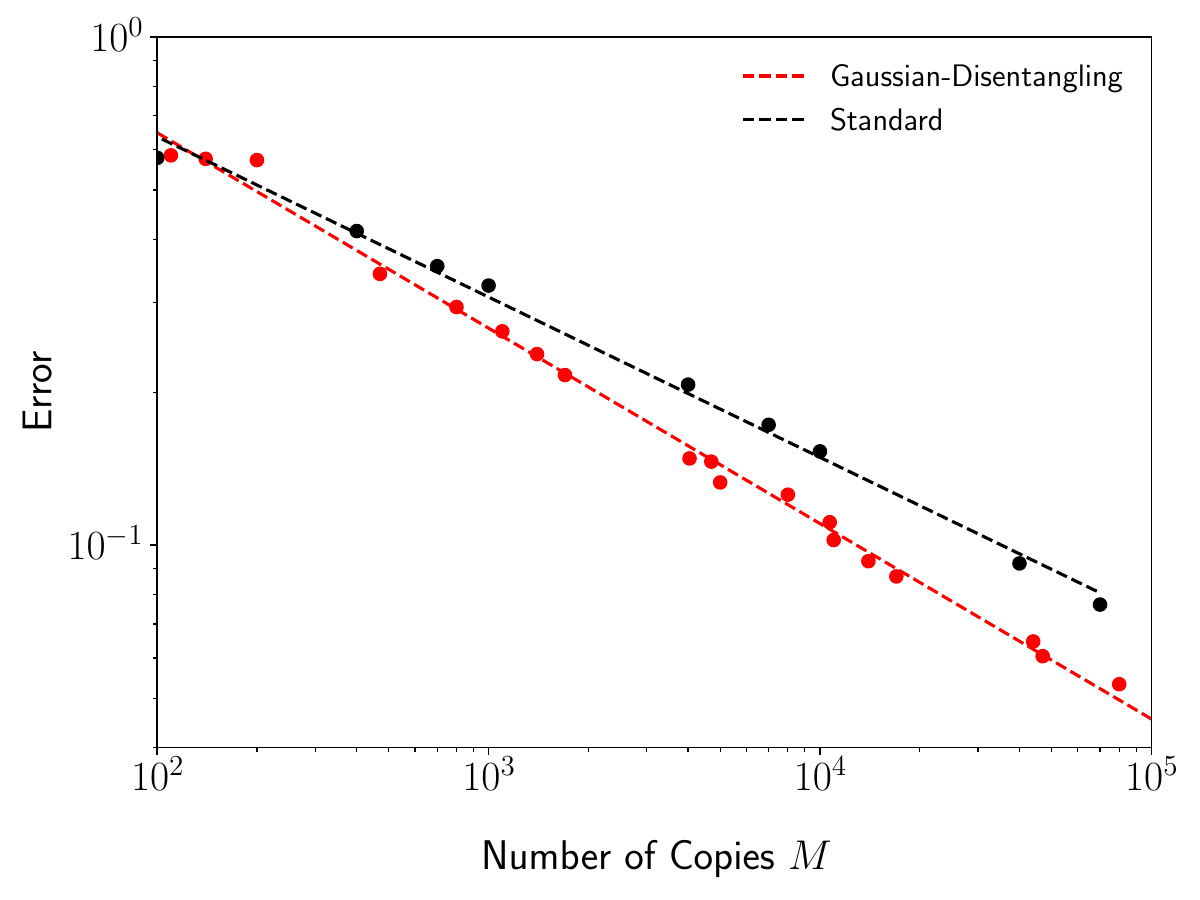}
			\caption{Performance of Gaussian-disentangling protocol versus direct tomography. Here, we numerically simulate the quantum state tomography process on a two-mode GE state with the true state being a thermal state correlated by a beam-splitter.  The error, defined as the trace distance between the true state and the reconstructed state obtained through direct tomography using the standard algorithm~\cite{lvovsky2009continuous}, is shown as a function of the number of copies $M$ used in tomography and is shown by black dots. The black dashed line represents the linear fit to the black dots with a slope $k_1=-0.38$. In contrast, the bottom envelope of  achievable errors in the proposed Gaussian disentangling algorithm, after minimising over possible sample number %$M_v$ 
				in estimating displacements and the covariance matrix, is given by red dots. Their linear fit has a slope $k_2=-0.32$.  
				\label{fig:2mode_thermal_state_tomography}
			}
		\end{figure}

		%\bigskip
		%{\noindent{\bf Learning pure states}}
		%\\
		The GE cost is operationally meaningful, as it directly connects to the generation procedure of an NGE state. Moreover, as the clear structure of entanglement indicates, the cost of learning NGE states can be quantified by the GE cost, as we show in the following.
		In particular, to leverage the Gaussian entangling procedure in defining the GE cost, we consider a learning algorithm based on the one for GE states, with the local states corresponding to the minimal subsystems of modes $\{\map R_k+1\}$ that cannot be further disentangled by any Gaussian unitary (see details on ``generalized GE states'' in Supplemental Note \ref{app:learning GE states}).
		%consisting of the following steps. (1) Estimate mean and covariance matrix via, e.g. homodyne and heterodyne measurement on $M_v$ copies of the states; For the rest of the remaining copies of state, perform the next steps. (2) Apply a disentangling Gaussian unitary; (3) Learn a postprocessing beam-splitter that makes the state separable. (4) Perform local state tomography for the final state.
		To evaluate the performance, we begin by numerically simulating the tomography process for two-mode GE states with nondegenerate symplectic eigenvalues. As shown in Fig.~\ref{fig:2mode_thermal_state_tomography}, the Gaussian‑disentangling method—optimized over the number of copies used to estimate both the displacement vector and covariance matrix (red)—achieves lower errors and a markedly better scaling than direct tomography (black) \cite{lvovsky2009continuous}. Details of the simulation is presented in Supplementary Note~\ref{app:numerical simulation of the tomography process}.
		% \begin{center}
			% \begin{tabular}{ |l |  }
				% \hline 
				% (1)Estimation of mean and covariance matrix;\\ 
				% \hline
				% (2)Apply a disentangling Gaussian unitary; \\ 
				% \hline 
				% (3)Learning a postprocessing beam-splitter;\\ 
				% \hline 
				% (4)Local state tomography for the product state.\\
				% \hline 
				% \end{tabular}
			% \end{center}
		% \begin{enumerate}
			% \item [(1)] Estimate covariance matrix and displacement with homodyne measurement; 
			% \item [(2)] Partially undo the effect of Gaussian unitary by counter-rotating;
			% \item [(3)] Identification of the beam-splitter operation that makes the state separable;
			% \item [(4)] Local tomography for the product state after a beam-splitter operation identified by step (3). 
			% \end{enumerate}
		
		In the scenario where the number of modes $m$ is large, the sample complexity is shown in Proposition \ref{lem:sample complexity upper bound} in Methods, which is an exponential function of $\max_k \map R_k$. {\color{black}It is worth noting that, for any family of states, the GE cost plays the role of prior information that guides the choice between two strategies: (i) applying a counter‑rotation followed by local tomography, or (ii) performing direct tomography on larger subspaces. The magnitude of the GE cost directly influences which strategy is preferable and, in turn, sets the overall sample complexity.  }

		%a concrete demonstration and analysis of this tomography protocol is shown in Supplementary Note \ref{app:learning GE states}, where we prove that the sample complexity of learning a pure GE state is a polynomial function of $n$. In contrast, tomography of all pure NGE states with the same GE cost has exponential overheads in sample complexity upper bound (see Methods). 
		
		%\QZ{move proposition to method}
		%\QZ{sample complexity is defined as : if I want to achieve global state error epsilon, how many samples do I need}
		%\QZ{upper bound: from the protocol, and you can decide on how to distribute epsilonj and epsilonv}
		%\QZ{lower bound: we define a tomography task which is easier: we give the information about the Gaussian unitary. Then you know it becomes a tomography task of a product state. then tomography error on global state $\epsilon$, the number of copies is $\max_j (1/\epsilon_j)^{R_j} $. Because of data-processing inequality, subsystem error $\epsilon_j\le \epsilon$. the problem we try to solve is $\min_{\{\epsilon_j\}}^\prime \max_j(1/\epsilon_j)^{R_j} $. It will be dominated by $\max R_j$---lower bound of}
		%\begin{align}M&\ge \min_{\{\epsilon_j\}: \sum \epsilon_j \ge \epsilon; \epsilon_j\le \epsilon} \max_j(1/\epsilon_j)^{R_j} \\&\ge \min_{\{\epsilon_j\}: \sum \epsilon_j \ge \epsilon;\epsilon_j\le \epsilon} (1/\epsilon_j)^{\max_j R_j} \\&\ge (1/\epsilon)^{\max_j R_j}\end{align}
		In general, one %can show that Eq.~(\ref{eq9_proposition5}) 
		can have a tight scaling with respect to the GE cost for the nondegenerate scenario:
		\begin{theorem}[Sample complexity for nondegenerate case]\label{lem:connection GE cost and learning} 
			Consider an energy‑constrained $m-$mode pure state in $\calH_{m,E}^2$,  whose GE cost is specified by an $l-$element vector $\map R$ and the function $\map R_f$. Meanwhile, its covariance matrix has non‑degenerate symplectic eigenvalues. The required sample complexity for tomography is: 
			%\begin{align}M=\textbf{\em poly} (m) \cdot \Theta \left(\epsilon^{-\max_k \map R_k-3}\right), \end{align}
			\begin{align}
				%M &= \textbf{\em poly} (m,\epsilon) \cdot \Theta \left(\epsilon^{-\max_k \map R_k}\right).\\
				M= %&\textbf{\em poly} (m,\epsilon,E) \cdot \Theta \left[\left(\frac{E_{\rm II}}{\epsilon}\right)^{\max_k \map R_k}\right]\\
				&\textbf{\em poly}\left(\ell,E,\frac{1}{\epsilon},\log \frac 1 \delta  \right)\cdot \Theta \left[\left(\frac{E}\epsilon\right)^{ \map R_{f}}\right]
			\end{align}
			where $\textbf{\em poly}\left(\ell,E,\frac{1}{\epsilon},\log \frac 1 \delta  \right)$ refers to a polynomial function of $\ell$, $E$, the error $\epsilon$ in trace distance, and failure probability. $\Theta$ denotes the big theta notation. 
			%{\color{blue}In addition, the sample complexity of tomography an arbitrary pure state with nondegenerate symplectic eigenvalues is:\begin{align}M=\textbf{poly} (m) \cdot \Theta \left(\epsilon^{-\max_k \map R_k-1}\right). \end{align}}
		\end{theorem}
		
		A detailed proof of Theorem \ref{lem:connection GE cost and learning} can be found in Supplementary Note \ref{app:proof of theorem 4}. {\color{black} In particular, we prove that the exponent of $E/\epsilon$ in the upper bound achieved by our protocol matches that of the lower bound demonstrated in Proposition \ref{lem:sample complexity upper bound}. However, a residual gap remains, scaling as  $\ell^{2 \map R_f}$. }%\QZ{The achievable part relies on a Gaussian-disentanlging tomography we design to make use of the GE structure. The basic idea is to first obtain an estimation of the covariance matrix of the state. Then, based on the GE entanglement structure as the prior knowledge, one can disentangle the original state via Gaussian unitary to reduce the tomography task to many smaller ones. }
		%\XZ{Show two mode example for mixed state with counter rotation of Gaussian unitary }
		%\XZ{If our protocol optimal}
		%\QZ{****have Pengcheng do some simulation of the tomography process, for GE error decays fast, and for NGE it is slow, and compare with the bound****}
		%\subsection{Computation assisted with $t$-doped circuit}
		Notably, the GE cost can be connected with the recently proposed $t$-doped bosonic Gaussian states \cite{mele2024learning}---states generated from unlimited multi-mode Gaussian unitaries and $t$ $\kappa$-local non-Gaussian unitaries. Our results extend this class to the NGE class measured by the GE cost, and the tomography overhead is directly related to the GE cost. In particular, any $t$-doped state can be extended into a GE state with $(\kappa t-1)$ ancillary modes. From Theorem \ref{lem:connection GE cost and learning}, the sample complexity of learning a GE state has an overhead $\epsilon^{-\kappa t-2}$ as its GE cost vector is $\map R=(\kappa t-1,0,\cdots)$.  
		
		%\QZ{numerical results to be added}

		\ 
		
		\section{Discussion}
		
		%{\noindent\bf \Large Discussion}\\
		In this work, we have identified a quantitative relationship between the nature of quantum correlations in bosonic systems and the complexity of quantum tomography, uncovering the role of genuine non-Gaussian entanglement---entanglement which cannot be produced by Gaussian protocols.
		
		We point out a few open problems. We have focused on pure states in the study of sample complexity in learning, while deferring the generalization to mixed state to future work. There, the disentangling procedure for Gaussian multi-mode channels is to be developed.
		%An open question is how to reverse Gaussian channel when learning mixed GE states. In addition, state reversion is also essential when defining NGE of formation. For instance, TMSV operation following with pure loss channel can be reversed with another TMSV operation \cite{shi2024entanglement}. Whether this phenomenon can be extended to arbitrary two-mode or multi-mode case is unknown.  
		Another open question is the quantification of the %NGE 
		genuine non-Gaussian entanglement for measurement, which represents a dual problem within the resource theory of NGE states. %For instance, the paradigmatic Boson sampling protocol~\cite{aaronson2011computational} has measurement bases evolving under Gaussian unitary. After a Gaussian unitary, single-mode photon number resolving detector are implemented. Thus, its measurement basis is in the class of GE.
		%A further potential extension of this discovery pertains to fermionic systems~\cite{hackl2021bosonic}. 
		{\color{black}
			In addition, an intriguing open question is how to design a tomography protocol whose time complexity—specifically, the number of search rounds required within each $\epsilon$-covering net—scales polynomially with the number of modes, while preserving the already efficient sample complexity.
		}
		At last, asymptotic conversion rate %\cite{hayashi2024generalized} 
		between NGE states and generalization to fermionic systems %~\cite{hackl2021bosonic}
		are also of interest. 
		
		%\XZ{We might discuss the interplay between Gaussian entangling and a randomization process, as an extension to discuss mixed states. Do some lemmas/axioms. }
		
		%Extension towards conditional state production can be considered in future works. \\[-0.5em]

		%Another open problem is the proof of any tripartite NGE state. 
		
		%\emph{Conclusions.--- } We investigated a hybrid resource theory that takes into consideration both non-Gaussianity and inseparability given states. In particular, we modeled every state generated by applying Gaussian protocols to separable states as free states. We proved that every pure free state can be expressed as a Gaussian unitary acting on a product state. Further, we demonstrate paradigmatic examples of resource states, i.e., the NGE state, such as NOON state and superposition of TMSV states. We also investigated free operations and the operations that preserve NGE states. We showed how to verify pure NGE states. More over, we explore two measures for the resourcefulness of a given state, one based on entanglement entropy subject to counter-rotation, and the other based on the system size of GE state extension. Finally, we show that the pure GE state can be learned efficiently. 
		%\\[-0.5em]
		
		\emph{Acknowledgement.--- } We thank Xun Gao, Haocun Yu, and Changhun Oh for useful comments on an earlier version of this manuscript. This project is supported by the NSF (2240641, 2350153, 2326746), ONR Grant No. N00014-23-1-2296, DARPA (HR0011-24-9-0362,HR0011-24-9-0453, D24AC00153-02) and AFOSR MURI FA9550-24-1-0349. This work was partially funded by an unrestricted gift from Google.  F.A.M.~acknowledges financial
		support by PRIN 2022 "Recovering Information in Sloppy QUantum modEls (RISQUE)", code 2022T25TR3, CUP E53D23002400006. U.C.~acknowledges funding from the European Union’s Horizon Europe Framework Programme (EIC Pathfinder Challenge project Veriqub) under Grant Agreement No.~101114899.
		\\[-0.5em]
		
		\ 
		
		%{\noindent\bf \Large Appendix}\\[0.5em]
		
		\appendix
		
		\section{Preliminary}
		
		%\noindent{\bf Preliminary %\ref{lem:pureGEstate} 
			%}\\
		%\noindent 
		{\color{black}An $m$-mode bosonic quantum system is completely specified by the statistical moments of its canonical quadrature operator vector  $\hat{\bs r} =[\hat q_1, \hat p_1,\cdots,\hat q_m,\hat p_m]^{\rm T}$. Its first and second moments are captured by the displacement vector $\bs \xi:=\Tr[\hat{\bs r} \hat\rho ]$ and the covariance matrix $
			V:=\frac 1 2 \Tr [\{(\hat{\bs r}-\bs \xi),(\hat{\bs r}-\bs \xi)^{\rm T}\}\hat \rho]$~\cite{Weedbrook_2012,serafini2017quantum}. The first-moment changes are modeled by the displacement operation $\hat D_{\bs \xi}=\exp \left(-i\bs \xi^T \Omega \hat {\bs r}\right)$ where $\bs \xi=(\xi_1,\cdots,\xi_{2m})$ denotes the vector of displacements, $\Omega=\bigoplus_{j=1}^m\left(\begin{matrix}
				0&1\\
				-1&0
			\end{matrix}\right)$ refers to the symplectic form. Then, given an arbitrary operator $\widehat X$, one can define the Wigner characteristic function $\chi_{\widehat X}(\bs \xi)=\Tr[\widehat X \hat D_{\bs \xi}]$ \cite{serafini2017quantum}. For notational simplicity, we will let $\chi_{kl}(\bs \xi)$ denote the characteristic function of the operator $|k\>\<\ell|$, where $\{|k\>,k=0,\cdots,\infty\}$ is the Fock basis. }\\[-0.5em]

		%define displacement and characteristic function\begin{align}\chi_{kl}(u)=\Tr[|\bs k\>\<\bs \ell| e^{-iu^T \Omega (\cdots,p_\ell,q_\ell,\cdots)^T/2} ]\end{align}
		
		\section{A useful lemma for the proof of Theorem \ref{lem:pureGEstate} }
		
		%\noindent{\bf A useful lemma for the proof of Theorem \ref{lem:pureGEstate} }\\
		%\noindent 
		
		The proof of Theorem \ref{lem:pureGEstate} is based on the following lemma:
		
		\begin{lemma}[Generalizing Lemma 2 in Ref.~\cite{mari2014quantum}]\label{lem:13_main} 
			The only states that remain pure under multi-mode pure loss are coherent states.
		\end{lemma}
		
		\noindent The detailed proof of this lemma is given in Supplementary Note \ref{app:proof of the pure state theorem}. \\[-0.5em]
		
		\section{Verfication of NGE states}
		
		%\noindent{\bf Verfication of NGE states}\\
		%\noindent 
		
		From Theorem~\ref{lem:pureGEstate}, we have the following corollary for  identification of NGE states:
		\begin{corollary}[Verification of pure NGE state]\label{cor:nge}
			Let $\ket{\psi}_{AB}$ be a bi-partite pure state and let $S$ be a symplectic matrix that diagonalizes its covariance matrix. 
			(1) If the symplectic eigenvalues are non-degenerate, and $\hat U^g_S\ket{\psi}_{AB}$ is not a product state, then $\ket{\psi}_{AB}$ is NGE. 
			(2) If there is no beam-splitter network $\hat U^g_O$ with orthogonal $O$ such that $\hat U^g_O \hat U^{g }_{S^{-1}}\ket{\psi}_{AB}$ is a product state, then $\ket{\psi}_{AB}$ is NGE.
		\end{corollary}
		
		\section{Additional example of NGE state}
		
		%\noindent{\bf Additional example of NGE state}\\
		%\noindent 
		
		In addition to the NGE state examples shown in the main text, there is another example as follows: 
		
		\noindent\emph{Example iii (two-mode arithmetic progression state): 
			The following state is NGE: 
			\begin{align}\label{NGE_state:example1}
				|\Psi\>_{k,\ell}=&\sum_{j,h=0}^\infty  c_{jh} \, |\ell+jk\>_A|\ell'+hk'\>_B,
			\end{align}
			where $k,k'\ge 2$ are integers, $\ell,\ell'\in \N$ are natural numbers, $\{c_{jh}\in\C\}$ are coefficients satisfying the normalization condition $\sum_{jh}|c_{jh}|^2=1$ and can not be written in the product form, i.e. $c_{jh}\neq c_j 'd_h'$, the symplectic eigenvalues of the state in Eq.~(\ref{NGE_state:example1}) are assumed to be different.  }

		The proof of Example iii is shown in Supplementary Note \ref{app:pure nge state}. A simple case of Example iii is the state $|\Psi\>=(\cosh r)^{-1}\sum_{j=0}^\infty \tanh^j r |3j\>|1+3j\>$. More generally, for any entangled bi-partite pure state, one can construct a non-Gaussian-entanglable version of it by changing the Fock basis $\{|j\>\}$ to $|\ell + j k\>$ for $\ell\ge 0$ and $k\ge 2$ ($\ell,k\in \Z$). \\[-0.5em]
		
		\section{Maximal fidelity in state  preparation}
		
		%\noindent{\bf Maximal fidelity in state  preparation}

		%\noindent 
		{\color{black} Consider a specific partition of the Hilbert space $\map H_{\bs A}=\otimes_{j=1}^n \map H_{A_j}$. Then, the  maximal fidelity %/ probability of success 
			of preparing a pure state $|\psi\>_{\bs A}\in \map H_{\bs A}$ from any GE state  $|\psi'\>_{\bs A}\in\map H_{\bs A}$ can be obtained as follows: }
		\begin{align}
			\map P_{\rm NG}(|\psi\>\<\psi|_{\bs A})&= \max_{\substack{|\phi\>_j\in \map H_{A_j}\\\forall j=1,\cdots,n\\\hat U^g\in \mathbb{GU}(\map H_{\bs A})}} \left|\<\psi|_{\bs A} \hat U^g \bigotimes_{j=1}^n |\phi_j\>\right|^2\\
			&\equiv \max_{\hat U^g\in \mathbb{GU}(\map H_{\bs A})}\left\|\hat U^{g\dag} |\psi\>\<\psi|_{\bs A}\hat U^g\right\|_\times %\\
			%&\le \max_{U^g\in \mathbb{GU}(\map H_{\bs A})}\left\|\left(U^{g\dag} |\psi\>\<\psi|_{AB}U^g\right)^{T_B}\right\|_\infty
		\end{align}
		where $\|\cdot\|_\times=\max_{\substack{|\phi\>_j\in \map H_{A_j}\\\forall j=1,\cdots,n}} \bigotimes_{j=1}^n\<\phi_j| \cdot \bigotimes_{k=1}^n|\phi_k\> $ refers to the cross norm. Note that the cross norm can be bounded by the operator norm as $\|\hat X\|_\times\le \|\hat X^{T_{\bs A_0}}\|_\infty, (\bs A_0\subset \bs A)$ which can be computed efficiently.\\[-0.5em]
		
		%\XZ{We may need to develop an SDP algorithm for the monotone.  }
		
		%\XZ{Otherwise, we may compare PNG with the bound with operator norm. }
		
		\section{Learning state with the same GE cost}
		
		%\noindent{\bf Learning state with the same GE cost}
		
		%\noindent 
		
		Regarding the states with the same GE cost, we have the following proposition: 
		\begin{proposition}[Sample complexity of large NGE states]\label{lem:sample complexity upper bound}
			Consider the set of $m$-mode pure states $\calH_{m,E}^2$ 
			with a GE cost vector $\map R=(\map R_1,\cdots,\map R_\ell)^T$ satisfying $\map R_f>7$.
			%$\sqrt{\Tr\left[\left(\sum_{j=1}^m a^\dag_j a_j+\frac m 2 I \right)^2|\psi\>\<\psi|\right]}\le m E_{\rm II}$. 
			%where $\otimes_j|\psi_j\>$ denotes the post-Gaussian-counter-rotation state, assumed without loss of generality to have zero displacement and a diagonal covariance matrix, that cannot be further disentangled. 
			Then, the tomography of such states will take $M$ copies that satisfy:
			\begin{align}
				%M\le&\,\underline\Omega \left[ m^{\frac 3 2}\left(1-\frac{\delta\epsilon^2}{m^{\frac 7 2}\ell}\right)\cdot \sum_{k=1}^\ell  \left(\frac{\epsilon}{2\ell}\right)^{-\map R_k-3}\right].\\
				&M\ge \Theta \left[\frac{1-\delta}{m\log E}\left(\frac{E}{\epsilon}\right)^{ \map R_f+1}\right],\label{eq10_proposition5}\\
				&M\le \mathcal O \left[\textbf{ \em poly}\left( \ell^{5\map R_f+11}, E^{2\map R_f+3},\right.\right.\nonumber\\
				&\left.\left. \left(\frac 1 {\epsilon}\right)^{2\map R_f+6},\log \frac 1 {\delta } , 3^{2 \map R_f+2}\right)\right]\label{eq9_proposition5} 
			\end{align}
			where $\delta$ denotes the failure probability, $\epsilon$ refers to the trace distance between the input and the reconstructed state. Note that we have $\ell=\mathcal O(m)$ by definition. %{\color{blue}maybe covariance matrix or forth moment for constraint}
		\end{proposition}
		
		A detailed proof of Eqs. (\ref{eq9_proposition5}) and (\ref{eq10_proposition5}) is shown in Supplementary Note \ref{app:connection between GE cost and learning}. In addition, it is shown that  the sample-complexity scaling is dominated by covariance-matrix estimation when $\map R_f$ is small (e.g. $\map R_f\le 7$). 
		\\[-0.5em]
		
		\section{Volume of the state space}
		
		%\noindent{\bf Volume of the state space}
		
		%\noindent 
		
		A crucial step in  the proofs of Theorem \ref{theo:sample complexity of GE state} and Proposition \ref{lem:sample complexity upper bound} is to show that Step (BS6) of the algorithm in Supplemental Note \ref{app:tomography algorithm} requires fidelity-witnesses on $ \textbf{exp}\left(\mathcal O\left[\textbf{poly}\left(m\right)\right]\right)$ passive-separable states, so that applying the union bound \cite{becker2024classical} yields a polynomial sample complexity (see Supplemental Notes \ref{app:connection between GE cost and learning} and \ref{app:proof of theorem 2 for pure GE states}). 
		
		%The number of target states  is closely related to the volume of the underlying state space. 
		Intuitively, given that the state is known to lie within a specific class, the problem of determining how many states can be distinguished by tomography reduces to quantifying the amount of classical information required to specify any state in that class to within an error $\epsilon$. The associated quantity can also be referred to as the state description complexity or volume of the state space, whose logarithm determines the number of classical bits required to transmit the information. More rigorously, the following definition is given:
		\begin{definition}[$\epsilon$-covering net]
			Let $\map S$ denote the set of $m$-mode pure states $|\psi\>$. Then, there exists a $\epsilon$-covering net $\map S_{\rm net}\subset\map S$, such that for every  $|\psi\>\in\map S$, there exists $|\psi'\>\in\map S_{\rm net}$ satisfying $1/2\||\psi\>\<\psi|-|\psi'\>\<\psi'|\|_1\le \epsilon$ where $\|X\|_1=\sqrt{\Tr[X^\dag X]}$ denotes the trace norm.
		\end{definition}
		
		Thus, the number of target states that require certification in tomography –referred to as the 'volume' of the state space –is determined by the cardinality of the corresponding  $\epsilon$-covering net. To visualize the upper bound, let us look at a more general case with GE states:
		\begin{corollary}[Volume of GE state space]\label{theo:computational complexity} Consider the set of $m$-mode pure GE states $\{|\psi\>\in \calH_{m,E}^1\}$ produced by applying a Gaussian unitary on the tensor product of single-mode states. 
			%$\{U_{\bs \alpha}U_S \otimes_{j=1}^m |\psi_j\>\}$ where $\{|\psi_j\>\}$ are single-mode local states.  %assumed without loss of generality to have zero displacement, a diagonal covariance matrix. The overall state satisfies $\psi\in \calH_{m,E}^1$. 
			%$\<\psi|\left(\sum_{j=1}^m a^\dag_j a_j+\frac {mI} 2\right)|\psi\>\le m E_{\rm I}$. 
			The size of the corresponding $\epsilon$-covering net is bounded by: 
			\begin{align}
				\left|\map S_{\rm ge,net}\right| \le&\,\textbf{\em exp}\left(\mathcal O\left[{\textbf{\em poly}} \left(m^3, E, \frac{1}{\epsilon^2}\right)\right]\right).\label{eq:epsilon covering net of GE}
			\end{align}
		\end{corollary}
		The proof of Corollary \ref{theo:computational complexity} is shown in Supplemental Note \ref{app:learning GE states}. In contrast, the size of $\epsilon$-covering net fulfilling the energy constraint $|\psi\>\in \calH_{m,E}^1$ is lower bounded as $\left|\map S_{\rm e,net}\right|\ge  \textbf{ exp}\left( \Theta
		%\mathcal O
		\left[\left(\frac{E}{12\epsilon^2}\right)^m\right]\right)$ \cite{mele2024learning}. 
		
		In our algorithm, we propose local fidelity witnesses to approximate the global fidelity witness for passive-separable states (see Fig.~\ref{fig:scheme_learning}). This requires slightly fewer observables than the 'volume' of passive-separable states, which is already smaller than that of GE states in Eq.~(\ref{eq:epsilon covering net of GE})—at the cost of a controllable global fidelity witness error (see Supplemental Note \ref{app:tomography algorithm} for details).\\[-0.5em]
		
		%Definition+motivation. then quote general case is exp. have a theorem say GE is exp poly. 

		%\XZ{Minor revision: highlight the moment/energy constraint in CV systems in suitable places.   }
		
		% \iffalse 
		
		% {\em Discussion on the NG entropy of mixed states} \XZ{I think we could remove this appendix since infinite additive noise can transform any state into separable. } The definitions of NG entropy and fidelity of GE state conversion are not straightforward for mixed states, since the transformation from a general mixed GE state to a separable state is unknown. Still, there exist certain types of Gaussian channels and its corresponding GE states for which the decoupling can be properly defined. For instance, when a Gaussian channel induces a transformation on covariance matrix as $V\to XVX^T+Y$ with $X$ being an invertible matrix, this Gaussian channel will produce GE states with a zero-value NG entropy after a minimization over all Gaussian channels (see Supplementary Note %\ref{app:reversion of mixed GE state} for details)
		% . In addition, GE states generated by two-mode squeezing operation followed by photon loss can be decoupled via another two-mode squeezed operation \cite{shi2024entanglement}. Therefore, one can directly apply counter rotation when defining NG entropy and success probability for such type of states.  
		% \fi 

		\section{Connection to universality}
		
		%\noindent{\bf Connection to universality}\\
		To produce multi-mode non-Gaussian states, a general approach is to enable inline multi-mode non-Gaussian interactions, which is challenging especially in a network setting. 
		Despite the universality~\cite{lloyd1999quantum} reduction of such interactions into Gaussian operations and single-mode non-Gaussian gates, the required concatenation of inline single-mode non-Gaussian gates and multiple rounds of Gaussian operations remains a conundrum for experimental realization. In this regard, it is intriguing to consider the class of Gaussian-entanglable (GE) states generated from a single round of multi-mode Gaussian interaction from separable local non-Gaussian states. Avoiding the most challenging inline non-Gaussian gates, GE states are more experimentally friendly.\\[-0.5em]

		\section{Related works}
		
		%\noindent{\bf Related works}\\
		So far, various attempts have been made to understand the interplay between entanglement and non-Gaussianity. From the perspective of measurement design, the paradigmatic  Boson sampling protocol \cite{knill2001scheme} employs passive Gaussian operations and single-photon detectors, demonstrating the potential for scalable photonic quantum computing. 
		More recently, a state-driven resource theory about non-Gaussian correlation has been proposed in Ref.~\cite{park2017quantifying}. Nevertheless, this resource theory is non-convex, hence being operationally constrained. %{\color{red} and fails to quantify non-Gaussian entanglement}. %We will provide a resource theory that is convex, operation-driven and quantifies genuine non-Gaussian entanglement. 
		Further, Refs.~\cite{sperling2019mode,chabaud2022holomorphic,chabaud2023resources,lopetegui2024detection} consider non-Gaussian entanglement based upon whether a state can be mapped to a separable state via a beam-splitter network. %our work generalizes the notion of passive separability to Gaussian-entanglable---GE states include all passive-separable states. 
		In addition to advancing understanding of GE states, there are also investigations into the implementation of non-Gaussian operations on Gaussian entangled states~\cite{navarrete2012enhancing,walschaers2017entanglement}. Lastly, Ref.~\cite{laha2024genuine} focuses on non-Gaussian entanglement in light-atom interactions and the formulation and resource theory of genuine non-Gaussian entanglement is not studied. \\[-0.5em]%, while the present work focuses on applying Gaussian operations on non-Gaussian separable states. 
		%Hyrbid resource theory? Ref.~\cite{chitambar2016}, \XZ{unifies the resource theories of entanglement and coherence, while our work analyses a joint behavior between non-Gaussianity and entanglement in an operational setting. }
		%Indeed, local operation, classical communication, and Gaussian entangled state are sufficient to produce any quantum state. Specifically, one can locally produce any bipartite state at one end and, after that, use Gaussian entanglement and operations to execute teleportation, thus producing any desired state. For the same reason, Gaussian protocol and non-Gaussian ancillas will be able to produce any quantum state because non-Gaussian ancilla and Gaussian unitary can be used to realize non-Gaussian gate and achieve universality. For example, cubic phase state plus Gaussian operations allow the implementation of cubic-phase gate~\cite{gottesman2001encoding}, which is universal when combined with Gaussian operations~\cite{lloyd1999quantum}.
		
		%the probability of heterodyne measurement is: \begin{align}p(\alpha,\beta|\hat \sigma_{AB} )&=\Tr[\hat \sigma_{AB}(|\underline{\alpha}\>\<\underline{\alpha}|_A\otimes |\underline{\beta}\>\<\underline{\beta}|_B)]\\&= \sum_{x,y} P_{x,y}\Tr \left[(\rho_{AB,y}\otimes |0\>\<0|^{\otimes K})\, U_{x}^\dag  (|\underline{\alpha}\>\<\underline{\alpha}|_A\otimes |\underline{\beta}\>\<\underline{\beta}|_B\otimes I_{\rm anc})U_x\right]\end{align}

		\addtocontents{toc}{\protect\setcounter{tocdepth}{-1}}
		
		\bibliographystyle{apsrev4-1}
		\bibliography{reference}

		\begin{widetext}

			\addtocontents{toc}{\protect\setcounter{tocdepth}{3}}

			\clearpage            
			\beginsupplement  
			
			\section*{Supplemental information}

			\section{Preliminaries}
			
			\subsection{General introduction to continuous-variable systems}\label{app:general introduction to CV systems}
			Continuous-variable quantum systems model continuous degrees of freedom, such as optical fields~\cite{walls2007quantum}, mechanical motions of ions~\cite{fluhmann2019encoding}, and microwave cavities~\cite{blais2020quantum}. Owing to their information-rich infinite-dimensional Hilbert space, these systems are crucial for various quantum information processing tasks. 
			For example, existing fiber communication protocols are all based on coherent-state encoding that occupies all levels of photon numbers of the optical field, which is also known to be the capacity-achieving encoding for fiber communication~\cite{giovannetti2014ultimate}; The majority of quantum sensing advantages has been relying on squeezed vacuum~\cite{lawrie2019quantum}---a continuous-variable quantum state, including the famous use case on Laser Interferometer Gravitational-Wave Observatory (LIGO)~\cite{ganapathy2023broadband}; Moreover, continuous-variable systems are also promising candidates for measurement-based quantum computing~\cite{larsen2021fault}, in particular with the development of the Gottesman-Kitaev-Preskill (GKP) codes, which are also capacity-achieving for quantum communication over fibers~\cite{gottesman2001encoding,brady2024advances}. More broadly speaking, a transmon qubit---a popular unit for quantum computing---is obtained by focusing on the two lowest energy levels of an infinite-dimensional continuous-variable system~\cite{kjaergaard2020superconducting}.
			
			Among the continuous-variable quantum states and operations, Gaussian states and operations~\cite{weedbrook2012gaussian,serafini2017quantum} are not only analytically tractable but also experimentally friendly, in particular for quantum optical systems. Gaussian operations such as beam-splitters, phase rotations, squeezing and homodyne/heterodyne detection are realized by well-developed off-the-shelf quantum optical components. They also model the communication links in a quantum network.
			However, such Gaussian resources are limited in their capabilities: they are not sufficient for entanglement distillation~\cite{eisert2002distilling,giedke2002characterization,fiuravsek2002gaussian,zhang2010distillation}, error correction~\cite{niset2009no}, loophole-free violation of Bell's inequality~\cite{banaszek1998nonlocality,banaszek1999testing,filip2002violation,chen2002maximal,nha2004proposed,invernizzi2005effect,garcia2005loophole,ferraro2005nonlocality}, and universal quantum computation~\cite{lloyd1999quantum,bartlett2002universal,ohliger2010limitations,menicucci2006universal}. Therefore, characterizing quantum states and operations beyond the Gaussian class is important for understanding the resource needed for these tasks. Towards this end, various types of resource theories of non-Gaussianity have been developed~\cite{marian2013relative,genoni2008quantifying,genoni2010quantifying,zhuang2018resource,takagi2018convex,albarelli2018resource,chabaud2020stellar,chabaud2021}. 
			
			In the following supplementary notes, we consider Gaussian protocol defined as follows: 
			\begin{definition}[Gaussian protocol]
				Gaussian protocol $\mathfrak G$ includes the following set of quantum operations: (1) Gaussian unitaries on the system or part of the system, (2) Partial trace, (3) Composition with ancillary vacuum states, (4)  Probabilistically applying the above operations. 
				\label{def_Gaussian_protocol}
			\end{definition}

			% In particular, Ref.~\cite{takagi2018convex} provides an operational-driven resource theory of non-Gaussianity that allows probabilistic Gaussian protocol to be free operations.

			% Recent quantum state engineering has progressed towards multi-mode continuous-variable quantum systems~\cite{diringer2024,brady2024advances} and towards a quantum network setting.  Indeed, multi-mode GKP states are superior to the single-mode GKP states in error correction~\cite{royer2022encoding,wu2023optimal,conrad2022gottesman,brady2024safeguarding}. 

			% For example, multi-mode GKP states are GE and can be conveniently engineered by applying Gaussian unitaries on single-mode GKP states~\cite{conrad2022gottesman,brady2024safeguarding}, reducing the challenge to preparing single-mode GKP states. 
			% To capture such an advantage in state engineering, we define the Gaussian-entanglable (GE) states, states that can be engineered by locally preparing non-Gaussian state and apply one round of Gaussian operations. 

			\subsection{Preliminary on Gaussian channels}\label{app:property of Gaussian unitary}
			
			Gaussian quantum channels are completely positive trace-preserving (CPTP) maps that transform all Gaussian states into Gaussian states. During the process, the displacement and covariance matrix of a Gaussian state will undergo transformations 
			\begin{align}\label{CPTP-Gaussian-moments}
				\begin{cases}
					\bs \xi&\to X \bs \xi+\bs d\\
					V&\to X V X^T +Y    
				\end{cases}
			\end{align}
			where $\bs d=(d_1,\cdots,d_{2m})^T$ is a real vector, $X$ and $Y=Y^T$ are real matrices satisfying the complete positivity condition $Y+i \Omega-iX \Omega X^T\ge 0$, $\Omega$ is the symplectic form \cite{serafini2017quantum}. In a Gaussian unitary process $\map U(\cdot)=U \cdot U^\dag$, the quadrature operators will be transformed as $U^\dag \bs r U =S\bs r+\bs d$, where $S$ is a symplectic matrix satisfying the condition  $S\Omega S^T=\Omega$, $\bs d$ is a real vector. Therefore, the displacement of a Gaussian state will be transformed as $\bs \xi\to S \bs \xi +\bs d$. Accordingly, the covariance matrix will be subject to a symplectic transformation $V\to SVS^T$. %Furthermore, as stated by Lemma \ref{lem:Bloch Messiah decomposition}, every Gaussian unitary may be broken down into the form of Eq.~(\ref{gaussian unitary}). 
			For more general input states, the following lemma is given: 
			
			\begin{lemma}\label{lem:covariance matrix of nonGaussian state plus Gaussian unitary}
				Consider a Gaussian unitary that transforms Gaussian states with an additional displacement $\bs d$ and a symplectic transformation $S$. Then, given an arbitrary (probably non-Gaussian) state with a displacement $\bs \xi$ and a covariance matrix $V$,  this Gaussian unitary will transform the state's displacement vector $\bs \xi $ into $S\bs\xi+\bs d $ and the covariance matrix $V$ into $SVS^T$.  
			\end{lemma}
			
			\begin{proof} Consider a state $U\rho U^\dag$ with $U$ being a Gaussian unitary. Its displacement is $\bs \xi'=\Tr[\bs r U \rho U^\dag ]=\Tr[(S\bs r +\bs d) \rho  ] = S \bs \xi+\bs d $. Further, its covariance matrix is $V:=\frac 1 2 \Tr [\{({\bs r}-\bs \xi'),({\bs r}-\bs \xi')^{\rm T}\}U\rho U^\dag ]=\frac 1 2 \Tr [\{(S{\bs r}-S\bs \xi),(S{\bs r}-S\bs \xi)^{\rm T}\}\rho  ]=SVS^T$. 
			\end{proof}

			Explicitly, any multi-mode Gaussian channel can be equivalently described in the dilated form: 
			\begin{lemma}\label{lem:Bloch Messiah decomposition} (Bloch Messiah decomposition \cite{braunstein2005squeezing})
				An arbitrary $m$-mode Gaussian channel $\map G$ can be expressed as follows: 
				\begin{align}
					\map G(\cdot)&=\Tr_{\rm anc} \left[U(\cdot \otimes |0\>\<0|^{\otimes m'})U^\dag\right]  \\
					U&= \left(\bigotimes_{j=1}^{m+m'} D(\alpha_j) \right)  U_2  \left(\bigotimes_{j=1}^{m+m'} S_{\rm q}(r_j) \right) U_1 \label{gaussian unitary}
				\end{align}
				where $U_1,U_2\in \text{\em U}(m+m')$ are $(m+m')$-mode beam-splitter operations with arbitrary phase rotation, $D(\alpha_j)=e^{\alpha_j a^\dag_j -\alpha_j^* a_j}$, with an amplitude $\alpha_j=(d_{2j}+id_{2j+1})/2$ defined with Eq.~(\ref{CPTP-Gaussian-moments}), denotes the displacement operator on the $j$-mode that leaves the covariance matrix unchanged, $S_{\rm q}(r_j)=e^{ {r_j}  ( a_j^2 - a_j^{\dag 2})/2}$ with $r_j\in \R$ denotes the single-mode squeezing operator on the $j$-mode, respectively. 
			\end{lemma}
			
			In phase space, the following lemma is given: 
			
			\begin{lemma}\label{lem:williamson}
				(Williamson theorem, uniqueness (Theorem 8.11 and Proposition 8.12 of \cite{de2006symplectic}))
				
				Let $V$ be a positive-definite symmetric real $2m\times 2m$ matrix, there exists $S\in {\rm Sp}(2m,\R)$ such that $S^TVS=\Lambda=\bigoplus \lambda_j I^{(2)}$.
				
				(i)The sequence $\lambda_1,\cdots,\lambda_n$ does not depend, up to a reordering of its items, on the choice of $S$ diagonalising $V$.
				
				(ii) Assume $S, S^\prime\in {\rm Sp}(2m,\R)$ such that $S^{\prime T}V S^\prime=S^T V S=\Lambda$ is the Williamson diagonal form. Then $S(S^\prime)^{-1}\in {\rm U}(m)$ is unitary (orthogonal and symplectic).

			\end{lemma}
			
			In addition, the following lemmas are given: 
			
			\begin{lemma}
				\label{lemma_nmode_Williamson}
				Consider a positive-definite symmetric $2m\times 2m$ matrix $V$. All its symplectic eigenvalues $\{\lambda_i\}$ are different: $\lambda_i\neq \lambda_j, \forall i\neq j$. Then for $S, S^\prime\in {\rm Sp}(2m,\R)$ such that $V=S^{\prime T}\Lambda S^\prime=S^T \Lambda S$ where $\Lambda=\bigoplus_j \lambda_j I^{(2)}$ is the Williamson diagonal form, we have $S(S^\prime)^{-1}=\bigoplus_j U_j$  where $U_j\in {\rm U}(1)$ corresponds to single-mode phase rotations.
			\end{lemma}
			
			\begin{proof}
				
				From Lemma~\ref{lem:williamson}(ii), the unitary can be written as: 
				\begin{align}
					U=S(S')^{-1}=\left(\begin{matrix}
						U_{11} & U_{12} & \cdots & U_{1m}\\
						U_{21} & U_{22} & \cdots & U_{2m}\\
						\vdots & \vdots  & \ddots & \vdots \\
						U_{m1} & U_{m2} & \cdots & U_{mm}
					\end{matrix}\right)
				\end{align}
				where $\{U_{ij}\}$ are $2\times 2$  matrices%and $I^{(2)}$ is a $2\times 2$ identity operator
				. Meanwhile, we can write the Williamson diagonal form as
				\begin{align}
					\Lambda=\left(\begin{matrix}
						\lambda_1 I^{(2)} & 0 & \cdots & 0\\
						0 & \lambda_2 I^{(2)} & \cdots & 0\\
						\vdots & \vdots  & \ddots & \vdots \\
						0 & 0 & \cdots & \lambda_m I^{(2)}
					\end{matrix}\right),
				\end{align}
				where $I^{(2)}$ represents the $2\times 2$ identity matrix. Since $V=S^{\prime T}\Lambda S^\prime=S^T \Lambda S$, we have the condition $U\Lambda U^\dag =\Lambda $, which indicates that $U\Lambda=\Lambda U$. Equivalently, we have: 
				\begin{align}
					\lambda_i U_{ij}=U_{ij} \lambda_j.
				\end{align}
				Therefore, if the eigenvalues are different, i.e. $\lambda_i\neq \lambda_j, \forall i,j$, then $U_{i\neq j}=0$. In this case, the unitary $U$ must be a block-diagonal matrix in the form $U=\bigoplus_{j=1}^m U_{jj}$, where each $U_j\in \text{U}(1)$ corresponds to a single-mode phase rotation operation.    
			\end{proof}
			
			\begin{lemma}\label{lem:degenerate_uniqueness}
				Consider a positive-definite symmetric $2m\times 2m$ matrix $V$. It has  $d<m$ different symplectic eigenvalues $\{\lambda_j\}_{j=1}^d$, where the $j$-th  symplectic eigenvalue $\lambda_j$ has degeneracy $m_j$ (satisfying the condition  $\sum_{j=1}^d m_j=m$). Then, the Williamson form is $\Lambda=\bigoplus_{j=1}^d \lambda_j I^{(2m_j)}$ where $I^{(2m_j)}$ denotes the $2m_j\times 2m_i$ identity matrix.
				For $S, S^\prime\in {\rm Sp}(2m,\R)$ such that $V=S^{\prime T}\Lambda S^\prime=S^T \Lambda S$, we have $S(S^\prime)^{-1}=\bigoplus_{j=1}^d U_j$, where $U_j\in {\rm U}(m_j)$ is a $2m_j \times 2m_j$ orthogonal symplectic matrix and corresponds to an $m_j$-mode beam-splitter network.
			\end{lemma}
			
			\begin{proof}
				Similar to the proof of Lemma~\ref{lemma_nmode_Williamson}, 
				the unitary
				\begin{align}
					U=S(S')^{-1}=\left(\begin{matrix}
						U_{11} & U_{12} & \cdots & U_{1d}\\
						U_{21} & U_{22} & \cdots & U_{2d}\\
						\vdots & \vdots  & \ddots & \vdots \\
						U_{d1} & U_{d2} & \cdots & U_{dd}
					\end{matrix}\right),
				\end{align}
				where the block form $U_{ij}$ has dimension $2m_i\times 2m_j$, and the
				the Williamson diagonal form as
				\begin{align}
					\Lambda=\left(\begin{matrix}
						\lambda_1 I^{(2m_1)} & 0 & \cdots & 0\\
						0 & \lambda_2 I^{(2m_2)} & \cdots & 0\\
						\vdots & \vdots  & \ddots & \vdots \\
						0 & 0 & \cdots & \lambda_d I^{(2m_d)}
					\end{matrix}\right).
				\end{align}
				
				Now the condition $U\Lambda=\Lambda U$ indicates that 
				\begin{align}
					\lambda_i U_{ij}=U_{ij} \lambda_j, \forall 1\le i,j \le d.
				\end{align}
				Therefore, we have off-diagonal terms being zero: $U_{ij}=0$ for $1\le i\neq j \le d$. We can conclude that the unitary $U$ is block diagonal in the form: $U=\bigoplus_{j=1}^d U_j$, where $U_j\in {\rm U}(m_j)$ is a $2m_j \times 2m_j$ orthogonal symplectic matrix and corresponds to an $m_j$-mode beam-splitter network.
			\end{proof}

			\section{Proof of Theorem~\ref{lem:pureGEstate}}
			\label{app:proof of the pure state theorem}

			\subsection{Proof that coherent state is the only state that remains pure after loss}\label{app:proof of coherent state is the only state that remains pure after pure loss}
			
			\begin{lemma}[Generalizing Lemma 2 in Ref.~\cite{mari2014quantum}]\label{lem:13} The only states that remain pure under multi-mode pure loss are coherent states. 
			\end{lemma}
			
			\begin{proof} Pure loss process can be  modeled as a beam-splitter operation jointly acting on the input and vacuum environment states \cite{serafini2017quantum}. Without loss of generality, let us consider an ($m+m'$)-mode beam-splitter unitary $V$ acting on an $m$-mode state $|\Psi\>$ with $m'$ vacuum states. The unitary $V$ transforms the field operator of the $j$-th mode ($j=1,\cdots,m+m'$) as follows: 
				\begin{align}
					V\,a_j\,V^\dag&=\sum_{k=1}^{m+m'} u_{kj} a_k 
				\end{align}
				where the coefficients $\{u_{jk}\}$ satisfies the normalization condition  $\sum_j u_{hj}u^*_{jk}=\delta_{hk}$.  Then, this input-output relation indicates that the output state of the channel has a Wigner characteristic function:
				\begin{align}
					\chi'\left(\{z_1,\cdots,z_m\}\right) =&  \chi \left(\left\{\sum_{h=1}^{m} u_{jh} z_h;\,j=1,\cdots,m \right\}\right) \exp\left(-\frac 1 2 \sum_{k=m+1}^{m+m'}\left|\sum_{l=1}^{m}u_{kl} z_l  \right|^2\right)
					\label{Wigner_S_out}
				\end{align}
				and that the output ancilla (vacuum at the input side) has the Wigner characteristic function 
				\begin{align}
					\chi_v'\left(\{z_{m+1},z_{m+m'}\}\right) =&  \chi \left(\left\{\sum_{h=m+1}^{m+m'} u_{jh} z_h;\,j=1,\cdots,m \right\}\right) \exp\left(-\frac 1 2 \sum_{k=m+1}^{m+m'}\left|\sum_{l=m+1}^{m+m'}u_{kl} z_l  \right|^2\right).
					\label{Wigner_V_out}
				\end{align}
				In the meantime, the overall Wigner characteristic function $\chi'_{\rm All}$ can also be obtained by the overall input-output relation
				\begin{align}
					\chi'_{\rm All}\left(\{z_1,\cdots,z_m,z_{m+1},\cdots,z_{m+m'}\}\right)=\chi \left(\left\{\sum_{h=1}^{m+m'} u_{ji} z_h;\,j=1,\cdots,m \right\}\right) \exp\left(-\frac 1 2 \sum_{k=m+1}^{m+m'}\left|\sum_{l=1}^{m+m'}u_{kl} z_l  \right|^2\right).
					\label{Wigner_SV_out}
				\end{align}
				
				The output of a beam-splitter operation is a product of a $m$-mode state and a $m'$-mode state implies that the overall Wigner characteristic function $\chi'_{\rm All}\left(\{z_1,\cdots,z_m,z_{m+1},\cdots,z_{m+m'}\}\right)=\chi'\left(\{z_1,\cdots,z_m\}\right) \chi_v'\left(\{z_{m+1},\cdots,z_{m+m'}\}\right)$ can be written as a product of reduced state Wigner characteristic function. Combining Eqs.~\eqref{Wigner_S_out}, \eqref{Wigner_V_out} and \eqref{Wigner_SV_out}, we have
				\begin{align}
					&\chi \left(\left\{\sum_{h=1}^{m} u_{jh} z_h;\,j=1,\cdots,m \right\}\right) \exp\left(-\frac 1 2 \sum_{k=m+1}^{m+m'}\left|\sum_{l=1}^{m}u_{kl} z_l  \right|^2\right) \times \nonumber\\
					&\chi \left(\left\{\sum_{i=m+1}^{m+m'} u_{jh} z_h;\,j=1,\cdots,m \right\}\right) \exp\left(-\frac 1 2 \sum_{k=m+1}^{m+m'}\left|\sum_{l=m+1}^{m+m'}u_{kl} z_l  \right|^2\right)\nonumber
					\\
					=&  \chi \left(\left\{\sum_{i=1}^{m+m'} u_{jh} z_h;\,j=1,\cdots,m \right\}\right) \exp\left(-\frac 1 2 \sum_{k=m+1}^{m+m'}\left|\sum_{l=1}^{m+m'}u_{kl} z_l  \right|^2\right).
				\end{align}
				
				If we define a function $w(\{z_1,\cdots\})=\chi(\{z_1,\cdots\})\exp\left(\frac 1 2 \sum_j |z_j |^2\right)$ we have: 
				\begin{align}
					&w \left(\left\{\sum_{i=1}^{m} u_{jh} z_h;\,j=1,\cdots,m \right\}\right)w \left(\left\{\sum_{h=m+1}^{m+m'} u_{jh} z_h;\,j=1,\cdots,m \right\}\right)\nonumber \\
					&\times \exp\left[-\frac 1 2 \sum_{k=1}^{m+m'}\left(\left|\sum_{l=1}^{m}u_{kl} z_l  \right|^2+\left|\sum_{l=m+1}^{m+m'}u_{kl} z_l  \right|^2\right)\right]\nonumber \\
					=& w \left(\left\{\sum_{h=1}^{m+m'} u_{jh} z_h;\,j=1,\cdots,m \right\}\right) \exp\left(-\frac 1 2 \sum_{k=1}^{m+m'}\left|\sum_{l=1}^{m+m'}u_{kl} z_l  \right|^2\right)\\
					=& w \left(\left\{\sum_{i=1}^{m+m'} u_{jh} z_i;\,j=1,\cdots,m \right\}\right) \exp\left[-\frac 1 2 \sum_{k=1}^{m+m'}\left(\left|\sum_{l=1}^{m}u_{kl} z_l  \right|^2+\left|\sum_{l=m+1}^{m+m'}u_{kl} z_l  \right|^2\right)\right]
				\end{align}
				where the last equation uses the relation $\sum_j u_{hj}u^*_{jk}=\delta_{hk}$. Therefore, we have: 
				\begin{align}
					w\left(\left\{z'_j;j=1,\cdots,m\right\}\right)w\left(\left\{z''_j;j=1,\cdots,m\right\}\right)=w\left(\left\{z'_j+z''_j;j=1,\cdots,m\right\}\right)\label{eq32}
				\end{align}
				where $z'_j=\sum_{h=1}^{m} u_{jh} z_h$ and $z''_j=\sum_{h=m+1}^{m+m'} u_{jh} z_h$. Further, the function $w(\{z_1,\cdots\})$ satisfies $w(\{-z_1,-z_2,\cdots\})=\overline{w(\{z_1,\cdots\})}$. The only solution satisfying these conditions is the exponential function, which indicates: 
				\begin{align}\label{eq33}
					\chi (\{z_1,\cdots\})=\exp\left[\sum_j z_j^* \alpha_j -z_j\alpha_j^*-\frac 1 2 |z_j|^2\right].
				\end{align}
				{\color{black} This can be quickly proved: First, Eq.~(\ref{eq32}) shows that $\omega(\{z_j\})$ is nowhere zero: if it vanished at any single point $\{z_j\}$, continuity would force it to vanish everywhere, contradicting Eq.~(\ref{eq32}). Because $\omega(\{z_j\})\neq 0$   for all arguments, we may write
					$\omega(\{z_j\})=\exp(\ln \omega(\{z_j\}) )$ where $\ln$ is the complex logarithm function for nonzero complex numbers. Moreover, Eq.~(\ref{eq32}) implies that $\ln \omega(\{z_j\})$ depends linearly on the variables $\{z_j\}$. Combining these observations as well as the constraint $w(\{-z_1,-z_2,\cdots\})=\overline{w(\{z_1,\cdots\})}$ yields Eq.~(\ref{eq33}).  }
				Finally, it can be verified that this is the characteristic function of a (multi-mode) coherent state.
			\end{proof}

			\subsection{Proof of Theorem~\ref{lem:pureGEstate}}

			In this note, we will prove Theorem~\ref{lem:pureGEstate}, which asserts that every pure GE state can be represented as a Gaussian unitary applied to a product state. Before showing the main proof, let us look at the following lemma: 
			
			\begin{lemma}
				\label{lemma:pure_GUP}
				Consider a pure state that can be expressed as $\ketbra{\Psi}_{AB}=\mathfrak G(|\psi\>\<\psi|_A\otimes |\phi\>\<\phi|_B)$, where $\mathfrak G$ is a Gaussian protocol as defined in Definition~\ref{def_Gaussian_protocol}. Then, the state $\ketbra{\Psi}_{AB}$ can be also expressed as a Gaussian unitary acting on a product state. 
			\end{lemma}
			
			\begin{proof} According to Definition~\ref{def_Gaussian_protocol}, any Gaussian protocols can be interpreted as a probabilistic implementation of Gaussian channels. If the output state is restricted to pure states, the Gaussian protocol can be equivalently expressed as a single multi-mode Gaussian channel, independent of any probabilistic process. Given the general unitary dilation theorem for bosonic Gaussian channels \cite{caruso2008multi}, we have:
				\begin{align}
					\ketbra{\Psi}_{AB}&= \Tr_{\rm anc} \left[U^g(|\psi\>\<\psi|_A\otimes |\phi\>\<\phi|_B \otimes |0\>\<0|_{\rm anc})U^{g\dag}\right].  
				\end{align}
				where $|0\>_{\rm anc}=|0\>^{\otimes m'}$ denotes the vacuum state of the ancillay modes, $U^g$ is a Gaussian unitary acting on $AB$ and ancilla that induces a symplectic transformation $\bs \xi\to S\bs \xi+\bs d$ in the phase space. 
				
				Meanwhile, we have a premise that the output ancilla must be a pure state.  Therefore, we have: 
				\begin{align}
					\ket{\Psi}_{AB}\ket{\zeta}_{\rm anc} =U^g \ket{\psi}_A\otimes \ket{\phi}_B \otimes \ket{0}_{\rm anc}.
				\end{align}
				Given Lemma \ref{lem:covariance matrix of nonGaussian state plus Gaussian unitary}, the covariance matrix of the whole system will be:
				\begin{align}
					\left(\begin{matrix}
						V_{AB}' & 0 \\
						0 & V_{\rm anc}' 
					\end{matrix}\right)= S\left(\begin{matrix}
						V_{AB} & 0 \\
						0 & I 
					\end{matrix}\right)S^T.
				\end{align}
				Suppose we apply a symplectic transform  on both sides to diagonalize the covariance matrix of $V_{AB}'$ and $V_{\rm anc}'$, we have
				\begin{align}
					\left(\begin{matrix}
						\Lambda_{AB}' & 0 \\
						0 & \Lambda_{\rm anc}' 
					\end{matrix}\right)= S'\left(\begin{matrix}
						\Lambda_{AB} & 0 \\
						0 & I 
					\end{matrix}\right)S^{\prime T},
				\end{align}
				where we have absorbed the symplectic transform that diagonalizes $V_{AB}$ into $S'$. Overall, $S'$ and $S$ are the same up to local Gaussian unitaries on AB and the ancilla. Given Lemma \ref{lem:williamson}, $S'$ must be associated with a multi-port beam-splitter operation on the entire system. Note that the state with covariance matrix $V_{AB}$,  $\Lambda_{AB}$, and $\Lambda_{AB}'$ are pure states. Then, from Lemma~\ref{lem:13}, either the input with $\Lambda_{AB}$ is a coherent state, or the beam-splitter acts trivially as a product on AB and ancilla. In both cases, the state can be written as a Gaussian unitary on a product state.
			\end{proof}

			Now, let us prove the following theorem: 
			
			\begin{theorem}
				An arbitrary pure bi-partite GE state can be written as: 
				\begin{align}\label{eqd5}
					|\phi\>_{AB}= U^g |\psi\>_A|\phi\>_B. 
				\end{align}
				where $U^g$ is a Gaussian unitary. 
			\end{theorem}

			\begin{proof}
				
				For an arbitrary pure state $\rho_{AB}$ that can be written in the form of  Eq.~\eqref{eq1} of the main text, we have
				\begin{align}
					0=H(\rho_{AB})\ge  \sum_{x,y} p_{x,y} H(\map G_x(\sigma_{AB,y})),
				\end{align}  
				where $H(\rho)=-\Tr[\rho \log_2 \rho ]$ is the von Neumann entropy. Due to the positivity of the von Neumann entropy, we have $H(\map G_x(\rho_{AB,y}))=0$ for all $x$ and $y$. In addition, equality of concavity is only achieved when all states are equal, i.e., $\map G_x(\rho_{AB,y})=\ketbra{\phi}_{AB}$ is independent of $x$ or $y$. Thus, we have $\ketbra{\phi}_{AB}=\map G(\sigma_{AB})$ where $\map G$ is a Gaussian channel.
				
				Similarly, when considering an arbitrary separable initial state in Eq.~(\ref{eq1}): $\sigma_{AB}
				=\sum_j p_j |\psi_j\>\<\psi_j|_A\otimes |\phi_j\>\<\phi_j|_B$, we have: 
				\begin{align}
					0=H(\map G(\rho_{AB}))\ge \sum_j p_j H(\map G(|\psi_j\>\<\psi_j|_A\otimes |\phi_j\>\<\phi_j|_B)),
				\end{align}
				which indicates $H(\map G(|\psi_j\>\<\psi_j|_A\otimes |\phi_j\>\<\phi_j|_B))=0$ for all $j$ and, equivalently, that the state $\map G(|\psi_j\>\<\psi_j|_A\otimes |\phi_j\>\<\phi_j|_B)$ must be a pure state independent of $j$. Therefore, every pure GE state can be expressed as $|\phi\>\<\phi|_{AB}=\map G(|\psi\>\<\psi|_A\otimes |\phi\>\<\phi|_B)$ where $\map G$ is a Gaussian channel. Finally, from Lemma~\ref{lemma:pure_GUP}, We have for pure GE state $|\phi\>_{AB}=U^g|\psi\>_A\otimes |\phi\>_B$ where $U^g$ is a Gaussian unitary. 
			\end{proof}
			
			Furthermore, the inverse of the Gaussian unitary $U^{g}$ (in Eq.~(\ref{eqd5})) will transform a pure GE state $U^g|\psi\>_A|\phi\>_B$ into a product state, whose covariance matrix is diagonal. From Lemma \ref{lem:williamson}, the symplectic transformation that diagonalizes a covariance matrix is unique, subject to a $\text{U}(n)$ operation. Therefore, for any symplectic matrix $S$ such that $SV_{AB}S^T=\Lambda$ is the Williamson diagonal form, we have the corresponding Gaussian unitary $U_S$ equivalent to $U^{ g}$ up to a  beam-splitter $U_O$ for orthogonal matrix $O$ and displacement operation $U_{\bs \alpha}$ with $\bs \alpha$ being a vector of displacements, i.e., $U^g=U_S U_O U_{\bs \alpha}$.
			
			\section{General properties of GE and NGE states}\label{app:general properties of GE states}
			
			\subsection{Multipartite definition of GE states }\label{app:multipartite definition of GE states}

			To facilitate further discussion on resource theory, let us generalise the notion of GE state to multipartite scenario between $n$ subsystems, each with $\{m_j,j=1,\cdots,n\}$ modes:
			
			\begin{definition}[multipartite Gaussian-entanglable state] \label{defi:multi GE state} Consider a $n$-partition of the Hilbert space $\map H=\map H_{A_1}\otimes \cdots \otimes\map H_{A_n}$. A quantum state $\rho_{A_1A_2\cdots A_n}$ is Gaussian-entanglable if it can be obtained by probabilistically applying a Gaussian channel $\map G_x\in \mathbb G(\map H)$ on a (general) ensemble of separable state as follows: 
				\begin{align}\label{multi GE state}
					&\rho_{A_1A_2\cdots A_n}= \sum_{x_1,x_2,\cdots,x_n} p_{x,y} \map G_{x} \left(\sigma_{A_1A_2\cdots A_n,y}\right),\\
					&\nonumber \sigma_{A_1A_2\cdots A_n,y}\in \mathbb{{SEP}}\left(\map  H|\left\{A_1|\cdots|A_n\right\}\right)
				\end{align}
				where $p_{x,y}$ is a joint probability distribution, $\mathbb{{SEP}}\left(\map H;\left\{A_1|\cdots|A_n\right\}\right)$ denotes the set of fully separable states that can be written as: $\sigma_{A_1A_2\cdots A_n}= \sum_\lambda  p_\lambda \bigotimes_{j=1}^n\sigma_{A_j,\lambda }$. Accordingly, the set of all GE states in terms of partition $\left\{A_1|\cdots|A_n\right\}$ is denoted as $\mathbb{{GE}}\left(\map H;\left\{A_1|\cdots|A_n\right\}\right)$. 
			\end{definition}
			\noindent In addition, we can extend  Theorem~\ref{lem:pureGEstate} of the main text as follows: 
			
			\begin{corollary}
				\label{lem:multipartite extension of theorem 3.5}A pure multipartite GE state can be expressed in the following form:
				\begin{align}
					|\Psi\>_{A_1A_2\cdots A_n}&=U^g\bigotimes_{j=1}^n |\psi_j\>_{A_j} 
				\end{align}
				where $U^g$ is a global Gaussian unitary on all subsystems. 
			\end{corollary}
			\noindent The proof of Corollary \ref{lem:multipartite extension of theorem 3.5} follows that of Theorem  \ref{lem:pureGEstate}, with the only difference being that it assumes the GE state is created from a fully separable state.

			\subsection{Linear correlation and independence}\label{app:Linear correlation and independence}
			
			%\QZ{motivated by classical side: Uncorrelatedness versus independence, \url{https://en.wikipedia.org/wiki/Uncorrelatedness_(probability_theory)}, \url{https://en.wikipedia.org/wiki/Misconceptions_about_the_normal_distribution}}
			In classical statistics, when the covariance of two real-valued random variables ${\rm cov}(X,Y)=\mathbb{E}[XY]-\mathbb{E}[X]\mathbb{E}[Y]=0$, the two random variables $X,Y$ are regarded as (linearly) uncorrelated. However, such (linear-)uncorrelatedness does not mean independence. In terms of Gaussian variables, uncorrelation and independence are equivalent. 
			
			In terms of quantum states, similarly, zero quadrature covariance, does not imply independence. {\color{black} For instance, a given two-mode state $\rho_{AB}$ has no linear correlation between two subsystems $A$ and $B$ when its covariance matrix is a block-diagonal matrix i.e., $V=V_A\oplus V_B$. On the other hand, the two subsystems $A$ and $B$ are independent if the state $\rho_{AB}$ is a product state, i.e. $\rho_{AB}=\rho_A\otimes \rho_B$. } For Gaussian states, uncorrelatedness and independence are equivalent. However, for convex mixtures of Gaussian states, uncorrelatedness and independence are not equivalent. In this work, we obtain a class of states beyond Gaussian states that allow the equivalence:
			\begin{remark}[Linear-uncorrelatedness] 
				For pure GE states where symplectic eigenvalues are non-degenerate, linear-uncorrelatedness and independence are equivalent.
			\end{remark}
			\noindent For mixed GE states, they are not equivalent. In addition, the NGE states in Example i and ii of the main text are such instances where they are linear independent, but two subsystems are dependent.

			\subsection{Conservation law of GE and NGE states }\label{app:onservation law of GE and NGE states}
			
			For GE states, a conservation law can be presented as the following proposition: 
			
			\begin{proposition}\label{lem:Gaussian preserves GE}
				Gaussian protocols (that preserves the number of modes) map GE states on $AB$ to GE states on $AB$. 
			\end{proposition}

			\begin{proof}
				An arbitrary Gaussian protocol acting on a GE state defined by Eq.~(\ref{eq1}) will produce the following state: $\rho_{AB}'=\sum_{x,y,z} q_{x,y,z} p_{x,y} \map G_{x,y,z}'\circ \map G_{x,y}\left(\sigma_{AB,y}\right)$ where $\map G$ and $\map G'$ refer to Gaussian channels that have the same number of input and output modes, $q_{x,y,z}$ and $p_{x,y}$ denote joint probability distributions, $\sigma_{AB,y}$ is a separable state. Given that $\map G_{x,y,z}'\circ \map G_{x,y}$ is also a Gaussian channel with the same number of input and output modes, and that $q_{x,y,z}p_{x,y}$ is still a probability distribution, we can conclude that Gaussian protocol will map a GE state to another GE state. 
			\end{proof}
			
			For NGE states, we have the following proposition: 
			
			\begin{proposition}[Gaussian unitary on NGE state] \label{lem:Gaussian unitary preserves NGE state}
				Gaussian unitary maps NGE states to NGE states. 
			\end{proposition}
			
			\begin{proof}
				Consider a NGE state $\rho_{AB}$ that can not be expressed in the form of Eq.~(\ref{eq1}). If there exists a Gaussian unitary $U^g$ that converts $\rho_{AB}$ into a GE state, we have $U^g\rho_{AB}U^{g\dag}=\sum_{x,y} p_{x,y} \map G_{x} \left(\sigma_{AB,y}\right)$, where $p_{x,y}$, $\Phi_x$, and $\sigma_{AB,y}$ are defined in Eq.~(\ref{eq1}). It implies $\rho_{AB}=\sum_{x,y} p_{x,y} \map G_{x}' \left(\rho_{AB,y}\right)$ where $\map G_{x}'(\cdot)=U^{g\dag}\map G_{x}(\cdot)U^g$ is a Gaussian protocol. Therefore, it contradicts the premise that the state $\rho_{AB}$ is NGE. 
			\end{proof}

			\section{Examples of pure NGE states}\label{app:pure nge state}
			
			\subsection{An example with non-degenerate symplectic eigenvalues}
			
			Consider a quantum state as follows: 
			\begin{align}\label{eqa1}
				|\Psi\>_{\ell,\ell',k,k'}=&\sum_{j,h=0}^\infty  c_{jh} \, |\ell+kj\>_A|\ell'+k'h\>_B,
			\end{align}
			where $k,k'\ge 2$ are integers, $\ell,\ell'\in \N$ are natural numbers, $\{c_{jh}\in\C\}$ are complex numbers satisfying the normalization condition $\sum_{jh}|c_{jh}|^2=1$. For the following derivation, we will denote $\<O\>:=\<\Psi|_{\ell,\ell',k,k'}O|\Psi\>_{\ell,\ell',k,k'}$ for simplicity. In addition, we will consider a natural unit $\hbar=2$, which indicates the position and momentum operators:
			\begin{align}
				\begin{cases}
					q_a=&a+a^\dag\\
					p_a=&i(a^\dag-a)\\
					q_b=&b+b^\dag\\
					p_b=&i(b^\dag-b)
				\end{cases},
			\end{align}
			where $a(b)$ and $a^\dag(b^\dag)$ are annihilation and creation operators of the first(second) mode, respectively. 
			
			\subsubsection{Symplectic eigenvalues}\label{subsec:symp}
			
			It can be quickly verified that its covariance matrix is block diagonal because: 
			\begin{align}\label{eqa2}
				&\<\Delta q_a\Delta q_b\>=\<\Delta p_a\Delta p_b\>=\<\Delta q_a\Delta p_b\>=\<\Delta p_a\Delta q_b\>=0
			\end{align}
			where $\Delta O$ is defined as $\Delta O=O-\<O\>$. Further, the symplectic eigenvalue can be computed by $|i\Omega V_j|$ up to a sign, where $V_j$ refers to the $j$-th sub-blocks of the covariance matrix. More explicitly, if we have $k=2$ and $k'>2$, the symplectic eigenvalues are: 
			\begin{align}
				\nu_1=&1+4\left(\<a^\dag a\>^2+\<a^\dag a \>-\<a^{\dag 2}\>\<a^2\>\right)\\
				=&1+4\left[\left(\sum_{j,h=0}^\infty |c_{jh}|^2(\ell+kh)\right)^2+\sum_{j,h=0}^\infty |c_{jh}|^2(\ell+kj)\nonumber \right.\\
				&\left.-\chi(k)\left|\sum_{j,h=0}^\infty c_{jh}c_{j+1,h}^*\sqrt{(\ell+kj+1)(\ell+kj+2)}\right|^2\right]\\
				\nu_2=&1+4\left(\<b^\dag b\>^2+\<b^\dag b \>-\<b^{\dag 2}\>\<b^2\>\right)\\
				=&1+4\left[\left(\sum_{j,h=0}^\infty |c_{jh}|^2(\ell'+k'h)\right)^2+\sum_{j,h=0}^\infty |c_{jh}|^2(\ell'+k'h)\nonumber \right.\\
				&\left.-\chi(k')\left|\sum_{j,h=0}^\infty c_{jh}c_{j,h+1}^*\sqrt{(\ell'+k'h+1)(\ell'+k'h+2)}\right|^2\right]\\
				\chi(x)=&\begin{cases}
					1&x=2\\
					0&x>2
				\end{cases}
			\end{align}
			Therefore, by choosing different values of $\{\ell,\ell',k,k'\}$ for a specific set of coefficients $\{c_{jh}\}$, we will have different symplectic eigenvalues of the covariance matrix of the state $|\Psi\>_{\ell.\ell',k,k'}$. For example, in the case with $k=k'>2$ and $c_{jh}=c_{jh}$, different symplectic eigenvalues can be induced when the condition $\ell\neq \ell'$ is satisfied.

			\subsubsection{Sufficient condition for entanglement}\label{subsec:ent}
			
			Consider the positive partial transposition (PPT) criterion of separability \cite{horodecki1997separability}. The state $|\Psi\>_{\ell.\ell',k,k'}$ is entangled if its partial transposition has negative eigenvalues. Explicitly, we have: 
			\begin{align}
				|\Psi\>\<\Psi|^{T_B}_{\ell.\ell',k,k'}&=\sum_{j,j',h,h'=0}^\infty  c_{jh} c_{j'h'}^*\,|\ell+kj\>\<\ell+kj'|_A\otimes  |\ell'+k'h'\>\<\ell'+k'h|_B .
			\end{align}
			Provided with the condition that $\{c_{jh}\}$ can not be decomposed into a product of two distributions, i.e., $c_{jh}\neq c_j' d_h'$, it has an entangled eigenvector: 
			\begin{align}
				|\psi\>=\frac{1}{\sqrt 2}\left(\gamma^2|\ell+kj\>|\ell'+k'h'\>-\gamma^{*2}|\ell+kj'\>|\ell'+k'h\>\right)
			\end{align}
			with $\gamma={|c_{j'h'}c^*_{jh}|}/{c_{j'h'}c^*_{jh}}$ and a negative eigenvalue: 
			\begin{align}
				E=- |c_{j'h'}c^*_{jh} | . 
			\end{align}
			
			\subsubsection{Sufficient condition for being a NGE state}

			Theorem \ref{lem:pureGEstate} of the main text indicates that any pure bi-partite EG state whose covariance matrix has different symplectic eigenvalues can be transformed into a product state with a unique Gaussian unitary operation, subject to local transformations. In addition, this Gaussian unitary operation coincides with the Gaussian unitary that block-diagonalize the covariance matrix. 
			
			Given the proof in Supplementary Notes \ref{subsec:symp} and \ref{subsec:ent}, it is possible to construct a bi-partite entangled state in the form of Eq.~(\ref{eqa1}) such that its symplectic eigenvalues are different. Meanwhile, the identity operation could be the only operation that diagonalises its covariance matrix. In this case, the state must be a NGE state.

			\subsection{Degenerate cases}
			\label{app:degenerate_NGE}

			\subsubsection{Superposition of two-mode squeezed vacuum states}
			
			Let us consider a specific case of Eq.~(\ref{eqa1}) with degenerate symplectic eigenvalues: 
			\begin{align}\label{eqe12}
				|\Psi\>=&\frac 1 {\sqrt {2+ 2\cosh^{-1} 2r }}\left[S_{{\rm q},2}(r)+S_{{\rm q},2}(-r)\right]|0\>_A|0\>_B
			\end{align}
			where $S_{{\rm q},2}(r)=\exp(rab -r a^\dag b^\dag ),(r>0)$ denotes the two-mode squeezing operator. Its covariance matrix is as follows: 
			\begin{align}
				V=&\lambda I^{(4)}\\
				\lambda =&8 \sinh ^4(r) \text{sech}(2 r)+1
			\end{align}
			Given Theorem \ref{lem:pureGEstate} in the main text, an arbitrary GE state can be written in the form $U_S |\psi\>_A|\phi\>_B$ where $U_S$ is a Gaussian unitary associated to a symplectic transformation $S$. If the state Eq.~(\ref{eqe12}) is a GE state, from Lemma \ref{lem:williamson}, we have: 
			\begin{align}\label{eqe15}
				U'|\Psi\>=|\psi\>_A|\phi\>_B,
			\end{align}
			where $U'\in\text{U}(2)$ denotes a two-mode beam-splitter operation that trivially acts on the two symmetric sub-blocks for the covariance matrix. More explicitly, a beam-splitter operation leads to the following transformation: 
			\begin{align}\label{eqe16}
				\begin{cases}
					U'aU^{'\dag}  & =\alpha a + \beta b\\
					U'bU^{'\dag}  &= -e^{i\theta }\beta^* a + e^{i\theta }\alpha^* b  \\
					U'|\Psi\>&=\frac 1 {\sqrt {2+2\cosh^{-1}2r}}\left[U'S_{{\rm q},2}(r)U^{'\dag} +U'S_{{\rm q},2}(-r)U^{'\dag} \right]|0\>_A|0\>_B. 
				\end{cases}
			\end{align}
			where $\alpha$ and $\beta$ are complex numbers satisfying $|\alpha|^2+|\beta|^2=1$, $\theta\in \R$ denotes a real number. The state $U'|\Psi\>$ can be rewritten as: 
			\begin{align}
				U'|\Psi\>=&  \frac{2}{\cosh r\sqrt {2+ 2\cosh^{-1} 2r }}\sum_{k=0}^\infty \frac{\tanh^{2k} r}{k!} \left[(\alpha^*a^\dag +\beta^*b^\dag )(-e^{-i\theta}\beta a^\dag+e^{-i\theta }\alpha b^\dag )\right]^{2k} |0\>_A|0\>_B\\
				=&  \frac{2}{\cosh r\sqrt {2+ 2\cosh^{-1} 2r }} \exp \left[\tanh^{2} r(\alpha^*a^\dag +\beta^*b^\dag )^2(-e^{-i\theta}\beta a^\dag+e^{-i\theta }\alpha b^\dag )^2\right]|0\>_A|0\>_B\\
				= & \frac{2}{\cosh r\sqrt {2+ 2\cosh^{-1} 2r }} \int \frac{\d^2 \gamma\d^2 \gamma'}{\pi^2}(|\underline\gamma\>\<\underline\gamma|_A\otimes |\underline\gamma'\>\<\underline\gamma'|_B)\nonumber \\
				&\times \exp \left[\tanh^{2} r(\alpha^*a^\dag +\beta^*b^\dag )^2(-e^{-i\theta}\beta a^\dag+e^{-i\theta }\alpha b^\dag )^2\right]|0\>_A|0\>_B\\
				= & \frac{2}{\cosh r\sqrt {2+ 2\cosh^{-1} 2r }} \nonumber \\
				&\times \int \frac{\d^2 \gamma\d^2 \gamma'}{\pi^2}\exp(-\frac {|\gamma|^2+|\gamma'|^2}2)\exp \left[\tanh^{2} r(\alpha^*\gamma^* +\beta^*\gamma^{'*} )^2(-e^{-i\theta}\beta \gamma^*+e^{-i\theta }\alpha \gamma^{'*} )^2\right]|\underline\gamma\>_A|\underline\gamma'\>_B.\label{eqh20}
			\end{align}
			where $|\underline{\gamma}\>:=\exp(\alpha a^\dag -\alpha^* a)|0\>$ denotes the coherent state that satisfies the relation $\int \frac{\d^2 \gamma}{\pi}|\underline\gamma\>\<\underline\gamma|=I$.   
			
			The state $U'|\Psi\>$ is a product state if and only if its joint distribution for $|\gamma\>_A$ and $|\gamma'\>_B$ can be expressed as a product of two independent distributions. That indicates: 
			\begin{align}
				\exp \left[\tanh^{2} r(\alpha^*\gamma^* +\beta^*\gamma^{'*} )^2(-e^{-i\theta}\beta \gamma^*+e^{-i\theta }\alpha \gamma^{'*} )^2\right]&=f(\gamma)f'(\gamma').
			\end{align}
			However, due to the quadratic form in the exponential function, it can not be separated by adjusting the values of the coefficients $\alpha$, $\beta$, and $\theta$. (see a similar case shown in the Lemma 2 of Ref.~\cite{mari2014quantum}. )
			
			Therefore, one can conclude that the state $|\Psi\>$ is NGE.

			\subsubsection{NOON state}
			
			Another case where the state in Eq.~(\ref{eqa1}) has degenerate symplectic eigenvalues is as follows: 
			\begin{align}
				|\Phi\>&= \frac{1}{\sqrt 2} \left(|N\>_A|0\>_B+|0\>_A|N\>_B\right),\ \ \ \  (N\ge  1, N\in \Z).
			\end{align}
			It is quick to prove that its covariance matrix is 
			\begin{align}
				V=(N+1)I^{(4)}. 
			\end{align}
			If the state $|\Phi\>$ is GE, from Theorem \ref{lem:pureGEstate} and Lemma \ref{lem:williamson}, we have the relation $U'|\Phi\>=|\psi\>_A|\phi\>_B$ with $U'$ being a beam-splitter operation. 
			
			Given the overcompleteness of coherent state $\int \frac{\d^2 \gamma}{\pi}|\underline\gamma\>\<\underline\gamma|=I$, we have: 
			\begin{align}
				U'|\Phi\>&=U'\frac{1}{\sqrt 2 }\int \frac{\d^2\gamma}{\pi}e^{-\frac{|\gamma|^2}{2}}\frac{\gamma^{*N}}{\sqrt{N!}}\left(|\underline\gamma\>_A|0\>_B+|0\>_A|\underline\gamma\>_B\right)\\
				&=\frac{1}{\sqrt 2 }\int \frac{\d^2\gamma}{\pi}e^{-\frac{|\gamma|^2}{2}}\frac{\gamma^{*N}}{\sqrt{N!}}\left(|\underline{\alpha^*\gamma} \>_A|\underline{\beta^*\gamma}\>_B+|\underline{-\beta \,e^{i\theta}\,\gamma}\>_A|\underline{\alpha\, e^{i\theta}\,\gamma}\>_B\right)\\
				&=\frac{1}{\sqrt 2 }\sum_{j,k=0}^\infty \int \frac{\d^2\gamma}{\pi}e^{-|\gamma|^2}\frac{\gamma^{*N}\gamma^{j+k}}{\sqrt{N!j!k!}}\left[\alpha^{*j}\beta^{*k}+(-\beta \,e^{i\theta})^j(\alpha\, e^{i\theta})^k\right]|j\>_A|k\>_B\\
				&=\frac{1}{\sqrt 2 }\sum_{k=0}^N \int_{s=0}^\infty \d s \, 2 s\,e^{-s^{2}}\frac{s^{2N}}{\sqrt{N!(N-k)!k!}}\left[\alpha^{*(N-k)}\beta^{*k}+(-\beta \,e^{i\theta})^{N-k}(\alpha\, e^{i\theta})^k\right]|N-k\>_A|k\>_B\\
				&=\frac{1}{\sqrt 2 }\sum_{k=0}^N \sqrt{\left(\begin{matrix}
						N\\ k
					\end{matrix}\right)}\left[\alpha^{*(N-k)}\beta^{*k}+(-\beta \,e^{i\theta})^{N-k}(\alpha\, e^{i\theta})^k\right]|N-k\>_A|k\>_B
			\end{align}
			where $\left(\begin{matrix}
				N\\k
			\end{matrix}\right):=\frac{N!}{(N-k)!k!}$ denotes the binomial coefficient, the coefficients $\alpha,\beta,\theta$ are defined in Eq.~(\ref{eqe16}). For simplicity of notation, let us define the coefficients $c_k:=\frac{1}{\sqrt 2 } \sqrt{\left(\begin{matrix}
					N\\k
				\end{matrix}\right)}\left[\alpha^{*(N-k)}\beta^{*k}+(-\beta \,e^{i\theta})^{N-k}(\alpha\, e^{i\theta})^k\right]$ for $k=0,\cdots,N$. Then, the state $U'|\Phi\>$ is a product state if its reduced state: 
			\begin{align}
				\Tr_B[U'|\Phi\>\<\Phi|U^{'\dag}]&=\sum_{j,k=0}^N c_j c_k^* \<k|j\>_B |N-j\>\<N-k|_A,\\
				&=\sum_{k=0}^N |c_k|^2  |N-k\>\<N-k|_A,\label{eqh30}
			\end{align}
			is pure. It indicates that the modulus of one of the coefficients $\{c_k\}$ is equal to one, while the other coefficients are zero due to the normalization condition $\sum_k |c_k|^2=1$. Explicitly, we have: 
			\begin{align}\label{eqe19}
				\left(\begin{matrix}
					N\\k
				\end{matrix}\right)&\left|\alpha^{*(N-k)}\beta^{*k}+(-\beta)^{N-k}\alpha^k e^{iN\theta}\right|^2=2,\\
				&\exists\, 0\le k\le N, \alpha,\beta\in\C, |\alpha|^2+|\beta|^2=1, \theta\in \R.\nonumber 
			\end{align}
			Here is a table of solutions for $N\le 2$:
			\begin{center}
				\begin{tabular}{ |c |c| c| c|c| }
					\hline
					\diagbox{N}{coefficients} & $k$ & $\alpha$ & $\beta$ &$\theta$\\ 
					\hline
					1 & 1 & $\frac{1}{\sqrt 2}$  &$\frac{1}{\sqrt 2}$& 0 \\  
					\hline
					2 & 1 & $\frac{1}{\sqrt 2}$  &$\frac{1}{\sqrt 2}$& $\frac \pi 2 $  \\
					\hline
				\end{tabular}
			\end{center}
			
			Moreover, by substituting coefficients $R=|\alpha|^2$ and $\theta'=\arg(\alpha)+\arg (\beta)+N\theta$ into Eq.~(\ref{eqe19}), we have: 
			\begin{align}\label{eqe30}
				\left(\begin{matrix}
					N\\k
				\end{matrix}\right)&\left\{R^{N-k}(1-R)^k+R^k(1-R)^{N-k}+2[R(1-R)]^{\frac N 2}(-1)^{N-k} \cos \theta'\right\}=2,\\
				& \exists\ 0\le k\le N,R\in [0,1],\theta'\in \R. \nonumber 
			\end{align}
			Given the law of cosines, the l.h.s of Eq.~(\ref{eqe30}) is upper bounded as follows: 
			\begin{align}
				f_{\rm up}=\left(\begin{matrix}
					N\\k
				\end{matrix}\right)
				\left(R^{\frac{N-k}2}(1-R)^{\frac k 2}+R^{\frac k 2}(1-R)^{\frac{N-k}2}\right)^2.
			\end{align}
			It can be proven that the function $f_{\rm up}$ is continuous and attains its maximum at $R=1/2$. Therefore, we have: 
			\begin{align}
				f_{\rm up}\le &\, 2^{2-N}\left(\begin{matrix}
					N\\k
				\end{matrix}\right)\\
				<&\, 2. 
			\end{align}
			where the second inequality is derived using the fact that the maximum value of binomial distribution $p(k)=2^{-N}\left(\begin{matrix}
				N\\k
			\end{matrix}\right)$ lowers monotonously as the value of $N$ increases and that it achieves the value $1/2$ when $N=2$. 
			
			Therefore, it is not possible to fulfill Eq.~(\ref{eqe30}) in the case with $N>2$. Hence, any NOON state with photon number $N>2$ is NGE.

			\section{NG entropy for pure states}\label{app:resource theory of NGE}
			
			In this Supplementary Note, we introduce the NG entropy for pure states.  \\[-0.5em]

			\begin{definition}\label{defiE1} (NG entropy for pure states). Given an arbitrary  $n$-partite pure state $|\psi\>_{A_1,\cdots A_n}\in\otimes_{j=1}^n \map H_{A_j}$, the genuine non-Gaussian entanglement is defined as follows: %via the convex-hull construction: 
				%\begin{align}\label{eq12ff}\map E_{\rm of,NG}(\rho )=\min_{\genfrac{}{}{0pt}{}{\{p_x, |\psi\>\<\psi|_{x}\}}{ {\rm s.t.\ }\sigma=\sum_{x,y} p_{x}|\psi\>\<\psi|_{y}}}\sum_{x} p_{x}\, \map E_{\rm NG}(|\psi\>\<\psi|_{x}),\end{align} where the minimization is taken over all possible pure-state decomposition of the state $\rho$, the function $\map E_{\rm NG}(|\psi\>\<\psi|_{A_1A_2,\cdots A_n})$ is defined as: 
				\begin{align}\label{eqi1}
					& \map E_{\rm NG}(|\psi\>\<\psi|)=\min_{U^g\in \mathbb{GU}\left(\otimes_{j=1}^n \map H_{A_j}\right)} \frac 1 n \sum_{j=1}^n  H \left(\Tr_{k\neq j}\left[U^g|\psi\>\<\psi| U^{g\dag} \right]\right)
				\end{align}
				with $ H(\rho)=-\Tr[\rho \log_2 \rho]$ being the von Neumann entropy  of the state $\rho$. 
			\end{definition}
			%\QZ{Prove $\mathbb{G}_E$ if and only if it is zero. Non-increasing under Gaussian unitaries.} 
			%For mixed state, we define the resource measure
			%\begin{align}\calD(\rho)=\min_{\sigma_{AB}\in {\rm GE}} \|\rho_{AB}-\sigma_{AB}\| \end{align} 
			%\QZ{Entanglement-based measure}
			
			%\QZ{what does except mean?}
			
			Then, the following remark is given: 
			\begin{remark}[Gaussian unitary for NG entropy] \label{lem: NGE entropy zero for free states} Given an arbitrary pure GE state $|\psi\>$, the resource measure $\map E_{\rm NG}$ satisfies the following criteria: (1) $\mathcal{E}_{\rm NG}(|\psi\>\<\psi|)=0$ if and only if $\psi$ is Gaussian-entanglable, (2) $\mathcal{E}_{\rm NG}(|\psi\>\<\psi|)$ is invariant under Gaussian unitary. (3) $\mathcal{E}_{\rm NG}(|\psi\>\<\psi|)$ is nonzero if the state $|\psi\>$ is NGE. 
			\end{remark}
			\begin{proof} Since the von Neumann entropy for pure states is zero and Corollary \ref{cor:nge} of Methods, we can infer that Remark \ref{lem: NGE entropy zero for free states} is valid for pure states. 
			\end{proof}

			Note that the NG entropy requires a minimization over all possible Gaussian unitaries, which is sometimes difficult to compute. For algebraic convenience, we introduce the following quantity as an estimate of NG entropy: \\[-0.5em]
			
			\begin{definition}\label{defiE3} (Computationally-simple NG entropy) Given an arbitrary  $n$-partite pure state $|\psi\>_{A_1,\cdots A_n} \in\otimes_{j=1}^n \map H_{A_j}$, the computationally-simple NG entropy (CS-NG entropy) is defined as follows: 
				\begin{align}
					& \overline{\map E}_{\rm NG}(|\psi\>\<\psi|)=\min_{\substack{U^g\in \mathbb{GU}_{|\psi\>}\left(\otimes_{j=1}^n \map H_{A_j}\right)} }\frac 1 n \sum_{j=1}^n  H \left(\Tr_{k\neq j}\left[U^g|\psi\>\<\psi| U^{g\dag} \right]\right)
				\end{align}
				with $\mathbb{GU}_{|\psi\>}\left(\otimes_{j=1}^n \map H_{A_j}\right)$ being the set of Gaussian unitaries that transforms $|\psi\>$ into a state with Williamson form of covariance matrix.  
			\end{definition}
			
			On this account, one can quickly identify that the following remark: 
			
			\begin{remark}
				The resource measure $\overline{\map E}_{\rm NG}$ satisfies the following criteria: (1) The CS-NG entropy of all GE state is zero. (2) It is invariant under Gaussian unitaries. (3) It has nonzero for any NGE state. 
			\end{remark}

			\subsection{NOON state}
			
			Consider a NOON state $|{\rm NOON}\>=(|N\>_A|0\>_B+|0\>_A|N\>_B)/\sqrt 2$ with $N\ge 3$. The NG entropy in Definitions \ref{defiE1} and \ref{defiE3} depend on a minimization over all Gaussian unitaries. Given Lemma \ref{lem:Bloch Messiah decomposition}, an arbitrary bi-partite Gaussian unitary can be written as: 
			\begin{align}\label{eqi2}
				U^g=\left[D(\alpha_1)\otimes D(\alpha_2)\right][e^{-i\theta_1 a^\dag a-i\theta_2 b^\dag b} ]e^{\theta(a^\dag b-b^\dag a)} e^{-i\theta_0 a^\dag a}\left[S_{{\rm q},1}(r_1)\otimes S_{{\rm q},1}(r_2)\right]U_1,\ \ \ (\alpha_1,\alpha_2\in \C;\theta,\theta_0,\theta_1,\theta_2\in [0,2\pi)]), 
			\end{align}
			where $U_1$ is a general beam-splitter operation with possible phase rotation, $S_{{\rm q},1}(r)=\exp(r/2(a^2-a^{\dag 2}))$. Note that here we already take into consideration all possible single-mode phase rotations in each implementation step.
			
			\iffalse 
			\begin{conjecture}\label{conjectureI3}
				The NG entropy is minimized by a Gaussian unitary defined in Eq.~(\ref{eqi2}) with $r_1=r_2$
			\end{conjecture}
			\fi 
			
			On this account, the NG entropy is defined as the average of von Neumann entropy for each local state of $U^g|{\rm NOON}\>$ (see Definition \ref{defiE1}). Given the following relations: 
			\begin{align}
				S_{{\rm q},1}(r)S_{{\rm q},1}(r')&=S_{{\rm q},1}\left(\frac{r+r'}{1+rr'}\right)\\
				e^{\theta(a^\dag b-b^\dag a)} \left[S_{{\rm q},1}(r)\otimes S_{{\rm q},1}(r)\right]=&\left[S_{{\rm q},1}(r)\otimes S_{{\rm q},1}(r)\right] e^{\theta_2\left[S_{{\rm q},1}^\dag(r)\otimes S_{{\rm q},1}^\dag(r)\right] \left(a^\dag b-b^\dag a\right)\left[S_{{\rm q},1}(r)\otimes S_{{\rm q},1}(r)\right] } \\
				=&\left[S_{{\rm q},1}(r)\otimes S_{{\rm q},1}(r)\right]  e^{\theta(a^\dag b-b^\dag a)}, 
			\end{align}
			and the fact that local operations does not change the entanglement entropy \cite{bennett1996concentrating}, 
			we can reduce the effective Gaussian unitary in minimization of Definitions \ref{defiE1}  by: 
			\begin{align}\label{eqi8}
				\overline U^g&= e^{\theta(a^\dag b-b^\dag a)}  \left[e^{-i\theta' a^\dag a} S_{{\rm q},1}(r)\otimes I\right]U_1,\ \ \theta,\theta\in[0,2\pi).
			\end{align}
			
			Given the difficulty of evaluating NG entropy with all possible Gaussian unitaries given by Eq.~(\ref{eqi8}), let us only look at the CS-NG entropy of a NOON state. From Supplementary Note \ref{app:degenerate_NGE}, we know that identity matrix already diagonalises the covariance matrix of the NOON state. Therefore, its CS-NG entropy just replies on the minimization over all possible beam-splitters due to Lemma \ref{lem:williamson}. From Eq.~(\ref{eqh30}), we know that the reduced state $\rho$ of a NOON state after beam-splitter operation is: 
			\begin{align}
				\rho&=\sum_{k=0}^N |c_k|^2  |N-k\>\<N-k|_A,
			\end{align}
			with coefficients $c_k:=\frac{1}{\sqrt 2 } \sqrt{\left(\begin{matrix}
					N\\k
				\end{matrix}\right)}\left[\alpha^{*(N-k)}\beta^{*k}+(-\beta \,e^{i\theta})^{N-k}(\alpha\, e^{i\theta})^k\right]$ for $k=0,\cdots,N$, $\alpha,\beta\in C$ satisfying $|\alpha|^2+|\beta|^2=1$, and $\theta\in \R$. The CS-NG entropy becomes: 
			\begin{align}
				\overline{\map E}_{\rm NG} (|{\rm NOON}\>\<{\rm NOON}|)&= - \max_{R\in [0,1],\theta'\in \R}  \sum_{k=0}^N |c_k|^2 \log_2 |c_k|^2 \\
				|c_k|^2&=\frac 1 2 \left(\begin{matrix}
					N\\k
				\end{matrix}\right)\left\{R^{N-k}(1-R)^k+R^k(1-R)^{N-k}+2[R(1-R)]^{\frac N 2}(-1)^{N-k} \cos \theta'\right\}.\label{eqi10}
			\end{align}
			%Given that the maximum of $|c_k|^2$ is given by $R=1/2$ as it is the only solution for $\d |c_k|^2/\d R=0$, its minimum should be achieved at $R=0,1$. In addition, the function $-x\log_2 x$ achieves its minimum at $x=0,1$. Thus,  we have: 
			%\begin{align}\overline{\map E}_{\rm NG} (|{\rm NOON}\>\<{\rm NOON}|)&= - 2\max_{R=0,1}  \sum_{k=0}^N |c_k|^2 \log_2 |c_k|^2\end{align}
			
			Thus, the CS-NG entropy can be numerically calculated by appropriately selecting values of $R$ and $\theta'$ in Eq.~(\ref{eqi10}). 
			
			\subsection{Superposition of TMSV states}
			
			From Supplementary Note 
			\ref{app:degenerate_NGE}, we know that the covariance matrix of the state in Eq.~(\ref{eqe12}) is already diagonalised. Thus, its CS-NG entropy only requires a minimization over all possible beam-splitter operations. Given by Eq.~(\ref{eqh20}), the reduced state has the following form: 
			\begin{align}
				\rho= &\frac{4}{\cosh^2 r\left(2+ 2\cosh^{-1} 2r \right)} \int \frac{\d^2 \gamma\d^2 \gamma'\d^2 \zeta\d^2 \zeta'\d^2 \xi}{\pi^5}\exp(-\frac{|\gamma|^2+|\gamma'|^2+|\zeta|^2+|\zeta'|^2}2)\nonumber \\
				&\times \exp \left[\tanh^{2} r(\alpha^*\gamma^* +\beta^*\gamma^{'*} )^2(-e^{-i\theta}\beta \gamma^*+e^{-i\theta }\alpha \gamma^{'*} )^2\right]\exp \left[\tanh^{2} r(\alpha\zeta +\beta\zeta^{'} )^2(-e^{i\theta}\beta^* \zeta+e^{i\theta }\alpha^* \zeta^{'} )^2\right]\nonumber \\
				&\times \<\underline\xi|\underline\gamma'\>\<\underline\zeta'|\underline\xi\>\cdot |\underline\gamma\>\<\underline\zeta|_A . 
			\end{align}
			Thus, one can quickly compute its von Neumann entropy via appropriate truncation in Hilbert space.  
			
			Here, we conduct a numerical evaluation of the NG entropy for both NOON state and the superposition of TMSV states. The numerical results as well as the entanglement entropy without counter-rotation are shown in Fig.~\ref{fig:NG entropy} of the main text.

			\section{GE cost in general cases} 
			\label{supp:GE_cost}
			In general, there exists a case where the mode number $m$ is larger than the number of partitions $n$. In this case, the vector representing the GE cost might not be presented in a sorted form as the mode number of different subsystems could be different. Still, one can define the GE cost in a decreasing order when each subsystem is of the same size.

			Then, the following definition is given: 
			
			\begin{definition}[{\color{black}Genuine non-Gaussian entanglement of formation} in more general scenarios]  Given an arbitrary $n$-partite state $\Psi\in\otimes_{j=1}^n \map H_{A_j}$, each subsystem having  $|A_j|$ number of modes, the genuine non-Gaussian entanglement of formation can be defined as the list of number of ancillary modes $\map R(\psi)= \{|A_1'|,\cdots,|A_n'|\}$ for each subsystem needed to generate the state:
				\begin{align}\label{GE cost 5.2}
					\map R(\psi) &=  \left(|A_1'|,|A_2'|,\cdots, |A_n'|\right)^T\\[0.5em]
					\emph{s.t.\ \ }\nonumber &\begin{cases}
						\widetilde\psi\in \mathbb{GE}\left(\bigotimes_{j=1}^n \map H_{A_j}\otimes \map H_{A_j'};\right.\\
						\left.\,\ \ \left\{A_1,A_1'|A_2,A_2'|\cdots|A_n,A_n'\right\}\right)\\[0.5em]
						\psi=\Tr_{A_j' } \left[\,\widetilde\Psi\,\right]
					\end{cases}.
				\end{align}
				Only in the scenario where the total Hilbert space has a partition into subspaces of the same size, i.e. $m_j=m_c,\forall j=1,\cdots,n$, the GE cost defined in Eq.~(\ref{GE cost 5.2}) can be represented in a decreasing order: $\map R^{\downarrow}(\psi)=(h_1,\cdots,h_n)^T\text{\em such\ that } h_1\ge h_2\ge \cdots,\ge h_n$. 
				
				Further, one can define the GE cost function as:
				\begin{align}
					\map {R}_{f}(\rho)&=\min_{\rho_{\rm ext}} \left\|\map R\right\|_\infty  
				\end{align}
				where the minimization is taken over all possible extensions $\rho_{\rm ext}$ or the original state $\rho$. 
			\end{definition}

			\section{Efficient learning of pure GE states }\label{app:learning GE states}
			
			Quantum state tomography refers to the procedure of obtaining a classical representation $\tilde \rho$ of an unknown state $\rho$, with the conditions: $\text{Pr}\left(\frac 1 2 \|\tilde \rho-\rho\|\ge \epsilon \right)\le \delta$, where $\epsilon,\delta \in (0,1)$ are small constants, $\|\cdot\|$ denotes some norm. The sample complexity corresponds to the number of state copies required for the tomography process. As specified in Definition \ref{defi:multi GE state}, an arbitrary $m$-mode GE state can be written as
			$\rho_{\bs A}= \sum_{x,y} p_{x,y} \map G_{x} \left(\sigma_{y}\right)$ over $\map H_{\bs A}$, where $\sigma_{y}= \sum_\lambda  p_{\lambda,y} \bigotimes_{j=1}^n\rho_{j,\lambda,y }$ 
			denotes fully separable states in terms of an $n\le m$ partition $\bs A=\{A_1,\cdots,A_n\}$, $\map G_x$ refer to a $m$-mode Gaussian channel depending on $6m^2+3m$ real parameters (for matrices $X$, $Y$, and displacement $\bs d$ defined in Eq.~(\ref{CPTP-Gaussian-moments})). Hence, the goal of a tomography process on GE states is to determine $|x||y| + |x|(6m^2+3m) + |y| |\lambda| D\, n $ parameters, where $|x|(|y|/|\lambda|)$ refers to the size of the parameter set for $x(y/\lambda)$, $D$ denotes the maximal number of parameters to describe local states $\rho_{j,\lambda,y}$ with bounded error. Here, we assume that the probability distribution $\{p_{x,y}\}$ can be learned with polynomial copies of the unknown GE states. With this premise, the total sample complexity will have the potential to scale as: $M=\textbf{poly} (n)$ since the states can be characterized by a polynomial number of parameters. On the other hand, unknown NGE states could be determined by more than $\textbf{exp}(n)$ parameters, which makes tomography a hard task.  
			
			In this Supplementary Note, we evaluate the sample complexity upper bound of any tomography algorithm for pure GE states. Given Lemmas  \ref{lem:covariance matrix of nonGaussian state plus Gaussian unitary}, \ref{lem:Bloch Messiah decomposition}, and Corollary \ref{lem:multipartite extension of theorem 3.5}, an arbitrary $n$-partite pure GE state can be written in the following form 
			\begin{align}
				|\psi\>=&\,U_{\bs \alpha'}  U_{S'}  \bigotimes_{j=1}^n |\psi_{{ \alpha_j''},j}\>\\
				=&\,U_{\bs \alpha'}  U_{S'}  \bigotimes_{j=1}^n U_{\alpha_j''}|\psi_{{ 0},j}\>\\
				=&\,U_{\bs \alpha}  U_{S'}  \bigotimes_{j=1}^n |\psi_{{ 0},j}\>\\
				=&\,U_{\bs \alpha}  U_{S}  \bigotimes_{j=1}^n |\psi_{j}\>,\label{eqf3}
			\end{align}
			where $U_{\bs \alpha}$ is an $m$-mode displacement operation with displacements $\bs \alpha=\bs \alpha'+S'(\alpha_1'',\cdots,\alpha_{m}'')^T$, %$U_O$ is a multi-port beam-splitter operation associated with a $2m\times 2m$ symplectic orthogonal matrix $O$, 
			$U_{S}$ is a Gaussian unitary associated with a $2m\times 2m$ symplectic matrix $S=S'\bigoplus_{j=1}^m S_j$ with $\{S_j^{-1}\}$ being the symplectic operation that diagonalize the covariance matrix of the $j$-th mode state $|\psi_{{ 0},j}\>$, $|\psi_{ \alpha_j'',j}\>\in \map H_{A_j}$ denotes the $j$-th local state with a displacement $ \alpha_j''$, $|\psi_{j}\>$ denotes a local state at $j$-the mode with zero displacement and a diagonal covariance matrix.

			Therefore, the goal of a tomography process for pure GE state is to determine the unitaries $\{U_{\bs \alpha}, U_{S}\}$ as well as the local states $\{|\psi_{j}\>\}$. Without loss of generality, we assume that the pure GE state $|\psi\>$ 
			%\in \mathbb {GE} (\otimes_{j=1}^n \map H_{A_j};\{A_1|A_2|\cdots |A_n\})$ 
			being measured consists of $m$ modes and has a limited amount of square-energy, i.e., 
			\begin{align}\label{eq87ccc}
				\sqrt{\Tr[\left(\sum_j a_j^\dag a_j+ mI/2\right)^2|\psi\>\<\psi| ]}\le mE_{\rm II}
			\end{align}
			with $E_{\rm II}\ge 0$ being a quantity larger than energy $E_{\rm I}:=1/m \Tr[\left(\sum_j a_j^\dag a_j+ mI/2\right)|\psi\>\<\psi| ]$ due to the concavity of the square-root function. %For simplicity of discussion, we assume that the number of partitions $n$ is equal to the number of modes $m$.  %In addition, the square-energy of each local state $|\psi_{{\bs 0},j}\>$ is denoted as $E_{{\rm II},j}$. %In the following contents, we will denote the local state $|\psi_{{\bs 0},j}\>$ as $|\psi_{j}\>$ for simplicity. 

			\begin{figure}[ht]
				\centering
				\includegraphics[width=0.7\linewidth]{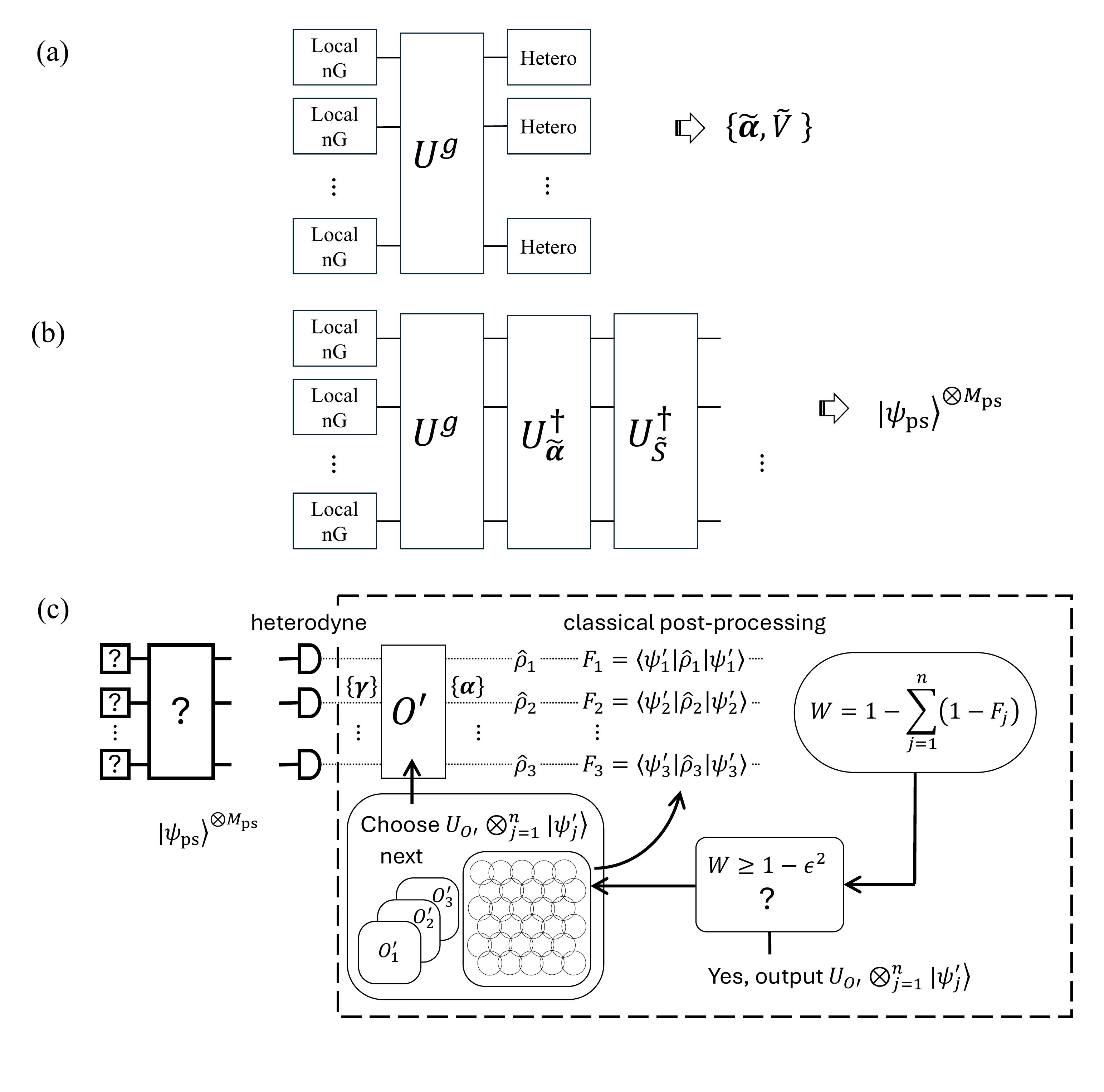}
				\caption{Schematic of the tomography process for pure GE states, which are expected to be generated by performing a Gaussian unitary $U^g$ on local non-Gaussian states. The learning process has three steps: (a) Estimate the displacement $\widetilde {\bs \alpha}$ and the covariance matrix $\widetilde V$ via heterodyne/homodyne measurement. (b) Based on the information of displacement and covariance matrix, implement a counter-rotating displacement operation $U_{\bs {\widetilde \alpha}}^\dag $ as well as a 
					Gaussian unitary $\widetilde U_{\widetilde S}^\dag$. The output state will be a passive-separable state $|\psi_{\rm ps}\>$. (c) Implement heterodyne measurements on each of the $m$ modes of the unknown passive-separable state to obtain outcomes $\{\gamma_k,k=1,\cdots,m\}$. From these data, conduct a basis rotation to obtain $\bs \alpha=O'\bs \gamma$ with $\bs\alpha=(\alpha_1,\cdots,\alpha_j)^T$. Then, estimate the reduced density operators $\{\hat \rho_j\}$. For each reduced  state, compute the global fidelity $F_j=\<\psi'_j|\hat \rho_j|\psi_j'\>$ and form the lower bound $W$ on the global fidelity. Here, the target local states $\{|\psi_j'\>\}$ as well as the target basis rotation $O'$ is randomly selected from those associated with the $\epsilon$-covering net of all energy-constrained passive-separable states.  Once the quantity $W$ is larger than $1-\epsilon^2$ with $\epsilon$ being a small number, output the reconstructed state $U_{O'}\otimes_{j=1}^n |\psi_j'\>$. 
					\label{fig:schematic of learning}
				}
			\end{figure}

			\subsection{Tomography algorithm for general pure GE states}\label{app:tomography algorithm}
			
			The schematic of a tomography algorithm for pure GE states is illustrated in Fig.~\ref{fig:schematic of learning}. Specifically, it consists of three steps:

			\begin{enumerate}
				\item[Step 1:] \textbf{Estimate covariance matrix and displacement with homodyne measurement.---} 
				
				\begin{framed}
					
					\begin{center}
						\textbf{Algorithm 1: estimation of covariance matrix and displacement}
					\end{center}
					
					\textbf{Input:} Accuracy $\epsilon_v$, failure probability $\delta_v$, square-energy constraint $E_{{\rm II}}$, $M_v$ copies of unknown $m$-mode pure GE states $|\psi\>$ fulfilling the constraint $\sqrt{\Tr [\left(\sum_{j=1}^m a_j^\dag a_j +m I/2 \right)^2|\psi\>\<\psi| ]}\le m E_{\rm II}$. 
					
					\textbf{Output:} An estimate of the covariance matrix $\widetilde V\in \R^{2m,2m}$ and the displacement $\widetilde {\bs \xi}\in \R^{2m}$.

					1.1: Perform joint and local \emph{homodyne} measurements \cite{braunstein2005quantum} for two quadrature operators $\bs r_j$ and $\bs r_k$ $(j\neq k)$ with observables $\widehat{\{{\bs r}_j, {\bs r}_k\}}$, $\hat{\bs r}_j$, and $\hat{\bs r}_k$. In addition, perform local \emph{heterodyne} measurements to estimate $\widehat{\{{\bs r}_{2j}, {\bs r}_{2j+1}\}}$ for $j=1,\cdots,m$. With multiple rounds of experiments, the $m(2m-1)$ off-diagonal entries of the covariance matrix $\{v_{jk}:=\<\{\bs r_j ,\bs r_k\}\>/2-\<\bs r_j\>\<\bs r_k\>|\,j\neq k\}$ can be estimated via constructing median-of means estimators of the expectation value.

					1.2: Conduct \emph{photon counting} directly on the $j$-th mode with an observable $\widehat{N}_j^{(1)}:=\widehat{a^\dag_j a_j}$. Then, implement a single-mode squeezing operation $S_1(r)=\exp[r(a^2-a^{\dag2})/2]$ on another copy of the state and measure the photon number of the $j$-th mode with an observable $\widehat{N}_j^{(2)}$. By classically solving the following equations:
					\begin{align}
						\<\bs r_{2j}^2\>+\<\bs r_{2j+1}^2\>=&\,4\<\widehat{N}_j^{(1)}\>+2,\\
						e^{-2r}\<\bs r_{2j}^2\>+e^{2r}\<\bs r_{2j+1}^2\>=&\,4\cosh 2r \<\widehat{N}_j^{(2)}\>+ 4 \sinh^2 r-\sinh 2r \left(\<\bs r_{2j}^2\>-\<\bs r_{2j+1}^2\>\right)+2,
					\end{align}
					one can obtain the values of $\<\bs r_{2j}^2\>$ and $\<\bs r_{2j+1}^2\>$. Given that $\<\bs r_{2j}\>$ and $\<\bs r_{2j+1}\>$ have already been estimated in Step 1.1, each diagonal element $\{v_{jj}=\<\bs r_j^2\>-\<\bs r_j\>^2\}$ of the covariance matrix, denoted by $\widetilde V'$, can be measured with one copy of the unknown state on average.

					1.3: \textbf{If} $\widetilde V' + \frac{\epsilon_v}{2}I +i\Omega\ge 0$, \textbf{then}
					
					\ \ \ \ \ \ \ \textbf{return} the estimated covariance matrix $\widetilde V=\widetilde V'+\frac{\epsilon_v}{2}I$, 
					
					\textbf{else} 
					
					\ \ \ \ \ \ \ Declare \emph{failure} and abort. 
					
					\textbf{end if}
					
				\end{framed}
				
				\iffalse

				Given that the average of quadratures $\{\bs \xi_j=\<\bs r_j\>\}$ of each mode can be estimated through a beam-splitter operation followed by homodyne measurement \cite{braunstein2005quantum}, the covariance matrix can be estimated by measuring a pair of the $m$  quadratures jointly. With four copies of states, it is possible to estimate the whole $2m\times 2m$ covariance matrix. 
				
				A single copy of the state allows the measurement of $m$ diagonal entries. Therefore, the total $m(2m+1)$ matrix elements can be estimated by using $N_v$ copies with: $N_v=4m\omega,$ when $n$ is even, and $N_v=\left(4m+4+\frac{2}{m-1}\right)\omega$ when $n$ is odd, where $\omega$ is the number of repeated experiments.

				On this ground, we can define $\delta_{jk}$ as the asymptotic root-mean-square error (RMSE) for each round of measurement estimating $v_{jk}$, which follows the relation: $\left|\widetilde v_{jk}-v_{jk}\right|=\delta_{jk}/\sqrt \nu $.
				
				\fi 
				
				In this step, each copy of the multi-mode state is only used once during the whole measurement procedure. This is because the post-measurement state, conditioned on the homodyne measurement result, has a partial covariance matrix of size $2(m-1)\times 2(m-1)$. In principle, this partial covariance matrix can be derived from the multipartite non-Gaussian correlation of the GE state, similar to the case of Gaussian states discussed in \cite{genoni2016conditional}. While, given the difficulty of describing the non-Gaussian correlation of a GE state, we propose to estimate a single off-diagonal element $v_{jk}$ of the covariance matrix with three copies of state, for estimating $\<\bs r_j\bs r_k\>$, $\<\bs r_j\>$, and $\<\bs r_k\>$, respectively. 
				
				Now, let us look at the sample complexity of this step. Given either Bernstein matrix inequality \cite{tropp2012user} or median-of-means estimation methods \cite{jerrum1986random}, one can achieve a success probability lower bound that converges to one exponentially as the ratio of the sample number to the variance of the parameters increases. Then, one can obtain the minimal number of copies needed to achieve a specific estimation accuracy and failure probability. Here, we can apply the Theorem S53 of Ref.~\cite{mele2024learning} %which is on energy-square constrained states with a condition $\sqrt{\Tr\left[\left(\sum_{j=1}^m a_j^\dag a_j+\frac m 2 I \right)^2\rho\right]}\le m E_{\rm II}$,
				to pure GE  states. The number of copies needed in the aforementioned estimation protocol is shown to be $M_v=(m+3)\left\lceil 68\log_2 \left(\frac{2(2m^2+3m)}{\delta_v}\right)\frac{200(8m^2 E_{\rm II}^2 +3m)}{\epsilon_v^2} \right\rceil$
				such that $\text{Pr}\left(\|\widetilde V-V\|_\infty \le \epsilon_v, \text{ and } \widetilde V+i\Omega\ge 0 \text{ and } \|\widetilde {\bs \xi}-\bs \xi\|_2 \le \frac{\epsilon_v}{20\sqrt{mE_{\rm II}}}\right)$ $\ge 1-\delta_v $ where $\|A\|_\infty$ denotes the operator norm, $\|\bs m\|_2=\sqrt{\bs m^T\bs m}$ refers to the Euclidean norm.

				\item[Step 2:] \textbf{Partially undo the effect of Gaussian unitary by counter-rotating.---} 
				
				Estimation of the displacement in the first step enables us to derive an estimate of the displacement operator $U_{\widetilde{\bs \alpha}}$ as shown in Eq.~(\ref{eqf3}). Furthermore, once the covariance matrix $\widetilde V$ is obtained, it is possible to calculate the symplectic matrix $\widetilde S$ that fulfills the requirement $\widetilde V= \widetilde S\left(\widetilde \Lambda \otimes I^{(2)}\right)\widetilde S^T$ using eigendecomposition  \cite{serafini2017quantum}: 
				\begin{align}\label{eqf15}
					i\Omega \widetilde V=&\widetilde W\left[\widetilde \Lambda \otimes\left(\begin{matrix}
						-1&0\\
						0&1
					\end{matrix}\right)\right] \widetilde W^{-1}\\
					\label{eqf16a}\widetilde S=&\Omega^T \widetilde W \left[I \otimes \frac{1}{\sqrt 2}\left(\begin{matrix}
						1&-i\\
						1&i
					\end{matrix}\right)\right]\Omega 
				\end{align}
				where $\widetilde \Lambda =\text{diag}(\widetilde\lambda_1,\cdots,\widetilde\lambda_m)$ is a $m\times m$ diagonal matrix fulfilling the condition $\det (i\Omega\widetilde V \pm \widetilde \lambda_j I )=0$, $\widetilde W=[\bs x_{1,+},\bs x_{1,-}\cdots,\bs x_{m,+},\bs x_{m,-}]$ is the eigenvector matrix of $i\Omega \widetilde V$ from the solution of $(i\Omega \widetilde V \pm \widetilde \lambda_j I)\bs x_{j,\pm}=0$. On the other hand, the Lemma \ref{lem:covariance matrix of nonGaussian state plus Gaussian unitary} indicates that the true value of the covariance matrix of the pure GE state $|\psi\>=U_{\bs \alpha}U_S \bigotimes_{j=1}^n |\psi_{j}\>$  is: 
				\begin{align}
					V%=&S\, \bigoplus_{j=1}^n D_j\, S^T \\
					=& S%\left(\bigoplus_{j=1}^n S_j\right)
					\left(\Lambda\otimes  I^{(2)}\right)%\left(\bigoplus_{j=1}^n S_j^T\right) 
					S^T\label{eqf7e}
				\end{align}
				where %$D_j$ is a $2m_j\times 2m_j$ covariance matrix for the state $|\psi_{\bs 0}\>_j$, $S_j$ is the matrix that symplectically diagonalises $D_j$, 
				$ \Lambda =\text{diag}(\lambda_1,\cdots,\lambda_m)$ is a $2m\times 2m$ diagonal matrix. Therefore, one can use the estimated symplectic matrix $\widetilde S$ as an approximation of $S\, %\left(\bigoplus_{j=1}^n S_j\right)
				$. Therefore, one can conduct the following step:

				%Given that the error of covariance matrix $\epsilon_v$ is assumed to be a small number, it can be approximately decomposed as follows: \begin{align}\left\|\widetilde V-V\right\|_\infty & \approx 2 \left\|\left(\widetilde S-S\right)\Lambda S^T  \right\|_\infty+\left\|S\left(\widetilde \Lambda -\Lambda \right) S^T \right\|_\infty\end{align}where $\Lambda =D \otimes \left(\begin{matrix}1&0\\0&1\end{matrix}\right)$ is obtained from $V=S\Lambda S^T$. 

				%Classically solve the equations for $\widetilde S$\begin{align}\begin{cases}\widetilde S\widetilde \Lambda \widetilde S^T=&\widetilde V\\\widetilde S \Omega\widetilde S^T=&\Omega\end{cases}.\end{align}
				\begin{framed}

					2.1: Apply the inverse of the Gaussian unitary, associated with $\widetilde S$, to all the left copies of the unknown pure GE states as follows: 
					\begin{align}\label{eqf17}
						\ket{\phi}=U_{\widetilde S}^\dag U_{\widetilde{\bs \alpha}}^\dag  |\psi\> .
					\end{align}
				\end{framed}
				Note that the operation $U_{\widetilde S}$ might not not unique. There might exist an extra degree of freedom in symplectic orthogonal rotation (see Lemma \ref{lem:williamson}). Thus, one needs to implement the following step.

				\item[Step 3: ] \textbf{Identification of the beam-splitter operation that makes the state separable.---}
				
				%\QZ{finally reconstrct the state by tensor, how much error will it have?}

				After applying the counter-rotation in Step 2, the state is reduced to a pure passive-separable state, which can be generated by applying a beam-splitter network to product states \cite{chabaud2023resources}. It is shown that this type of states can be efficiently simulated on a classical computer~\cite{chabaud2022holomorphic, chabaud2023resources}. Similarly, we will show that they can be learned efficiently. 
				
				Before delving into the proof of the efficient learning process, let us first look at the structure of the state space. As shown in Fig.~\ref{fig:epsilon_ball}, the set of GE states includes a subset that consists of pure GE states with constrained moments. Within this set of states, one can construct smaller subsets in which a single representative state can approximate any other state in the subset with an error no greater than $\epsilon$. Here we name the minimal set of representative states as the ``$\epsilon$-covering net.'' More specifically, we have the following proposition:

				\begin{figure}[t]
					\centering
					\includegraphics[width=0.5\linewidth,trim=50 200 50 200,angle=0,clip]{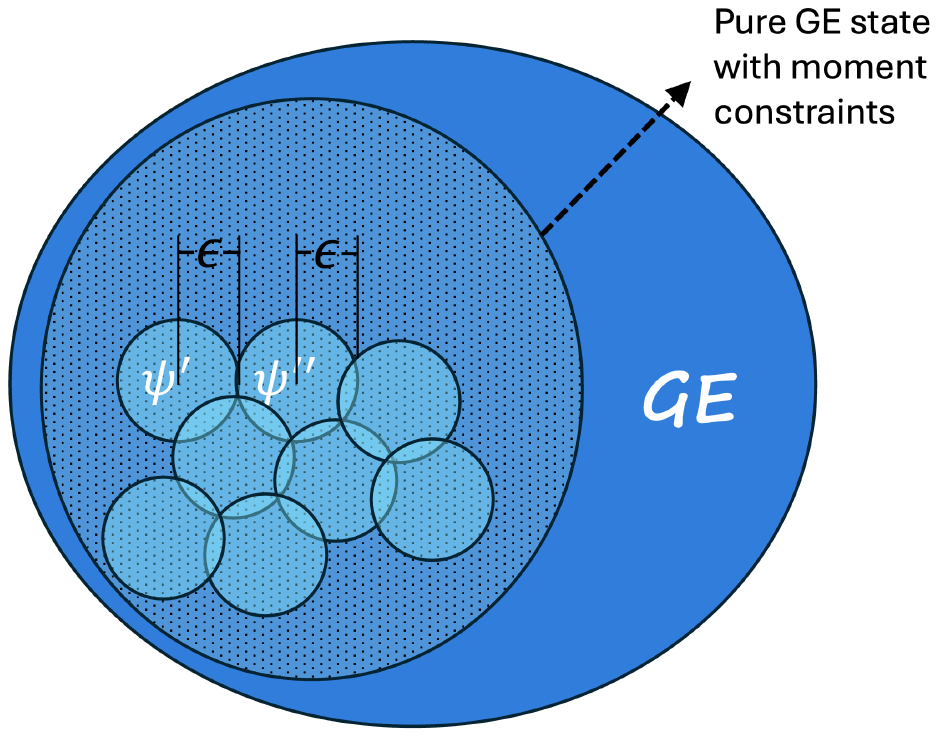}
					\caption{ $\epsilon$-covering net for pure GE states with moment constraints. Here, the state space of interest is covered by smaller subsets--``$\epsilon$ balls''--each of which allows any state within it to be approximated by a representative state with an error no greater than $\epsilon$. 
						\label{fig:epsilon_ball}
					}
				\end{figure}

				\begin{proposition}[$\epsilon$-covering net of pure passive-separable states] \label{prop:epsilon covering net}
					Consider the set of $m$-mode pure states $\map S_{\rm o}$ assumed to have the form  $|\psi_{\rm ps}\>:=U_O \bigotimes_{j=1}^n |\psi_j\>\in\bigotimes_{j=1}^n \map H_{A_j}$ where $U_O$ is a beam-splitter operator, $|\psi_j\>:=\sum_{ \{\ell_1,\cdots,\ell_{|A_j|}\}=\bs 0}^\infty  c_{j\bs\ell}|\ell_1,\cdots,\ell_{|A_j|}\>$ denotes the $|A_j|$-mode local pure states in the $j$-th subsystem. The overall state satisfies the energy constraint $\Tr [\sum_{\ell=1}^m a_\ell^\dag a_\ell  |\psi_{\rm ps}\>\<\psi_{\rm ps}| ]\le m  N_{\rm I}^*$. Then, a $\epsilon$-covering net $\map S_{\rm net,o}\subset \map S_{\rm o}$ can be established based on approximation of the coefficients $\{c_{j\bs \ell},j=1,\cdots,m;\ell_k=1,\cdots,K;k=1,\cdots,|A_j|\}$ and the entries $\{O_{jk}\}$ of the symplectic orthogonal matrix $O$. More precisely, given an arbitrary state $|\psi_{\rm ps}\>\in \map S_{\rm o}$, there exists a state $|\psi'_{\rm ps}\>=U_{O'} \bigotimes_{j=1}^m |\psi_j'\>\in \map S_{\rm net,o}$ such that the following conditions are fulfilled:  
					\begin{align}
						\begin{cases}
							\left|c_{j\bs\ell}-c_{j\bs \ell}'\right|&\le \epsilon_1,\forall j=1,\cdots,m;\ell_k=0\cdots,K;k=1,\cdots,|A_j|\\
							\left|O_{jk}-O_{jk}'\right|&\le \epsilon_2,\forall j,k=1,\cdots,2m
						\end{cases}.\label{eq103aa}
					\end{align}
					The trace distance between states $|\psi_{\rm ps}\>$ and $|\psi'_{\rm ps}\>$ is: 
					\begin{align}
						\frac 1 2 \left\||\psi_{\rm ps}\>\<\psi_{\rm ps}|-|\psi_{\rm ps}'\>\<\psi_{\rm ps}'|\right\|_1\le & \sqrt{\pi m \left(\sqrt 6+\sqrt{10}+5\sqrt 2 m \right)\left(mN_{\rm I}^*+1\right)\epsilon_2} \nonumber \\
						&+\sqrt{\frac{nm N_{\rm I}^*}{K+1}}+ \sum_{j=1}^n(K+1)^{|A_j|/2} \,\epsilon_1+\frac {(K+1)^{|A_j|} }2\epsilon_1^2,\label{eq104}
					\end{align}
					where $\|X\|_1=\Tr\sqrt{X^\dag X}$ denotes the trace norm. 
					
					If one specifically chooses $\frac \epsilon 3 \simeq\sqrt{\pi m \left(\sqrt 6+\sqrt{10}+5\sqrt 2 m \right)\left(mN_{\rm I}^*+1\right)\epsilon_2} \simeq  \sqrt{\frac{nm N_{\rm I}^*}{K+1}}\simeq  \sum_{j=1}^n(K+1)^{|A_j|/2}\,\epsilon_1\gg \sum_{j=1}^n \frac {(K+1)^{|A_j|}} 2\epsilon_1^2
					$, we have: %{\color{blue} Ulysse: it could be $m^2$ instead of $m^3$}
					\begin{align}
						\left|\map S_{\rm net,o}\right|
						=&\left[\mathcal O\left(\frac{n^{2} N_{\rm I}^{*1/2}}{\epsilon}\right)\right]^{{\textbf{\em poly}} \left(n^{4A_{\max}+5}, N_{\rm I}^{*A_{\max}+1}, \left(\frac{1}{\epsilon}\right)^{2A_{\max}},3^{2A_{\max}}\right)},
					\end{align}
					where $A_{\max}=\max_j |A_j|$ is the maximal size of subsystems. 
					
					In addition, when the covering net follows Eq.~(\ref{eq103aa}), the related passive-local observable 
					\begin{align}
						\widehat O_j=U^\dag_{O'}\left(|\psi_j'\>\<\psi_j'|\bigotimes_{k\notin A_j} I_k\right)U_{O'}, j=1,\cdots,n
					\end{align}
					has a total number: 
					\begin{align}
						\left|\map S_{\rm net,\widehat O}\right|
						=&\left[\mathcal O\left(\frac{n^{2} N_{\rm I}^{*1/2}}{\epsilon}\right)\right]^{{\textbf{\em poly}} \left(n^{4A_{\max}+4}, N_{\rm I}^{*A_{\max}+1}, \left(\frac{1}{\epsilon}\right)^{2A_{\max}},3^{2A_{\max}}\right)}.
					\end{align}
				\end{proposition}
				
				\begin{proof}
					The trace distance between states $|\psi_{\rm ps}\>$ and $|\psi'_{\rm ps}\>$ is: %{\color{blue} comment from Francesco: maybe we could start from single-mode sample complexity. Then, it might simplify the proof and achieve optimality. }
					\begin{align}
						&\frac 1 2 \left\||\psi_{\rm ps}\>\<\psi_{\rm ps}|-|\psi_{\rm ps}'\>\<\psi_{\rm ps}'|\right\|_1\nonumber \\\le&  \frac 1 2 \left\||\psi_{\rm ps}\>\<\psi_{\rm ps}|-U_{O'} \bigotimes_{j=1}^n |\psi_j\>\<\psi_j|U_{O'}^\dag \right\|_1+\frac 1 2 \left\|U_{O'} \bigotimes_{j=1}^n |\psi_j\>\<\psi_j|U_{O'}^\dag-|\psi'_{\rm ps}\>\<\psi'_{\rm ps}|\right\|_1\\
						\le&  \sqrt{\frac \pi 2 \left(\sqrt 6+\sqrt{10}+5\sqrt 2 m \right)\left(mN_{\rm I}^*+1\right)\left\|(O')^{-1}O-I\right\|_F}+\frac 1 2 \left\|\bigotimes_{j=1}^n |\psi_j\>\<\psi_j|-\bigotimes_{j=1}^n |\psi_j'\>\<\psi_j'|\right\|_1\\
						\le&  \sqrt{\frac \pi 2 \left(\sqrt 6+\sqrt{10}+5\sqrt 2 m \right)\left(mN_{\rm I}^*+1\right)\left\|O-O'\right\|_F}+\frac 1 2 \sum_{j=1}^n \left\| |\psi_j\>\<\psi_j|-|\psi_j'\>\<\psi_j'|\right\|_1\label{eq103ss}\\
						\le&\sqrt{\pi m \left(\sqrt 6+\sqrt{10}+5\sqrt 2 m \right)\left(mN_{\rm I}^*+1\right)\epsilon_2} + \sum_{j=1}^n \sqrt{\frac{N_j}{K+1}}+ (K+1)^{|A_j|/2} \,\epsilon_1+\frac {(K+1)^{|A_j|} }2\epsilon_1^2\label{eq112}\\
						\le&\sqrt{\pi m \left(\sqrt 6+\sqrt{10}+5\sqrt 2 m \right)\left(mN_{\rm I}^*+1\right)\epsilon_2} +  \sqrt{\frac{nm N_{\rm I}^*}{K+1}}+ \sum_{j=1}^n(K+1)^{|A_j|/2} \,\epsilon_1+\frac {(K+1)^{|A_j|} }2\epsilon_1^2\label{eq113}
					\end{align}
					where the first inequality follows from the triangle inequality, the second inequality is derived using a bound derived from the diamond norm (see Corollary S11 in Ref.~\cite{becker2021energy})%that is equivalent to the more refined completely bounded norm for unitaries  \cite{devetak2006multiplicativity,gupta2015multiplicativity} {\color{blue} Does this imply optimality? }
					, the relation $\|\ln O\|_F\le \frac \pi 2 \|O-I\|_F$, and the fact that trace norm is invariant under unitary operation,  $\left\|X\right\|_F
					= \sqrt{\sum_{jk} \left|X_{jk}\right|^2}$ denotes the Hilbert-Schmidt norm, the third inequality is obtained by the relation $\|UX\|_F=\|X\|_F$ for arbitrary unitary $U$ and by applying the following relations:
					\begin{align}
						\left\|\bigotimes_{j=1}^n |\psi_j\>\<\psi_j|-\bigotimes_{j=1}^n |\psi_j'\>\<\psi_j'|\right\|_1&=\left\|\sum_{j=1}^n \bigotimes_{k=1\atop k<j}^n |\psi_k'\>\<\psi_k'|\otimes \left(|\psi_j\>\<\psi_j|-|\psi_j'\>\<\psi_j'|\right)\bigotimes_{\ell=1\atop \ell>j}^n |\psi_\ell\>\<\psi_{\ell}|\right\|_1\\
						&\le \sum_{j=1}^n\left\| \bigotimes_{k=1\atop k<j}^n |\psi_k'\>\<\psi_k'|\otimes \left(|\psi_j\>\<\psi_j|-|\psi_j'\>\<\psi_j'|\right)\bigotimes_{\ell=1\atop \ell>j}^n |\psi_\ell\>\<\psi_{\ell}|\right\|_1\\
						&=\sum_{j=1}^n\left\| |\psi_j\>\<\psi_j|-|\psi_j'\>\<\psi_j'|\right\|_1.\label{eq107aa}
					\end{align} 
					%approximating each local states as $|\psi'_j\>\<\psi'_j|=|\psi_j\>\<\psi_j|+\Delta_{\psi_j}$,  applying triangle inequality, and omitting the higher order terms $\mathcal O(\Delta_{\psi_j}^2)$: 
					%\begin{align}\frac 1 2 \left\|\bigotimes_{j=1}^n |\psi_j\>\<\psi_j|-\bigotimes_{j=1}^n |\psi_j'\>\<\psi_j'|\right\|_1&= \frac 1 2 \left\| \sum_{j=1}^m \Delta_{\psi_j}\bigotimes_{k=1\atop k\neq j}^n |\psi_k\>\<\psi_k|+\sum_{j=1}^m\mathcal O(\Delta_{\psi_j}^2)\right\|_1\\&\le \frac 1 2 \left\| \sum_{j=1}^m \Delta_{\psi_j}\bigotimes_{k=1\atop k\neq j}^n |\psi_k\>\<\psi_k|\right\|_1+\frac 1 2 \sum_{j=1}^m\left\|\mathcal O(\Delta_{\psi_j}^2)\right\|_1\\&\le \frac 1 2 \sum_{j=1}^m \left\| |\psi_j\>\<\psi_j|-|\psi_j'\>\<\psi_j'|\right\|_1+\frac 1 2 \sum_{j=1}^m\left\|\mathcal O(\Delta_{\psi_j}^2)\right\|_1\\&\approx \frac 1 2 \sum_{j=1}^m \left\| |\psi_j\>\<\psi_j|-|\psi_j'\>\<\psi_j'|\right\|_1.\end{align}
					%In practice, matrix elements of local states' density matirx can be estimated through heterodyne measurement \cite{chabaud2021efficient}. 
					Further, Eq.(\ref{eq112}) is obtained because the error of reconstructing the $j$-th local state in Eq.(\ref{eq103ss}) is bounded as follows: 
					\begin{align}
						&\frac 1 2 \left\| |\psi_j\>\<\psi_j|-|\psi_j'\>\<\psi_j'|\right\|_1\nonumber \\
						=& \frac 1 2 \left\| \sum_{\bs\ell,\bs\ell'=\bs0}^\infty c_{j\bs\ell}c_{j\bs\ell'}^* |\bs\ell\>\<\bs\ell'|-\sum_{\bs\ell\bs\ell'=\bs 0}^{\{K,\cdots,K\}} \left(c_{j\bs\ell}c_{j\bs\ell'}^*+\epsilon_{1,j\bs\ell}c_{j\bs\ell'}^*+\epsilon_{1,j\bs\ell'}^*c_{j\bs\ell}+\epsilon_{1,j\bs\ell}\epsilon_{1,j\bs\ell'}^* \right)|\bs\ell\>\<\bs\ell'|\right\|_1 \\
						%&= \frac 1 2 \left\| \sum_{\ell\ell'=K+1}^\infty c_{j\ell}c_{j\ell'}^* |\ell\>\<\ell'|-\sum_{\ell\ell'=0}^K \left(\epsilon_{1,j\ell}c_{j\ell'}^*+\epsilon_{1,j\ell'}^*c_{j\ell}+\epsilon_{1,j\ell}\epsilon_{1,j\ell'}^* \right)  |\ell\>\<\ell'|\right\|_1\\
						\le &\frac 1 2 \left\| \sum_{\bs\ell\bs\ell'=\bs 0}^\infty c_{j\bs\ell}c_{j\bs\ell'}^* |\bs\ell\>\<\bs\ell'|-\sum_{\bs\ell\bs\ell'=\bs 0}^{\{K,\cdots,K\}} c_{j\bs\ell}c_{j\bs\ell'}^*|\bs\ell\>\<\bs\ell'|\right\|_1+ \left\|\sum_{\bs\ell\bs\ell'=\bs0}^{\{K,\cdots,K\}} \epsilon_{1,j\bs\ell}c_{j\bs\ell'}^*|\bs\ell\>\<\bs\ell'|\right\|_1+\frac 1 2 \left\|\sum_{\bs\ell\bs\ell'=0}^{\{K,\cdots,K\}} \epsilon_{1,j\bs\ell}\epsilon_{1,j\bs\ell'}^*|\bs\ell\>\<\bs\ell'|\right\|_1\label{eq125}\\
						\le & \sqrt{\frac{N_j}{K+1}}+ \sqrt{\sum_{\bs\ell=\bs0}^{\{K,\cdots,K\}}\left|\epsilon_{1,j\bs\ell}\right|^2\sum_{\bs k=\bs 0}^{\{K,\cdots,K\}} \left|c_{j\bs k}\right|^2}+\frac 1 2 \sum_{\bs\ell=\bs 0}^{\{K,\cdots,K\}}\left|\epsilon_{1,j\bs\ell}\right|^2\label{eq127up}\\
						%\le & \sqrt{\frac{|A_j|N_{\rm I}^*}{K+1}}+ \sqrt{\sum_{\bs\ell=\bs0}^{\{K,\cdots,K\}}\left|\epsilon_{1,j\bs\ell}\right|^2\sum_{\bs k=\bs 0}^{\{K,\cdots,K\}} \left|c_{j\bs k}\right|^2}+\frac 1 2 \sum_{\bs\ell=\bs 0}^{\{K,\cdots,K\}}\left|\epsilon_{1,j\bs\ell}\right|^2\label{eq127up}\\
						%&\le \frac 1 2 \left\| \sum_{\ell\ell'=K+1}^\infty c_{j\ell}c_{j\ell'}^* |\ell\>\<\ell'|\right\|_1+ \sqrt{K}\,\epsilon_1+\frac K 2\epsilon_1^2\\
						%&= \frac 1 2 \Tr\left[ \left(\sum_{\ell\ell'=0}^\infty c_{j\ell}c_{j\ell'}^* |\ell\>\<\ell'|\right) \left(\sum_{k=K+1}^\infty|k\>\<k|\right)\right]+ \sqrt{K}\,\epsilon_1+\frac K 2\epsilon_1^2\\
						\le &\sqrt{\frac{N_j}{K+1}}+ (K+1)^{|A_j|/2} \,\epsilon_1+\frac {(K+1)^{|A_j|} }2\epsilon_1^2\label{eq127}
						%\le &\sqrt{\frac{|A_j|N_{\rm I}^*}{K+1}}+ (K+1)^{|A_j|/2} \,\epsilon_1+\frac {(K+1)^{|A_j|} }2\epsilon_1^2\label{eq127}
					\end{align}
					where $\bs\ell=\ell_1,\cdots,\ell_{|A_j|}$ denotes the vector of Fock basis of the $j$-th subsystem, $\{\epsilon_{1,j\bs\ell}=c_{j\bs\ell}-c_{j\bs\ell'}\}$ denote the errors satisfying the condition $|\epsilon_{1,j\bs\ell}|\le \epsilon_{1},\forall j=1,\cdots,m;\bs\ell,\bs\ell'=\bs0,\cdots,\{K,K\cdots,K\}$, Eq.~(\ref{eq125}) is derived using the triangle inequality, the first term of Eq.~(\ref{eq127up}) follows from the gentle measurement lemma (see Lemma 9.4.2 of Ref.~
					\cite{wilde2013quantum}) and 
					Markov inequality \cite{vershynin2018high} for pure-state photon statistics: 
					\begin{align}
						\Tr[|\psi_j\>\<\psi_j| \left(I_{A_j}-\sum_{\bs \ell=\bs 0}^{\{K,\cdots,K\}} |\bs \ell\>\<\bs \ell|\right)]&\le 1-\prod_{\ell\in A_j}\left[1-\frac{\Tr[|\psi_j\>\<\psi_j| a_\ell^\dag a_\ell ]}{K+1}\right]\\
						&\approx \sum_{\ell \in A_j } \frac{\Tr[|\psi_j\>\<\psi_j| a_\ell^\dag a_\ell ]}{K+1}\\
						&\le \frac{N_j}{K+1},\label{eq121}
						%&\le \frac{|A_j|N_{\rm I}^*}{K+1},\label{eq121}
					\end{align}
					where $N_j$ denotes the  photon number of the local state $|\psi_j\>$, Eq.~(\ref{eq127}) is derived using the definition of $\epsilon_1$ and the relation $\sum_{\bs k=1}^{\{K,\cdots,K\}}|c_{j\bs k}|^2\le 1$. Eq.~(\ref{eq113}) is derived from the Cauchy-Schwarz inequality $  \sum_{j=1}^m \sqrt{ A_j}\le \sqrt{m \sum_{j=1}^m A_j}$.

					Now, let us look at the size of the $\epsilon$-covering net $\map S_{\rm net,o}$. It can be shown that each matrix elements of $O$ has a range $O_{jk}\in [-1,1]$ \cite{serafini2017quantum}. In addition, the fact that any single-mode pure state $|\psi_j\>:=\sum_{\bs\ell=\bs0}^\infty c_{j\bs\ell}|\bs\ell\>$ must satisfy the constraint ${\rm Re}c_{j\bs\ell},{\rm Im}c_{j\bs\ell}\in [-1,1],\forall j,\ell$ to fulfill the normalization condition $\<\psi_j|\psi_j\>=1$. Therefore, the net $\map S_{\rm net,o}$ has a size: 
					\begin{align}
						\left|\map S_{\rm net,o}\right|=&\left(\frac 1 {\epsilon_1}\right)^{2\sum_{j=1}^n(K+1)^{|A_j|}}\left(\frac 1 {\epsilon_2}\right)^{4m^2},
					\end{align}
					if we divide the range $[-1,1]$ of each real coefficients into $1/\epsilon_1(1/\epsilon_2)$ sub‑intervals of width $2\epsilon_1(2\epsilon_2)$ and simply identify which sub‑interval each quantity falls into.
					
					Note that we assume that each matrix element of $O$ and each coefficient $\{c_{j\ell}\}$ is measured to ensure the bound in Eq.~(\ref{eq112}) is satisfied. In practice, this may not be necessary, as several constraints, e.g., $O^{-1}=O^\dag$, on the matrix $O$ and local states
					$|\psi_j\>$ can reduce the number of independent parameters. Without loss of generality, we can re-parametrize the r.h.s. of Eq.~(\ref{eq112}) by a total error $\epsilon$, and choose 
					\begin{align}
						\frac \epsilon 3 &\simeq\sqrt{\pi m \left(\sqrt 6+\sqrt{10}+5\sqrt 2 m \right)\left(mN_{\rm I}^*+1\right)\epsilon_2} \simeq  \sqrt{\frac{nm N_{\rm I}^*}{K+1}}\simeq \sum_{j=1}^n (K+1)^{|A_j|/2} \,\epsilon_1\gg \sum_{j=1}^n\frac {(K+1)^{|A_j|} }2\epsilon_1^2.\label{eq122aa}
					\end{align}
					Then, we have: 
					\begin{align}
						\begin{cases}
							K+1&\le  \left(\frac{\sqrt{nm N_{\rm I}^*}}{\epsilon_1}\right)^{\frac{2}{A_{\max}+1}}\\
							%K+1&\simeq \left(\frac{\sqrt{|A_j|N_{\rm I}^*}}{\epsilon_1}\right)^{\frac 2 {|A_j|+1}},\forall j =1,\cdots,n\\
							\frac{2}{\epsilon_1}&\le  
							2\left(\frac{3n}{\epsilon}\right)^{A_{\max}+1} \left(nm N_{\rm I}^*\right)^{\frac{A_{\max}}{2}}
						\end{cases}
					\end{align}
					where $A_{\max}=\max_{j} |A_j|$ is the maximal size of  subsystems. 
					
					Further, the number of epsilon balls is given ($n=\mathcal O(m)$): 
					\begin{align}
						\left|\map S_{\rm net,o}\right|\le  &\left(\left(\frac{3n}{\epsilon}\right)^{A_{\max}+1} \left(nm N_{\rm I}^*\right)^{\frac{A_{\max}}{2}}\right)^{2\sum_{j=1}^n \left(nmN_{\rm I}^*\right)^{\frac {|A_j|} {A_{\max}+1}}\left(\left(\frac{3n}{\epsilon}\right)^{A_{\max}+1} \left(nmN_{\rm I}^*\right)^{\frac{A_{\max}}{2}}\right)^{\frac{2|A_j|}{A_{\max}+1}}}\nonumber \\
						&\times \left(\frac {9\pi m \left(\sqrt 6+\sqrt{10}+5\sqrt 2 m \right)\left(mN_{\rm I}^*+1\right)} {\epsilon^2}\right)^{4m^2}\\
						=&\left[\mathcal O\left(\frac{n^{2} N_{\rm I}^{*1/2}}{\epsilon}\right)\right]^{{\textbf{poly}} \left(n^{4A_{\max}+5}, N_{\rm I}^{*A_{\max}+1}, \left(\frac{1}{\epsilon}\right)^{2A_{\max}},3^{2A_{\max}}\right)}.
					\end{align}
					
					Similarly, when considering the set of passive-local observables 
					\begin{align}
						\widehat O_j=U^\dag_{O'}\left(|\psi_j'\>\<\psi_j'|\bigotimes_{k\notin A_j} I_k\right)U_{O'}, j=1,\cdots,n
					\end{align}
					the condition Eq.~(\ref{eq103aa}) indicates that the number of observables is: 
					\begin{align}
						\left|\map S_{\rm net,\widehat O}\right|=&\sum_{j=1}^n\left(\frac 1 {\epsilon_1}\right)^{2(K+1)^{|A_j|}}\left(\frac 1 {\epsilon_2}\right)^{4m^2}.
					\end{align}

					By imposing the assumption in Eq.~(\ref{eq122aa}), we have:
					\begin{align}
						\left|\map S_{\rm net,\widehat O}\right|\le  &\sum_{j=1}^n\left(\left(\frac{3n}{\epsilon}\right)^{A_{\max}+1} \left(nm N_{\rm I}^*\right)^{\frac{A_{\max}}{2}}\right)^{2 \left(nmN_{\rm I}^*\right)^{\frac {|A_j|} {A_{\max}+1}}\left(\left(\frac{3n}{\epsilon}\right)^{A_{\max}+1} \left(nmN_{\rm I}^*\right)^{\frac{A_{\max}}{2}}\right)^{\frac{2|A_j|}{A_{\max}+1}}}\nonumber \\
						&\times \left(\frac {9\pi m \left(\sqrt 6+\sqrt{10}+5\sqrt 2 m \right)\left(mN_{\rm I}^*+1\right)} {\epsilon^2}\right)^{4m^2}\\
						=&\left[\mathcal O\left(\frac{n^{2} N_{\rm I}^{*1/2}}{\epsilon}\right)\right]^{{\textbf{poly}} \left(n^{4A_{\max}+4}, N_{\rm I}^{*A_{\max}+1}, \left(\frac{1}{\epsilon}\right)^{2A_{\max}},3^{2A_{\max}}\right)}.
					\end{align}
				\end{proof}

				Without loss of generality, one extends the above discussion to pure GE states. Here, for simplicity of discussion, we consider a simpler scenario where the number of partition $n$ is equal to number of modes $m$: 
				
				\begin{theorem}[$\epsilon$-covering net of pure GE states ($n=m$)] \label{theo:epsilon covering net of GE state}
					Consider the set of $m$-mode pure states $\map S_{\rm ge}$ assumed to have the form $|\psi\>:=U_{\bs \alpha } U_S \bigotimes_{j=1}^m |\psi_j\>$ where $U_{\bs \alpha }$ denotes an $m$-mode displacement operator with $\bs\alpha=(\alpha_1,\cdots,\alpha_m)^T$, $U_S$ is an $m$-mode Gaussian unitary, $|\psi_j\>:=\sum_{\ell=0}^\infty c_{j\ell}|\ell\>$ denotes local pure states in the $j$-th subsystem. The state $|\psi\>$ satisfies the energy constraint $\<\psi|(\sum_{j=1}^m a_j^\dag a_j +m/2 I)|\psi\>\le mE_{\rm I}$. 
					%$\Tr [\left(\sum_{j=1}^m a_j^\dag a_j+\frac m 2 I \right)  |\psi\>\<\psi| ]\le  m E_{\rm I}$ and $\Tr[\left(a^\dag_ja_j+\frac I 2\right) |\psi_j\>\<\psi_j|]\le E_{\rm I} ,\forall j=1,\cdots,m$. 
					Then, a $\epsilon$-covering net $\map S_{\rm net,ge}\subset \map S_{\rm ge}$ can be established based on approximation of the coefficients $\{c_{j\ell},j=1,\cdots,m;\ell=1,\cdots,K\}$, the entries $\{S_{jk}\}$ of the symplectic orthogonal matrix $S$, and the displacements $\{\alpha_j\}$. More precisely, given an arbitrary state $|\psi\>\in \map S_{\rm ge}$, there exists a state $|\psi'\>=U_{\bs\alpha'}U_{S'} \bigotimes_{j=1}^m |\psi_j'\>\in \map S_{\rm net,ge}$ such that the following conditions are fulfilled:  
					\begin{align}
						\begin{cases}
							\left|c_{j\ell}-c_{j\ell}'\right|&\le \epsilon_1,\forall j=1,\cdots,m;\ell=0\cdots,K\\
							\left|S_{jk}-S_{jk}'\right|&\le \epsilon_2,\forall j,k=1,\cdots,2m\\
							\left|\alpha_{j}-\alpha_{j}'\right|&\le \epsilon_3,\forall j=1,\cdots,m
						\end{cases}.
					\end{align}
					The trace distance between states $|\psi\>$ and $|\psi'\>$ is: 
					\begin{align}
						\frac 1 2 \left\||\psi\>\<\psi|-|\psi'\>\<\psi'|\right\|_1\le & \sin \left(\min\left\{\sqrt m\, \epsilon_3\left(\sqrt{mN_{\rm I}'}+\sqrt{mN_{\rm I}'+1}\right),\frac \pi 2 \right\}\right)\nonumber \\
						&+\sqrt{2m\sqrt{2m}\left(\sqrt 6+\sqrt{10}+5\sqrt 2 m \right)\left(mN_{\rm I}^*+1\right)\epsilon_2}\left(\sqrt \pi+\sqrt 2 (r_s+2)\right)\sqrt{r_s+2}\nonumber \\
						&+m\sqrt{\frac{N_{\rm I}^*}{K+1}}+ m\sqrt{K+1}\,\epsilon_1+m\frac {K+1} 2\epsilon_1^2,
					\end{align}
					where the coefficients $\{r_s,r_a,N_{\rm I}^*,N_{\rm I}'\}$ are defined by $\|S\|_\infty \le r_s$, $|\alpha_j|\le r_a$, $\Tr [\left(\sum_{j=1}^m a_j^\dag a_j\right)  \bigotimes_{k=1}^m|\psi_k\>\<\psi_k| ]\le  m N_{\rm I}^*$ and $\Tr [\left(\sum_{j=1}^m a_j^\dag a_j\right)  U_S\bigotimes_{k=1}^m|\psi_k\>\<\psi_k|U_S^\dag  ]\le  m N_{\rm I}'$. 
					
					\iffalse 
					\begin{align}
						\begin{cases}
							N_{\rm I}&\to \mathcal O(E_{\rm I})\\
							N_{\rm I}'&\to \mathcal O(E_{\rm I})\\
							r_s&\to \mathcal O(\sqrt{4mE_{\rm I}})\\
							r_a&\to \mathcal O(\sqrt{E_{\rm I}})
						\end{cases}.
					\end{align}
					\fi 
					
					If one specifically chooses the four dominating terms of the above error are of the same level $\sim\epsilon/4$, %and assumes that $N_{\rm I}\approx N_{\rm I}=\mathcal O(1)=r_s\approx r_a$
					we have: %{\color{blue} Ulysse: it could be $m^2$ instead of $m^3$}
					\begin{align}
						\left|\map S_{\rm net,g}\right|=
						&
						\left[\mathcal O\left(m^{3} ,E_{\rm I}^{\frac 3 2 },\frac{1}{\epsilon}\right)\right]^{{\textbf{\em poly}} \left(m^3, E_{\rm I}, \frac{1}{\epsilon^2}\right)}.
					\end{align}
				\end{theorem}
				
				\begin{proof}
					Here, we follow a similar proof strategy as in Proposition~\ref{prop:epsilon covering net}. Consider the set of $m$-mode pure states $\map S_{\rm g}$ assumed to have the form $|\psi\>:=U_{\bs \alpha } U_S \bigotimes_{j=1}^m |\psi_j\>$ where $U_{\bs \alpha }$ denotes an $m$-mode displacement operator with $\bs\alpha=(\alpha_1,\cdots,\alpha_m)^T$ and $|\alpha_j|\le r_a$, $U_S$ is an $m$-mode Gaussian unitary with $\|S\|_\infty \le r_s$, $|\psi_j\>:=\sum_{\ell=0}^\infty c_{j\ell}|\ell\>$ denotes local pure states in the $j$-th subsystem satisfying the energy constraints $\Tr [\left(\sum_{j=1}^m a_j^\dag a_j\right)  \bigotimes_{k=1}^m|\psi_k\>\<\psi_k| ]\le  m N_{\rm I}^*$ and $\Tr [\left(\sum_{j=1}^m a_j^\dag a_j\right)  U_S\bigotimes_{k=1}^m|\psi_k\>\<\psi_k|U_S^\dag  ]\le  m N_{\rm I}'$. The trace distance between a given GE state $|\psi\>\in \map S_{\rm g}$ and the reconstructed state $|\psi'\>\in \map S_{\rm net,g}$ can be bounded as follows:
					\begin{align}
						&\frac 1 2 \left\||\psi\>\<\psi|-|\psi'\>\<\psi'|\right\|_1\nonumber \\\le&  \frac 1 2 \left\||\psi\>\<\psi|-U_{\bs \alpha'}U_{S} \bigotimes_{j=1}^m |\psi_j\>\<\psi_j|U_{S}^\dag U_{\bs \alpha'}^\dag \right\|_1\nonumber \\
						&+\frac 1 2 \left\|U_{\bs \alpha'}U_{S} \bigotimes_{j=1}^m |\psi_j\>\<\psi_j|U_{S}^\dag U_{\bs \alpha'}^\dag -U_{\bs \alpha'}U_{S'} \bigotimes_{j=1}^m |\psi_j\>\<\psi_j|U_{S'}^\dag U_{\bs \alpha'}^\dag \right\|_1\nonumber \\
						&+\frac 1 2 \left\|U_{\bs \alpha'}U_{S'} \bigotimes_{j=1}^m |\psi_j\>\<\psi_j|U_{S'}^\dag U_{\bs \alpha'}^\dag -|\psi'\>\<\psi'|\right\|_1\\
						=&  \frac 1 2 \left\||\psi\>\<\psi|-U_{\bs \alpha'}U_{S} \bigotimes_{j=1}^m |\psi_j\>\<\psi_j|U_{S}^\dag U_{\bs \alpha'}^\dag \right\|_1
						+\frac 1 2 \left\|U_{S} \bigotimes_{j=1}^m |\psi_j\>\<\psi_j|U_{S}^\dag  -U_{S'} \bigotimes_{j=1}^m |\psi_j\>\<\psi_j|U_{S'}^\dag \right\|_1\nonumber \\
						&+\frac 1 2 \left\|\bigotimes_{j=1}^m |\psi_j\>\<\psi_j| -\bigotimes_{j=1}^m|\psi'_j\>\<\psi'_j|\right\|_1\\
						\le&  \sin \left(\min\left\{
						%\frac{\|\bs d-\bs d'\|_2}{ 2}
						\sqrt{\sum_{j=1}^m\left|\alpha_j-\alpha_j'\right|^2}
						\left(\sqrt{mN_{\rm I}'}+\sqrt{mN_{\rm I}'+1}\right),\frac \pi 2 \right\}\right)\nonumber \\
						&+\sqrt{\sqrt{2m}\left(\sqrt 6+\sqrt{10}+5\sqrt 2 m \right)\left(mN_{\rm I}^*+1\right)}\left(\sqrt \pi+\sqrt 2 (r_s+2)\right)\sqrt{r_s+2}\sqrt{\left\|S-S'\right\|_F}\nonumber \\
						&+\sum_{j=1}^m\sqrt{\frac{N_j}{K+1}}+ m\sqrt{K}\,\epsilon_1+m\frac K 2\epsilon_1^2\\
						\le&  \sin \left(\min\left\{\sqrt m\, \epsilon_3\left(\sqrt{mN_{\rm I}'}+\sqrt{mN_{\rm I}'+1}\right),\frac \pi 2 \right\}\right)\nonumber \\
						&+\sqrt{2m\sqrt{2m}\left(\sqrt 6+\sqrt{10}+5\sqrt 2 m \right)\left(mN_{\rm I}^*+1\right)\epsilon_2}\left(\sqrt \pi+\sqrt 2 (r_s+2)\right)\sqrt{r_s+2}\nonumber \\
						&+m\sqrt{\frac{ N_{\rm I}^*}{K+1}}+ m\sqrt{K+1}\,\epsilon_1+m\frac {K+1} 2\epsilon_1^2\label{eq122}
					\end{align}
					where the first inequality is obtained by applying the triangle inequality, the second inequality follows from Eq.~(3) and Theorem 6 of Ref.~\cite{becker2021energy} with $r$ being defined by $\|S\|_\infty \le r_s$, together with Eqs.~(\ref{eq107aa}) and (\ref{eq127}) with $\{N_j\}$ being the photon number of each local states, $\|\bs m\|_2=\sqrt{\bs m^T\bs m}$ represents the Euclidean norm, %$\bs d=(2{\rm Re}\alpha_1,2{\rm Im}\alpha_1,\cdots )^T , (\bs d'=(2{\rm Re}\alpha_1',2{\rm Im}\alpha_1',\cdots )^T )$ is the displacement vector for the quadratures. 
					the last inequality is derived using the Cauchy-Schwarz inequality as well as the 
					definition of the errors $\epsilon_2$ and $\epsilon_3$. 
					
					Given the H\"older's inequality $\|S\|_F^2 \le \|S\|_\infty\cdot \Tr|S|$ and $\Tr|S|\le 2m\|S\|_\infty$, we have the relation $\sum_{jk=1}^{2m} |S_{jk}|^2\le 2m\, r_s^2$. Therefore, the net $\map S_{\rm net,g}$ has a size: 
					\begin{align}
						\left|\map S_{\rm net,g}\right|&= \left(\frac 1 {\epsilon_1}\right)^{2m(K+1)}\left(\frac {\sqrt{2m} \,r_s} {\epsilon_2}\right)^{4m^2}\left(\frac{r_a}{\epsilon_3}\right)^{2m}.
					\end{align}
					%{\color{blue} for $\epsilon_2$, maybe after estimating coefficients whose sum is close to one, the search region for others is small. }
					
					Without loss of generality, we assume that the first and second lines, as well as the first and second terms in the third line of Eq.~(\ref{eq122}), all have the same error level of $\epsilon/4$. Then, we have the net size: 
					\begin{align}
						\left|\map S_{\rm net,g}\right|=&\left(\frac{16m^2\sqrt{N_{\rm I}^*}}{\epsilon^2}\right)^{\frac{32m^3N_{\rm I}^*}{\epsilon^2}}\nonumber \\
						&\times\left(\frac {32 m\sqrt{2m} \left(\sqrt 6+\sqrt{10}+5\sqrt 2 m \right)\left(mN_{\rm I}^*+1\right)\left(\sqrt \pi +\sqrt 2 (r_s+2)\right)^2(r_s+2)\sqrt{2m}\,r_s} {\epsilon^2}\right)^{4m^2}\nonumber \\
						&\times\left(\frac{4r_a\sqrt m\left(\sqrt{mN_{\rm I}'}+\sqrt{mN_{\rm I}'+1}\right)}{\epsilon}\right)^{2m}.%\\
						%=&\left[\mathcal O\left(\frac{m^{2} }{\epsilon},\sqrt{N_{\rm I}},\sqrt{N_{\rm I}'},r_s^{2},r_a\right)\right]^{{\textbf{poly}} \left(m^3, N_{\rm I}, \frac{1}{\epsilon^2}\right)}.
					\end{align}
					
					Given Lemma \ref{lem: unify energy}, which will be presented later, we can substitute the moment constraints as follows: 
					\begin{align}
						\begin{cases}
							N_{\rm I}^*&\to \mathcal O(E_{\rm I})\\
							N_{\rm I}'&\to \mathcal O(E_{\rm I})\\
							r_s&\to \mathcal O(\sqrt{4mE_{\rm I}})\\
							r_a&\to \mathcal O(\sqrt{E_{\rm I}})
						\end{cases}.
					\end{align}
					
					Finally, we have the following size of $\epsilon$-covering net with a simple energy constraint $\<\psi|(\sum_{j=1}^m a_j^\dag a_j +m/2 I)|\psi\>\le mE_{\rm I}$: 
					\begin{align}
						\left|\map S_{\rm net,g}\right|=&
						\left[\mathcal O\left(m^{3} ,E_{\rm I}^{\frac 3 2 },\frac{1}{\epsilon}\right)\right]^{{\textbf{poly}} \left(m^3, E_{\rm I}, \frac{1}{\epsilon^2}\right)}.
					\end{align}
					
				\end{proof}
				
				Note that moment constraints for pure GE states can be bounded in various ways. In particular, the following lemma applies when the "lowest weight state" $|\phi\>$ is replaced by .
				
				\begin{lemma}[Extending Lemma S48 of \cite{mele2024learning}]\label{lem: unify energy}
					Consider an $m$-mode pure state $|\psi\>:=U_{\bs \alpha}U_S|\phi\>$, where $U_{\bs \alpha}$ is a displacement operator, $U_S$ is a Gaussian unitary, the state $|\phi\>$ has zero displacement and a diagonal covariance matrix $\Lambda$. One can define the following energy constraints: 
					\begin{align}
						\begin{cases}
							E_{\rm I}&=\frac 1 m\<\psi|\left(\sum_{j=1}^m a^\dag_j a_j +\frac m 2 I\right)|\psi\>\\
							%|\alpha_j|&\le  r_a\\
							%\|S\|_\infty &\le  r_s\\ 
							N_{\rm I}^*&\ge \frac 1 m \Tr [\left(\sum_{j=1}^m a_j^\dag a_j\right)  |\phi\>\<\phi| ]\\
							N_{\rm I}'&\ge \frac 1 m \Tr [\left(\sum_{j=1}^m a_j^\dag a_j\right)  U_S|\phi\>\<\phi|U_S^\dag  ] .
						\end{cases}
					\end{align}
					Then, we have the following relations: 
					\begin{align}
						\begin{cases}
							E_{\rm I}&\ge \frac 1 {4m} \|S\|^2_\infty + \frac 1 m \sum_{j=1}^m |\alpha_j|^2\\
							E_{\rm I}-\frac 1 2 &\ge N_{\rm I}'\ge N_{\rm I}^*
						\end{cases}\label{eq127aa}
					\end{align}
				\end{lemma}
				\begin{proof}
					The total energy of the state $|\psi\>$ will be:  
					\begin{align}
						E_{\rm I}=\frac 1 m\<\psi|\left(\sum_{j=1}^m a^\dag_j a_j +\frac m 2 I\right)|\psi\>=\frac 1 {4m} \Tr[V]+ \frac 1 m\sum_{j=1}^m |\alpha_j|^2
					\end{align}
					where the equation follows from  Lemma S48 of Ref.~\cite{mele2024learning}. Given the following relation:
					\begin{align}
						\Tr[V]&=\Tr[S\Lambda S^T]\\
						&=\sum_{k\ell=1}^{2m} S_{k\ell}^2 \Lambda_{\ell\ell}\\
						&\ge \sum_{k\ell=1}^{2m} S_{k\ell}^2 \\
						&=\Tr[SS^T]\\
						&=\|S\|^2_F\\
						&\ge \|S\|_\infty^2 
					\end{align}
					where $\{S_{k\ell}\in \R\}$ and $\{\Lambda_{k\ell}=d_{k}\delta_{k\ell},d_k\ge 1\}$ denote the matrix elements of $S$ and $\Lambda$, respectively, we have the first line of Eq.~(\ref{eq127aa}). Moreover, given that: 
					\begin{align}
						\Tr[V]&\ge \Tr [S^TS\Lambda ]\\
						&= \Tr \left[\frac{S^\dag S+(S^\dag S)^{-1}}{2}\Lambda \right]\\
						&\ge \Tr [\Lambda]
					\end{align}
					where the equation is derived by using the relations $S^T=S^\dag$, $S^\dag S=(S^\dag S)^{-1}$, and the inequality $(S^\dag S+(S^\dag S)^{-1})/2\ge I$, the energy of state $|\phi\>$ will be $N_{\rm I}'=\Tr[V]/(4m)-1/2\ge N_{\rm I}^*=\Tr[\Lambda]/(4m)-1/2$. Thus, we have the second line of Eq.~(\ref{eq127aa}).  
				\end{proof}

				Based on Proposition \ref{prop:epsilon covering net}, the global fidelity between any pure state with bounded energy and any candidate state in the set $\map S_{\rm o}$ can be efficiently estimated. A detailed algorithm is given as follows:
				
				\begin{framed}
					
					\begin{center}
						\textbf{Algorithm 2: fidelity witness of pure passive-separable states }
					\end{center}
					
					\textbf{Input:} (a) Multiple  copies of an unknown $m$-mode pure state $|\psi_{\rm ps}\>:=U_O \bigotimes_{j=1}^n |\psi_j\>\in\bigotimes_{j=1}^n\map H_{A_j}$ where $U_O$ is a beam-splitter operator, $|\psi_j\>\in \map H_{A_j}$ denotes local pure states in the $j$-th subsystem satisfying the energy constraint $\Tr [ a_j^\dag a_j  |\psi_{\rm ps}\>\<\psi_{\rm ps}| ]\le N_{\rm I}^*,\forall j=1,\cdots,m$, respectively. (b) Classical information on a target state $|\psi'_{\rm ps}\>:=U_{O'} \bigotimes_{j=1}^m |\psi_j'\>$ where $\{| \psi_j'\>:=\sum_{\bs\ell=\bs0}^{\{K,\dots,K\}} c_{j\bs\ell}'|\bs\ell\>,j=1,\cdots,m\}$ denotes the reconstructed local states that are truncated and could be unnormalized.
					
					\textbf{Output:} An estimate of global fidelity $F=|\<\psi_{\rm ps}|\psi_{\rm ps}'\>|^2$ with error $\epsilon$ and probability of failure $\delta$.

					3.1: Perform multi-mode heterodyne measurement on the input state $|\psi_{\rm ps}\>$ and obtain the sample $\gamma_j$ for $j=1,\cdots,m$.
					
					3.2: Compute $\bs \alpha=U_{ O'}  \bs \gamma$ with $\bs \alpha=(\alpha_1,\cdots,\alpha_m)^T$ and $\bs \gamma=(\gamma_1,\cdots,\gamma_m)^T$. For the $j$-th local system $A_j$, the post-measurement state is given by 
					$\bigotimes_{\ell\in A_j}|\alpha_\ell\>\<\alpha_\ell|$, occurring with probability $p_j=\Tr[\otimes_{\ell\in A_j}|\alpha_\ell\>\<\alpha_\ell|\rho_j]$, where the local state is defined as  $\rho_j:=\Tr_{\{A_k\}\atop k\neq j}[U_{O'}^\dag |\psi_{\rm ps}\>\<\psi_{\rm ps}|U_{O'}]$. Here we denote $y_j=(\cdots,2{\rm Re}\alpha_\ell,2{\rm Im}\alpha_\ell,\cdots)^T$ with $\ell\in A_j$ as the vector of the corresponding quadrature values of $\{\alpha_\ell,\ell\in A_j\}$.  
					
					3.3: Reconstruct the local fidelity: 
					\begin{align}
						F_j&= \<\psi_j'|\widehat \rho_j|\psi_j'\>= \sum_{\bs k\bs \ell=\bs 0}^{\{K,\cdots,K\}} c_{j\bs k}^{'*}\, c_{j\bs \ell}' \,\widehat \rho_{j,\bs k\bs \ell}\\
						\widehat \rho_{j,\bs k\bs \ell}&= \int_{\|u\|_2^2\le \infty } \frac{\d^{2|A_j|} u}{\pi} e^{  \|u\|_2^2/8-iu^T \Omega y_j/2 } \Tr[|\bs k\>\<\bs \ell| e^{-iu^T \Omega (\cdots,p_h,q_h,\cdots)^T/2} ]
					\end{align}
					where $\{\widehat \rho_{j,\bs k\bs \ell}\}$ denote the approximated entries of density matrix $ \rho_{j}$ of the $j$-th local state \cite{becker2024classical}, $q_h=a_h+a_h^\dag$ and $p_h=i(a^\dag_h-a_h)$ with $h\in A_j$ refer to the position and momentum operators of the $j$-th subsystem. 
					
					3.4: Compute the fidelity witness estimate 
					\begin{align}
						W&=1-\sum_{j=1}^n \left(1-F_j \right) .
					\end{align}
					
					3.5: Given the relation $W\le F\le \frac 1 n \sum_{j=1}^n F_j$ (see Lemma 2 in  Ref.~\cite{chabaud2021efficient}), we have the estimate of global fidelity: 
					\begin{align}
						\widetilde F\in \left[F-n\epsilon_{\rm lo},F+\epsilon_{\rm lo}\right], 
					\end{align}
					where $\epsilon_{\rm lo}$ is the estimation error for estimating each local fidelity $F_j$. 
				\end{framed}
				
				Furthermore, the following theorem is given: 
				
				\begin{theorem}[Efficient local tomography of passive-separable states] \label{theo:learn passive-separable states} Given $M$ copies of an unknown $m$-mode pure state $|\psi_{\rm ps}\>:=U_O \bigotimes_{j=1}^n |\psi_j\>$ where $U_O$ is a beam-splitter operator, $|\psi_j\>$ denotes local pure states in the $j$-th subsystem. The overall state satisfies the energy constraint $\Tr [ \sum_{\ell=1}^m a_\ell^\dag a_\ell |\psi_{\rm ps}\>\<\psi_{\rm ps}| ]\le m N_{\rm I}^*,\forall j=1,\cdots,m$. Then, shadow tomography of the state $|\psi\>$ requires a sample complexity
					\begin{align}
						M_{\rm ps}&= \mathcal O\left[\textbf{\em poly}\left( n^{5A_{\max}+6}, N_{\rm I}^{*2A_{\max}+1},\left(\frac 1 {\epsilon_{\rm ps}}\right)^{2A_{\max}+4},\log \frac 1 {\delta_{\rm ps} }, 3^{2A_{\max}} \right) \right], 
					\end{align}
					with $\epsilon_{\rm ps}$ being the error in trace distance, $\delta_{\rm ps}$ being the failure probability. 
				\end{theorem}
				
				\begin{proof}
					Based on Theorem 3 of Ref.~\cite{becker2024classical}, by replacing the observable as follows: 
					\begin{align}
						O_j=U^\dag_{O'}\left(|\psi_j'\>\<\psi_j'|\bigotimes_{k\notin A_j} I_k\right)U_{O'},
					\end{align}
					the local fidelity  $F_j=\<\psi_j'|\Tr_{A_k\neq A_j} [U_{O'}^\dag |\psi_{\rm ps}\>\<\psi_{\rm ps}|U_{O'}]|\psi'_j\>$ can be estimated to within an error $\epsilon_{\rm lo}$, with the following sample complexity:
					\begin{align}
						M&= \mathcal O\left[\textbf{poly} \left(\frac 1 {\epsilon^2_{\rm lo}}, \left(A_{\max}N_{\rm I}^{**}\right)^{A_{\max}}, \log \left(\frac 1 {\delta_{\rm ps}} \right),\log L\right) \right],
					\end{align}
					where $N_{\rm I}^{**}$ is the photon number per mode for the local state $\Tr_{A_k\neq A_j} [ |\psi_{\rm ps}\>\<\psi_{\rm ps}|]$, $L$ denotes the total number of observables (see Proposition \ref{prop:epsilon covering net})
					\begin{align}
						L=&\left[\mathcal O\left(\frac{n^{2} N_{\rm I}^{*1/2}}{\epsilon_{\rm ps}}\right)\right]^{{\textbf{poly}} \left(n^{4A_{\max}+4}, N_{\rm I}^{*A_{\max}+1}, \left(\frac{1}{\epsilon_{\rm ps}}\right)^{2A_{\max}},3^{2A_{\max}}\right)},
					\end{align}
					$\delta_{\rm ps}$ is the failure probability of $M$ experiments. Note that the local energy $A_{\max}N_{\rm I}^{**}$ can be upper bounded by $mN_{\rm I}^{*}$ due to the relation: $|A_j|N_{\rm I}^{**}%=&\Tr\left[\left(\sum_{\ell\in A_j}a^\dag_\ell a_\ell\right)\Tr_{A_k\neq A_j} [ |\psi_{\rm ps}\>\<\psi_{\rm ps}|]\right]\\
					=\Tr\left[\left(\sum_{\ell\in A_j}a^\dag_\ell a_\ell\right)|\psi_{\rm ps}\>\<\psi_{\rm ps}|\right]\le \Tr\left[\left(\sum_{\ell=0 }^m a^\dag_\ell a_\ell\right)|\psi_{\rm ps}\>\<\psi_{\rm ps}|\right]\le m N_{\rm I}^*$. 
					
					%\begin{align}\left|\map S_{\rm net,o}\right|=&\left[\mathcal O\left(\frac{n^{3/2} N_{\rm I}^{1/2}}{\epsilon}\right)\right]^{{\textbf{\em poly}} \left(n^{2A_{\max}+1}, N_{\rm I}^{A_{\max}+1}, \left(\frac{1}{\epsilon}\right)^{2A_{\max}},A_{\max}^{A_{\max}}\right)},.\end{align}
					
					Finally, one can identify the state $|\psi'_{\rm ans}\>$ with a global fidelity:
					\begin{align}
						\widetilde F\ge 1-n\,\epsilon_{\rm lo}  .
					\end{align}
					Using Fuchs-van de Graaf inequality \cite{fuchs1999cryptographic}, one have: 
					\begin{align}
						\frac 1 2 \left\||\psi_{\rm ps}\>\<\psi_{\rm ps}|-|\psi'_{\rm ans}\>\<\psi'_{\rm ans}|\right\|_1&\le \sqrt{1-\widetilde F}\\
						&\le \sqrt{n\,\epsilon_{\rm lo}}, 
					\end{align}
					where $|\psi'\>_{\rm ans}$ denotes the target state that maximizes the global fidelity.
					
					By re-parameterization $\epsilon_{\rm ps}=\sqrt{n\epsilon_{\rm lo}}$, and replacement of the value of $L=|\map S_{\rm net,o}|$ into $M$, one has Theorem \ref{theo:learn passive-separable states}. 
				\end{proof}

			\end{enumerate}

			Finally, one can reconstruct the pure GE state $|\psi\>$ based on the aforementioned tomography outcomes as follows: 
			\begin{align}
				\left|\widetilde \psi\right\>&=U_{\widetilde{\bs \alpha}} U_{\widetilde S} \left|\widetilde \psi_{\rm ps}\right\>,
			\end{align}
			where $U_{\widetilde{\bs \alpha}} U_{\widetilde S} $ and $U_{\widetilde{\bs \alpha}} U_{\widetilde S}$ are estimated from step 1, $\left|\widetilde \psi_{\rm ps}\right\>=U_{\widetilde O} \bigotimes_{j=1}^n \left|\widetilde \psi_j\right\>$ denotes the reconstructed passive-separable state from Step 3. Accordingly, the correct value of the symplectic orthogonal matrix is denoted by $O$.

			The error total error of learning pure GE state $|\psi\>$ (defined in Eq.~(\ref{eqf3})) is:
			\begin{align}
				2\epsilon:=&\left\|\left|\widetilde \psi\right\>\left\<\widetilde \psi\right|-|\psi\>\<\psi|\right\|_1 \\
				=& \left\| U_{\widetilde{\bs \alpha}} U_{\widetilde S} \left|\widetilde \psi_{\rm ps}\right\>\left\<\widetilde \psi_{\rm ps}\right|U_{\widetilde S}^\dag  U_{\widetilde{\bs \alpha}}^\dag  - U_{\bs \alpha} U_S   \left(\bigotimes_{j=1}^n|\psi_j\>\<\psi_j| \right) U_S^\dag U_{\bs \alpha}^\dag  \right\|_1 \\
				=& \left\|    \left|\widetilde\phi_{\rm ps}\right\>\left\<\widetilde\phi_{\rm bs}\right|       -U_{\widetilde S}^\dag U_{\widetilde{\bs \alpha}}^\dag  U_{\bs \alpha} U_S \left(\bigotimes_{j=1}^n |\psi_j\>\<\psi_j|  \right) U_S^\dag U_{\bs \alpha}^\dag U_{\widetilde{\bs \alpha}} U_{\widetilde S}  \right\|_1 \label{eqf22aa}\\
				\le &  \left\|    \left|\widetilde\phi_{\rm ps}\right\>\left\<\widetilde\phi_{\rm bs}\right|      -U_{ O}\left(\bigotimes_{j=1}^n  |\psi_j\>\<\psi_j|\right) U_{ O}^\dag    \right\|_1 \nonumber \\
				&+  \left\|    U_{ O}\left(\bigotimes_{j=1}^n  |\psi_j\>\<\psi_j|\right) U_{ O}^\dag        -U_{\widetilde S}^\dag U_{ S} \left(\bigotimes_{j=1}^n  |\psi_j\>\<\psi_j|\right) U_{ S}^\dag U_{\widetilde S}  \right\|_1\nonumber  \\
				&+  \left\|  U_{\widetilde S}^\dag U_{ S} \left(\bigotimes_{j=1}^n  |\psi_j\>\<\psi_j|\right) U_{ S}^\dag U_{\widetilde S}      -U_{\widetilde S}^\dag U_{\widetilde{\bs \alpha}}^\dag  U_{\bs \alpha} U_S \left(\bigotimes_{j=1}^n |\psi_j\>\<\psi_j|  \right) U_S^\dag U_{\bs \alpha}^\dag U_{\widetilde{\bs \alpha}} U_{\widetilde S} \right\|_1  
				\label{eqf23aa}\\
				= &  \left\|    \left|\widetilde\phi_{\rm ps}\right\>\left\<\widetilde\phi_{\rm bs}\right|      -U_{ O}\left(\bigotimes_{j=1}^n  |\psi_j\>\<\psi_j|\right) U_{ O}^\dag    \right\|_1 \nonumber \\
				&+  \left\|    U_{\widetilde S}U_O\left(\bigotimes_{j=1}^n  |\psi_j\>\<\psi_j|\right)U_O^\dag U_{\widetilde S}^\dag          - U_{ S} \left(\bigotimes_{j=1}^n  |\psi_j\>\<\psi_j|\right) U_{ S}^\dag  \right\|_1\nonumber  \\
				&+  \left\|   U_{ S} \left(\bigotimes_{j=1}^n  |\psi_j\>\<\psi_j|\right) U_{ S}^\dag      - U_{\widetilde{\bs \alpha}}^\dag  U_{\bs \alpha} U_S \left(\bigotimes_{j=1}^n |\psi_j\>\<\psi_j|  \right) U_S^\dag U_{\bs \alpha}^\dag U_{\widetilde{\bs \alpha}}  \right\|_1 \label{eqf24aa}\\
				\le & \, 2\epsilon_{\rm ps} %+\map G \left( N_{{\rm I}}'', \left\| O-\widetilde O  \right\|_\infty \right)
				+2\sqrt{\sqrt{2m}\left(\sqrt 6 +\sqrt{10}+5\sqrt 2 m\right)\left(mN_{\rm I}^* +1\right)}\left(\sqrt \pi +\sqrt 2(r_s+2)\right)\sqrt{r_s+2} \sqrt{\left\|\widetilde S O-S\right\|_F}\nonumber \\
				&+  2\sin \left(\left(\sqrt{mN_{{\rm I}}'}+\sqrt{mN_{{\rm I}}' +1}\right)\sqrt{\sum_{j=1}^m |\widetilde \alpha_j-\alpha_j|^2}\right)\label{eqf25aa}%\\\le & 2\epsilon_{\rm ps} +\map G \left( mN_{{\rm I}}^*,\|S^{-1}\|_{\infty}\left\|S\bigoplus_{j=1}^n S_j O -   \widetilde S\right\|_\infty \right)+  \frac{\sqrt{mN_{{\rm I}}'}+\sqrt{mN_{{\rm I}}' +1}}{20 \sqrt{E_{\rm II}}}\epsilon_v\label{eqf26aa}
			\end{align}
			where Eqs. (\ref{eqf22aa}) and (\ref{eqf24aa}) are derived by the fact that the one-norm $\|\cdot\|_1$ in invariant under unitary transformation, Eq.~(\ref{eqf23aa}) is obtained by using the triangle inequality, %the first term of Eq.~(\ref{eqf23aa}) refers to the difference between the reconstructed state from local state tomography and its ideal state $\rho_i$ at the corresponding step, the second term denotes the difference between $\rho_i$ and an ideal reconstructed state $\rho_{ii}$ with only error in beam-splitter identification, the third term represents the difference between $\rho_{ii}$ and an ideal reconstructed state $\rho_{iii}$ with error in both beam-splitter identification and 
			the first term of Eq.~(\ref{eqf25aa}) arises from the definition of the error introduced in Step 3, the second term of Eq.~(\ref{eqf25aa}) is from Theorem 6 of Ref.~\cite{becker2021energy} with $\|S\|_\infty\le r_s $, the definitions of energy are $m N_{{\rm I}}^*=\Tr\left[\left(\sum_{j=1}^m a^\dag_j a_j \right)\left(\bigotimes_{j=1}^n |\psi_j\>\<\psi_j|\right) \right]$ and $m N_{{\rm I}}'=\Tr\left[\left(\sum_{j=1}^m a^\dag_j a_j \right)U_{S} \left(\bigotimes_{j=1}^n |\psi_j\>\<\psi_j|\right)U_{S}^\dag \right]$. 
			
			Further, one can connect the above bound to the errors in Step 1 with the following relations: 
			\begin{align}
				\left\|\widetilde S O -S \right\|_F&\le \sqrt {2m} \left\|\widetilde S O -S \right\|_\infty \\
				&=\sqrt {2m} \left\|O^T\widetilde S^T  -S^T \right\|_\infty\\
				&=\sqrt {2m} \left\|\Omega O^T\widetilde S^T \Omega^T -\Omega S^T \Omega^T\right\|_\infty\\
				&=\sqrt {2m} \left\| O^{-1}\widetilde S^{-1}  - S^{-1} \right\|_\infty\\
				&\le 4\sqrt {2m} \left(\sqrt{\left\|V\right\|_\infty\left\|V^{-1}\right\|_\infty } + \frac{\sqrt{m^3 \|V\|_\infty /\|V^{-1}\|_\infty  }}{2\Delta}\right)\left\|V^{-1/2}\right\|_\infty \sqrt{\left\|\widetilde V-V\right\|_\infty }\\
				&\le 4\sqrt {2m} \left(\sqrt{4m E_{\rm I} } + \frac{2m^{\frac 5 2 }E_{\rm I}}{\Delta}\right) \sqrt{\epsilon_v}\\
				\sqrt{\sum_{j=1}^m |\widetilde \alpha_j-\alpha_j|^2}&= \frac 1 2 \|\widetilde {\bs\xi}-\bs \xi\|_2\le \frac{\epsilon_v}{40\sqrt{mE_{\rm II}}}
			\end{align}
			where the second inequality arises from the perturbation bound for symplectic transformation (Proposition 4.1 of Ref.~\cite{idel2017perturbation}), the third inequality follows from the relations $\|V\|_\infty \le \Tr [V]\le 4 mE_{\rm I}$ and $(4mE_{\rm I})^{-1}\le \|V^{-1}\|_\infty \le \|V^{-\frac 1 2 }\|_\infty \le 1$, $\Delta =\min_{k\neq j\atop \lambda_k\neq \lambda_j } \left|\lambda_j-\lambda_k\right|$ is the nonzero symplectic eigen-gap of the covariance matrix $V$.

	By using Lemma \ref{lem: unify energy} and the relation $E_{\rm II}\ge E_{\rm I}$, we have the following approximation: 
	\begin{align}
		\begin{cases}
			E_{\rm I}&\to \mathcal O(E_{\rm II})\\
			N_{\rm I}^*&\to \mathcal O(E_{\rm II})\\
			N_{\rm I}'&\to \mathcal O(E_{\rm II})\\
			r_s&\to \mathcal O(\sqrt{4mE_{\rm II}})\\
			\Delta &\to \mathcal O(1)
		\end{cases}.
	\end{align}

	Finally, the total error is: 
	\begin{align}\label{total error}
		\epsilon\le &\epsilon_{\rm ps} 
		+\sqrt{\sqrt{2m}\left(\sqrt 6 +\sqrt{10}+5\sqrt 2 m\right)\left(mN_{\rm I}^* +1\right)}\left(\sqrt \pi +\sqrt 2(r_s+2)\right)\sqrt{r_s+2} \nonumber \\
		&\times \sqrt{4\sqrt {2m} \left(\sqrt{4m E_{\rm I} } + \frac{2m^{\frac 5 2 }E_{\rm I}}{\Delta}\right) \sqrt{\epsilon_v}}\nonumber \\
		&+  \left(\sqrt{mN_{{\rm I}}'}+\sqrt{mN_{{\rm I}}' +1}\right)\frac{\epsilon_v}{40\sqrt{mE_{\rm II}}}\\
		=&  \epsilon_{\rm ps} + \mathcal O \left(m^{3.5 },E_{\rm II}^{1.75}, \Delta^{-0.5},\epsilon_v^{0.25} \right) .\label{eq181}
	\end{align}
	
	Now, let us look at the total number of copies needed in learning a pure GE state. %In  identifying the beam-splitter operation with uniform sampling for each matrix element of $O$, the experiment round  can be bounded as: \begin{align}\nu &=\frac{4m^2}{\left(\min_{jk} \left|\widetilde O_{jk}-O_{jk}\right| \right)}\\&\ge \frac{4m^2 }{\min_{\widetilde O} \|\widetilde O-O\|_\infty}     \\&\gtrapprox \frac{4m^2 }{\map G^{-1}\left(mN_{{\rm I}}'',2\sum_{j=1}^n \epsilon_j \right)} \label{eqf41aa}\end{align}where Eq.~(\ref{eqf41aa}) uses the result of \cite{becker2021energy} and Eqs. (\ref{eqf27aa})-(\ref{eqf29aa}), $x=\map G^{-1} \left( N,x' \right)$ is the inverse function of  $x'=\map G \left( N,x \right)$ defined in Eq.~(\ref{eqf30aa}), $N_{{\rm I}}''$ is a photon number defined by $mN_{{\rm I}}''=\Tr\left[\left(\sum_{j=1}^m a^\dag_j a_j \right)\bigotimes_{j=1}^n U_{S_j}^\dag |\psi_{\bs 0}\>\<\psi_{\bs 0}|_jU_{S_j}\right]$. Here, for algebraic convenience, we assume that the energy $E_{{\rm II},j}'$ is a constant in different rounds of the experiment.   
	%Therefore, 
	The total number of copies needed is: 
	\begin{align}
		M=&M_v+ M_{\rm ps} \label{eq126ppp}\\
		=&(m+3)\left\lceil 68\log_2 \left(\frac{2(2m^2+3m)}{\delta_v}\right)\frac{200(8m^2 E_{\rm II}^2 +3m)}{\epsilon_v^2} \right\rceil\nonumber \\
		&+  \mathcal O\left[\textbf{poly}\left( n^{5A_{\max}+6}, N_{\rm I}^{*2A_{\max}+1},\left(\frac 1 {\epsilon_{\rm ps}}\right)^{2A_{\max}+4},\log \frac 1 {\delta_{\rm ps} }, 3^{2A_{\max}} \right) \right], 
		\label{total copies for learning}
	\end{align}
	where $A_{\max}$ is the maximal size of subsystems that can not be further decomposed by Gaussian unitary. 
	
	At last, the total probability of failure is: 
	\begin{align}
		\delta &=1- (1-\delta_v)(1-\delta_{\rm ps}) \\
		&\sim \delta_v+\delta_{\rm ps}\label{total failure probability}.
	\end{align}
	
	\subsection{Numerical simulation of the tomography process with a finite number of modes}\label{app:numerical simulation of the tomography process}
	
	We conducted a numerical simulation of the tomography process for two-mode GE states with the true state being a two-mode thermal state correlated by a beam-splitter. Prior to the beam-splitter operation, the uncorrelated thermal states have photon numbers of 0.2 and
	0.3 in each mode, respectively.
	
	In the simulation experiment, the infinite-dimensional Hilbert space is truncated to a finite subspace of dimension ten for each mode. The error is defined by the trace distance between the true state and the reconstructed state obtained from the tomography process. Direct tomography and local tomography in our protocol are simulated mostly following the standard $R\rho R$  algorithm~\cite{lvovsky2009continuous}, except that the measurement operators are random homodyne operators.  Within each round of  $R\rho R$ algorithm, twenty iterations are performed. To account for randomness in the measurement operators and Gaussian gate $U^g$, multiple instances are simulated to compute the mean error. In the Gaussian-Disentangling protocol, the total number of copies includes those required to estimate the covariance matrix and displacement, as well as those needed for local tomography. Various combinations of copy usage are tested and plotted with cross markers in Fig.~\ref{fig:2mode_thermal_state_tomography_total}. Finally, the lower envelope of the achievable error can be obtained after multiple rounds of the Gaussian-Disentangling protocol (see the red dashed line in Fig.~\ref{fig:2mode_thermal_state_tomography_total}). %\PC{In the Gaussian-Disentangling scheme, the number of copies includes the ones needed to estimate the covariance matrix and the ones that needed to perform $R\rho R$  tomography. Different copies combinations are experimented and plotted in the Fig~, whereas Gaussian-Disentangling line is obtained as the lower envelop. }
	
	As shown in Fig.~\ref{fig:2mode_thermal_state_tomography_total}, both a lower error value and a clear scaling advantage over direct tomography are identified. This result indicates that the proposed Gaussian disentangling protocol offers advantages even in the scenario where the number of modes is finite.

	\begin{figure}[t]
		\centering
		\includegraphics[width=0.5\linewidth,trim=5 5 5 5,angle=0,clip]{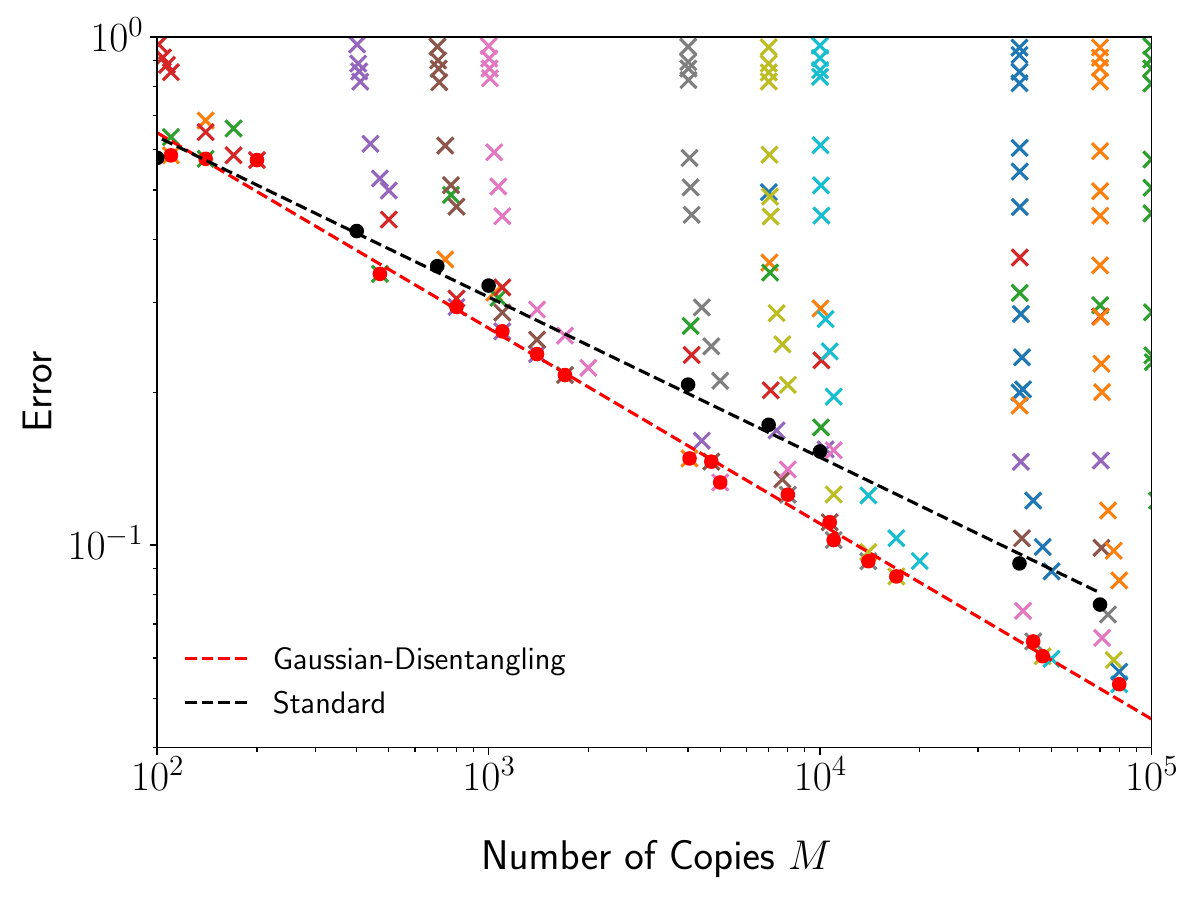}
		\caption{
			Comparative advantage of the proposed Gaussian-disentangling protocol over direct tomography. Here, we numerically simulated the quantum state tomography process for a two-mode GE state, where the true state is a thermal state correlated through a beam-splitter. Before the beam-splitter operation, the uncorrelated thermal states have photon numbers of 0.2 and 0.3 in each mode, respectively. The error, defined as the trace distance between the true state and the reconstructed state obtained through direct tomography using the standard algorithm~\cite{lvovsky2009continuous}, is shown as a function of the number of copies $M$ used in tomography and is shown by black dots. The black dashed line represents the linear fit to the black dots with a slope $k_1=-0.38$. The error achieved by the proposed Gaussian-disentangling protocol, for a specific sample number $M_v$ used to estimate displacements and the covariance matrix, is represented by cross markers of the same color. At last, all these errors form a lower envelope, shown as red dots connected by a red dashed line. A linear fit to this lower envelope yields a slope of $k_2 = -0.32$.
		}\label{fig:2mode_thermal_state_tomography_total}
	\end{figure}

	%Numerical simulations are conducted on a two-mode bosonic system to directly compare the performance of NGE tomography and the $R\rho R$ tomography algorithm~\cite{lvovsky2009continuous} across the entire system (see Fig.~\ref{fig:2mode_thermal_state_tomography} in the main text). The local non-Gaussian states are thermal states with mean photon numbers of 0.2 and 0.3, respectively, while the Gaussian gate $U^g$ is implemented as a random beam splitter. The bosonic Hilbert space is truncated to a maximum photon number of 10. For the $R\rho R$ algorithm, 20 iterations are performed. To account for randomness in the measurement operators and Gaussian gate $U^g$, multiple instances are simulated to compute the mean error.

	\subsection{Proof of Proposition \ref{lem:sample complexity upper bound} for large NGE states}\label{app:connection between GE cost and learning}
	
	In this Supplementary Note, we build up the connection between the GE cost and the sample complexity for a general NGE state. In general, the pure state space  $\bigotimes_{j=1}^m\map H_j$ can be divided into subsystems marked by $A_1,\cdots,A_n$. A general $m$-mode NGE state $\psi$ could belong to the set of GE states $\mathbb{GE} \left(\bigotimes_{j=1}^m\map H_j;\{B_{1}|\cdots|B_{\ell}\}\right)$ where $B_k,k=1,\cdots,\ell$ represents the union of subsystems $\{A_1,\cdots,A_n\}$ such that $\cup_{k=1}^\ell B_k=\cup_{j=1}^n A_j$ and $B_k\cap B_{k'}=\emptyset$. 
	
	For simplicity of analysis, let us consider a case with $|A_j|=1,\forall j=1,\cdots,n$ and $n=m$. For any pure NGE state $|\psi\>$, 
	one can consider an extension of the subsystem $B_k$ by adding $(|B_k|-1)$ vacuum ancillary states to its first mode. Then, one can apply an $(|B_k|-1)$-mode swap gate --- a specific Gaussian operation --- to both the $j$-th subsystem except the first mode and its corresponding ancillay states. The resulting state will be a $(2n-\ell)$-mode GE state regarding the initial partition $\{A_1|\cdots|A_n\}$. Given Definition \ref{defi:GE cost} of the main text, this NGE state has a GE cost:    
	\begin{align}
		\map R &=  \left(|B_1'|-1,|B_2'|-1,\cdots, |B_\ell'|-1\right)^T,
	\end{align}
	where $\{B_1',B_2',\cdots,B_\ell'\}$ is a permutation of $\{B_1,B_2,\cdots,B_\ell\}$ satisfying the condition $|B_1'|\ge |B_2'|\ge \cdots\ge |B_\ell'|$. 
	
	\subsubsection{Upper bound}
	
	The upper bound of the sample complexity can be given by any tomography protocol. Note that the state $\psi$ is a GE state regarding the system partition $\{B_1|\cdots|B_\ell\}$. If $\psi$ is a pure state with finite energy moment $E_{\rm II}$, we can apply the conclusions from Note \ref{app:tomography algorithm}. Without losing the generality, we assume the two terms in Eq.~(\ref{eq181}) are of the same level, i.e., $\frac \epsilon 2\sim \epsilon_{\rm ps} \sim  \mathcal O \left(m^{3.5 },E_{\rm II}^{1.75}, \Delta^{-0.5},\epsilon_v^{0.25} \right)$. Then, with Eq.~(\ref{total failure probability}), the first term of Eq.~(\ref{total copies for learning}) can be rewritten as $\mathcal O \left[\textbf{poly}\left(m, E_{\rm II},\frac 1 \epsilon,\log \frac 1 \delta \right)\right]$. In the case with $B_{\max}>7$, the scaling of sample complexity is dominated by the second term in Eq.~(\ref{total copies for learning}). In this case, we have a sample complexity upper bound of learning $\psi$: 
	\begin{align}
		\left.M\right|_{B_{\max}>7}\le &\,\mathcal O \left[\textbf{ poly}\left( \ell^{5B_{\max}+6}, E_{\rm II}^{2B_{\max}+1},\left(\frac 1 {\epsilon_{\rm ps}}\right)^{2B_{\max}+4},\log \frac 1 {\delta_{\rm ps} } , 3^{2B_{\max}}\right) \right]\label{eqk3}%\\
		%&\mathcal O\left[\textbf{\em poly}\left( n^{5A_{\max}+6}, N_{\rm I}^{*2A_{\max}+1},\left(\frac 1 {\epsilon_{\rm ps}}\right)^{2A_{\max}+4},\log \frac 1 {\delta_{\rm ps} },A_{\max}^{A_{\max}+1}, 3^{2A_{\max}} \right) \right]
	\end{align}
	%where $\epsilon_{\rm lo}:=\min_j \epsilon_j$ denotes the minimal error, $\delta_{\rm lo}:=\min_j \delta_j$ is the minimal failure probability of local tomography, respectively, the average energy moment after disentangling $E_{{\rm II},j}$ defined by Eq.~(\ref{eq101aaa}) is bounded $E_{{\rm II},j}\le\mathcal O(E_{\rm II})$ defined by Eq.~(\ref{eq87ccc}), the photon number after disentangling $N_{\rm I}''$ defined in Eq.~(\ref{eqf41aa}) is  bounded $N_{\rm I}''\le\mathcal O(E_{\rm I})$ without loss of generality, $\underline\Omega$ represents the big Omega notation.
	where $B_{\max}=\max_k |B_k|$ is the maximal size of subsystems that cannot be decomposed by a Gaussian unitary, $N_{\rm I}^*$ is substituted by $E_{\rm II}$ due to the relation $N_{\rm I}^*\le E_{\rm I}\le E_{\rm II}$ (see Lemma \ref{lem: unify energy}). 
	%Meanwhile, any state set with a GE cost atom-wisely larger than $\map R$ will require a sample complexity lower bound in Eq.~(\ref{eqk3}). 
	Furthermore, given that the second and third terms of Eq.~(\ref{total error}) vanish exponentially faster with the same samples, %and that we are computing an upper bound based a specific design of tomography algorithm, 
	we can assume: 
	\begin{align}
		\epsilon\approx \epsilon_{\rm ps}.  
	\end{align}
	Without loss of generality, %as the relation $\nu\approx \underline \Omega (m^{9/2}E_{\rm I}/(\ell^2\epsilon_{\rm lo}^2))$ holds and 
	we can omit the failure probability of estimating the covariance matrix and  
	have: 
	\begin{align}
		\delta\approx \delta_{\rm ps}. 
	\end{align}
	Thus, we have: 
	\begin{align}
		%M\le &\,\underline\Omega  \left[ m^{\frac 3 2}\left(1-\frac{\delta\epsilon^2}{m^{\frac 7 2}\ell}\right)\cdot \sum_{k=1}^\ell  \left(\frac{\epsilon}{2\ell}\right)^{-|B_k|-2}\right].\\
		M\le &\,\mathcal O \left[\textbf{ poly}\left( \ell^{5B_{\max}+6}, E_{\rm II}^{2B_{\max}+1},\left(\frac 1 {\epsilon}\right)^{2B_{\max}+4},\log \frac 1 {\delta } ,3^{2B_{\max}} \right) \right]\label{eql5},
	\end{align}
	in the conditions $m=\mathcal O(n)$ and $B_{\max}>1$. 
	%Here, one can also substitute the quantity $E_{\rm I}$ by $E_{\rm II}$ due to the relation $E_{\rm I}\le E_{\rm II}$. 
	
	Note that the $m$-mode pure GE states in the tomography protocol must be constrained by the following conditions: 
	\begin{align}
		\begin{cases}
			\sqrt{\Tr\left[\left(\sum_{j=1}^m a^\dag_j a_j+\frac m 2 I \right)^2|\psi\>\<\psi|\right]}&\le m E_{\rm II}\\
			\Tr\left[\left(\sum_{j=1}^m a^\dag_j a_j\right)|\psi_{\rm ps}\>\<\psi_{\rm ps}|\right]&\le m N_{\rm I}^*
		\end{cases},
	\end{align}
	where $|\psi_{\rm ps}\>$ denotes the post-counter-rotation state. Given Lemma \ref{lem: unify energy}, the second condition above can be further simplified as the first condition due to the relation $N_{\rm I}^*=\mathcal O(E_{\rm II})$. 
	%or a more restrictive condition: \begin{align}\Lambda & \ge \left(2 N_{\rm I}^*+1\right)I, \end{align}where $\Lambda$ is the diagonal covariance matrix of $\otimes_j |\psi_j\>$ defined by $V=S\Lambda S^T$. 
	
	In the case where $B_{\max}\le 7$, the first term of Eq.~(\ref{total copies for learning}) will dominate the scaling of sample complexity.

	\subsubsection{Lower bound}

	The sample complexity of state tomography can be lower bounded by a more favorable scenario where the multi-mode Gaussian unitary is known. With this premise, the tomography task becomes tomography of all local states $|\psi_k\>$ for $k=1,\cdots,\ell$ as one can conduct a perfect counter rotation. Using Theorem S26 in Ref.~\cite{mele2024learning} and assume an average energy moment $E_{\rm II}$ per mode for each local state as a sufficient condition due to $\sqrt{\<\left(\sum_j a^\dag_j a_j +m/2\right)^2\>}\le \sum_j \sqrt{\<(a_j^\dag a_j+1/2)^2\>}$, the sample complexity in tomography of state $|\psi_k\>$ is bounded as: 
	%\begin{align}M_k&=\Theta \left(\epsilon_j^{-|B_k|-2}\right).\end{align}
	\begin{align}
		M_k&=\Theta \left[\frac{1-\delta_k}{|B_k| \log E_{{\rm II}}}\left(\frac{E_{{\rm II}}}{\epsilon_k}\right)^{|B_k|}\right],
	\end{align}
	where $\epsilon_k$ denotes the error, $\delta_k$ refers to the failure probability of reconstructing the state $|\psi_k\>$, respectively. 
	
	The total number of sample complexity is lower bounded as: 
	%\begin{align}M&\ge \max_k M_k \\&\ge \min_{\substack{\{\epsilon_k,k=1,\cdots,\ell\}\\\sum_{k=1}^\ell\epsilon_k\ge \epsilon\\ \epsilon_k\le \epsilon,\forall k=1,\cdots,\ell}}\max_k \Theta \left(\epsilon_k^{-|B_k|-2}\right)\label{eql8}\\&\ge \min_{\substack{\{\epsilon_k,k=1,\cdots,\ell\}\\\sum_{k=1}^\ell\epsilon_k\ge \epsilon\\ \epsilon_k\le \epsilon,\forall k=1,\cdots,\ell}}\Theta \left(\epsilon_k^{-\max_k  |B_k|-2}\right)\\&\ge \Theta \left(\epsilon^{-\max_k  |B_k|-2}\right)\label{eql10}\end{align}
	\begin{align}
		M&\ge  \max_k M_k \\
		&\ge \min_{\substack{\{\epsilon_k,k=1,\cdots,\ell\}\\\sum_{k=1}^\ell\epsilon_k\ge \epsilon\\ \epsilon_k\le \epsilon,\forall k=1,\cdots,\ell}}\max_k \Theta \left[\frac{1-\delta_k}{|B_k|\log E_{{\rm II}}}\left(\frac{E_{\rm II}}{\epsilon_k}\right)^{|B_k|}\right]\label{eql8}\\
		&\ge \min_{\substack{\{\epsilon_k,k=1,\cdots,\ell\}\\\sum_{k=1}^\ell\epsilon_k\ge \epsilon\\ \epsilon_k\le \epsilon,\forall k=1,\cdots,\ell}}\Theta \left[\frac{1-\delta_k}{n\log E_{{\rm II}}}\left(\frac{E_{\rm II}}{\epsilon_k}\right)^{\max_{k'} |B_{k'}|}\right]\\
		&\ge \Theta \left[\frac{1-\delta}{m\log E_{{\rm II}}}\left(\frac{E_{\rm II}}{\epsilon}\right)^{B_{\max}}\right]\label{eql10}
	\end{align}
	where the restriction $\sum_{k=1}^\ell\epsilon_k\ge \epsilon$ is given by Eq.~(\ref{eq107aa}), $\epsilon_k\le \epsilon,\forall k=1,\cdots,\ell$ is given by the data processing inequality, the last inequality also uses the deta-processing inequality for failure probability $\delta\ge \delta_j$.

	Finally, it is quick to extend the discussion from the case where the number of modes is equal to the number of partitions $m=n$ to $m>n$. The scaling of the sample complexity bounds will remain the same. 
	
	\subsection{Proof of Theorem \ref{theo:sample complexity of GE state} for pure GE states}\label{app:proof of theorem 2 for pure GE states}
	
	Here we consider the tomography of pure GE states $\{|\psi\>\}$ generated by applying a Gaussian unitary $U^g$ to a product of single-mode states $\{\otimes_{j=1}^m |\psi_j\>\}$. Without losing generality, we assume that each local state has zero displacement and a diagonal covariance matrix. Meanwhile, the state satisfies a single square-energy constraint: 
	\begin{align}
		\sqrt{\Tr\left[\left(\sum_{j=1}^m a^\dag_j a_j+\frac m 2 I \right)^2|\psi\>\<\psi|\right]}&\le m E_{\rm II}.
	\end{align}
	
	Then, one can obtain a lower bound of sample complexity by assuming that the Gaussian unitary $U^g$ is known and imposing a sufficient condition that each local state has a square energy upper bound $E_{\rm II}$ per mode due to $\sqrt{\<\left(\sum_j a^\dag_j a_j +m/2\right)^2\>}\le \sum_j \sqrt{\<(a_j^\dag a_j+1/2)^2\>}$. {\color{black}  Similar to the proof of Theorem S26 in Ref.~\cite{mele2024learning}, one can construct an $\epsilon$-packing net for the set of product states $\{\otimes_{j=1}^m|\psi_j\>\}$, with each state of the product having a second moment of the energy bounded by $E_{\rm II}$. Using the lower bound of Holevo information for all possible states (see Lemma S14 of Ref.~\cite{mele2024learning}) and the upper bound of Holevo information proportional to number of copies, one can quickly establish a lower bound for the sample complexity of learning product states $\{\otimes_{j=1}^m|\psi_j\>\}$ that is proportional to the number of modes $m$:
		\begin{align}
			M&\ge \Theta \left[\frac{1-\delta}{\log E_{{\rm II}}}\frac{mE_{\rm II}}{\epsilon}\right]. 
		\end{align}
	}
	
	On the other hand, the upper bound can be achieved by applying our protocol introduced in Note \ref{app:learning GE states} with the exception that the photon number of local states $|\psi_j\>,\forall j$ is bounded by a positive number $N_j$ with $\sum_j N_j=mN_{\rm I}^*\le mE_{\rm I}\le mE_{\rm II}$ where the first inequality follows from $E_{\rm II}\ge E_{\rm I}$, the second inequality uses 
	Lemma \ref{lem: unify energy}. Thus, the error of  for Step 3 will be: 
	\begin{align}
		\frac 1 2 \left\||\psi_{\rm ps}\>\<\psi_{\rm ps}|-|\psi_{\rm ps}'\>\<\psi_{\rm ps}'|\right\|_1%\le & \sqrt{\pi m \left(\sqrt 6+\sqrt{10}+5\sqrt 2 m \right)\left(mN_{\rm I}^*+1\right)\epsilon_2} \nonumber \\&+\sum_{\{N_j\}\atop N_j\ge 0,\sum_{k}N_k=mN_{\rm I}^*}\sqrt{\frac{ N_j}{K+1}}+ m(K+1)^{1/2} \,\epsilon_1+\frac {m(K+1) }2\epsilon_1^2\\
		\le & \sqrt{\pi m \left(\sqrt 6+\sqrt{10}+5\sqrt 2 m \right)\left(mE_{\rm II}+1\right)\epsilon_2} \nonumber \\
		&+m\sqrt{\frac{ E_{\rm II}}{K+1}}+ m(K+1)^{1/2} \,\epsilon_1+\frac {m(K+1) }2\epsilon_1^2.
	\end{align}
	%where the second inequality follows from the Cauchy-Schwarz inequality $  \sum_{j=1}^m \sqrt{ A_j}\le \sqrt{m \sum_{j=1}^m A_j}$.

	If we assume the first, second, and the third terms of the above equation has the same level of contribution to error $\sim \epsilon/3$, we have: 
	\begin{align}
		\begin{cases}
			K+1&=\frac{\sqrt{E_{\rm II}}}{\epsilon_1}\\
			\frac{2}{\epsilon_1}&=\frac{18m^2E_{\rm II}^{\frac 1 2 }}{\epsilon^2}
		\end{cases}.
	\end{align}
	Thus, the number of observables in the tomography will be
	\begin{align}
		L&= m\left(\frac{18m^2E_{\rm II}^{1/ 2 }}{\epsilon^2}\right)^{\frac{18m^2E_{\rm II}}{\epsilon^2}}\left(\frac {18\pi m \left(\sqrt 6+\sqrt{10}+5\sqrt 2 m \right)\left(mE_{\rm II}+1\right)} {\epsilon^2}\right)^{4m^2}\\
		&=\left[\mathcal O\left(\frac{m^{3/2} E_{\rm II}^{1/2}}{\epsilon}\right)\right]^{{\textbf{poly}} \left(m^{2}, E_{\rm II}, \left(\frac{1}{\epsilon}\right)^{2}\right)}.
	\end{align}

	Thus, the tomography algorithm Step 3 for this GE states will have a sample complexity: 
	\begin{align}
		M_{\rm ps}&= \mathcal O\left[\textbf{poly}\left(m^5, E_{\rm II}^2,\left(\frac 1 {\epsilon_{\rm ps}}\right)^6,\log \frac 1 {\delta_{\rm ps}} \right)\right].
	\end{align}
	In this mean time, by assuming the second term of Eq.~(\ref{eq181}) has a level $\sim\epsilon/2$, we know that the copies estimating the covariance matrix $M_v$ is a polynomial function of $m$, $E_{\rm II}$, and $\epsilon$. Therefore, we have Theorem \ref{theo:sample complexity of GE state} of the main text. 
	
	\subsection{Proof of Theorem \ref{lem:connection GE cost and learning} for the nondegenerate case}\label{app:proof of theorem 4}
	
	If the symplectic eigenvalues of the input NGE or generic GE state (defined in Note \ref{app:connection between GE cost and learning}) are nondegenerate, we can omit the fidelity-witness procedure in Step 3 of Note \ref{app:tomography algorithm}. Instead, one can directly conduct local tomography algorithm as the state will be fully disentangled in Step 2. Under this assumption, the overall tomography error in Eq.~(\ref{eq181}) is modified as follows:
	\begin{align}
		\epsilon \le & \, \frac 1 2 \left\|\bigotimes_{j=1}^\ell \left|\widetilde \psi_j\right\>\left\<\widetilde \psi_j\right|-\bigotimes_{j=1}^\ell |\psi_j\>\<\psi_j|\right\|_1   + \mathcal O \left(m^{3.5 },E_{\rm II}^{1.75}, \Delta^{-0.5},\epsilon_v^{0.25} \right) \\
		\le &\frac 1 2 \sum_{j=1}^\ell\left\| \left|\widetilde \psi_j\right\>\left\<\widetilde \psi_j\right|-|\psi_j\>\<\psi_j|\right\|_1 + \mathcal O \left(m^{3.5 },E_{\rm II}^{1.75}, \Delta^{-0.5},\epsilon_v^{0.25} \right)\\
		=&\sum_{j=1}^\ell \epsilon_j+\mathcal O \left(m^{3.5 },E_{\rm II}^{1.75}, \Delta^{-0.5},\epsilon_v^{0.25} \right).\label{eq206}
	\end{align}
	where the second inequality follows from Eq.~(\ref{eq107aa}), $\{\left|\widetilde \psi_j\right\>\}$ refer to the reconstructed local states, $\{\epsilon_j\}$ denote the error of local tomography for states $\{|\psi_j\>\}$, respectively. 
	
	Without loss of generality, we can consider a tomography protocol with  $\ell\epsilon_j\sim m^{3.5}E_{\rm II}^{1.75}\epsilon_v^{0.25} \sim \mathcal O(\epsilon)$. Then, the sample complexity in Eq.~(\ref{total copies for learning}) will be upper bounded as: 
	\begin{align}
		M\le &(m+3)\left\lceil 68\log_2 \left(\frac{2(2m^2+3m)}{\delta_v}\right)\frac{200(8m^2 E_{\rm II}^2 +3m)}{\epsilon_v^2} \right\rceil+ \sum_{j=1}^\ell \mathcal O \left[\left(\frac{E_{{\rm II},j}}{\epsilon_j}\right)^{|B_j|}\right]\\
		\le &(m+3)\left\lceil 68\log_2 \left(\frac{2(2m^2+3m)}{\delta_v}\right)\frac{200(8m^2 E_{\rm II}^2 +3m)}{\epsilon_v^2} \right\rceil+ \mathcal O \left[\left(\frac{m\ell E_{{\rm II}}}{ \epsilon}\right)^{B_{\max}}\right]\\
		=& \textbf{poly}\left(\ell,E_{\rm II},\frac{1}{\epsilon},\log \frac 1 \delta  \right)\cdot \Theta \left[\left(\frac{E_{\rm II}}\epsilon\right)^{B_{\max}}\right]
	\end{align}
	where the second terms of the first inequality comes from the optimized algorithm in Ref.~\cite{mele2024learning}, $\{E_{{\rm II},j}\}$ denote the square energy of the local states satisfying the upper bound $E_{{\rm II},j}\le m E_{\rm II}$, $\{|B_j|\}$ denote the number of modes for $j$-th subsystem with its maximal value $\max_j |B_j|=B_{\max}$. 
	
	Combining with the lower bound is shown in Eq.~(\ref{eql10}), we have Theorem \ref{lem:connection GE cost and learning} of the main text.

\end{widetext}

\end{document}